\documentclass[useAMS,usenatbib]{mnras}
\usepackage[english]{babel}
\usepackage[latin1]{inputenc}
\usepackage{amssymb}
\usepackage[fleqn]{amsmath}
\usepackage{times}
\usepackage{graphicx}
\usepackage{epstopdf}
\usepackage{wasysym}
\usepackage{xcolor}
\usepackage{hyperref}

\def\aap{A\&A}
\def\apj{ApJ}

\def\apjl{ApJ}
\def\mnras{MNRAS}
\def\araa{ARA\&A}

\def\apss{Ap\&SS}      
\def\apjs{ApJS}
\def\pasp{PASP}

\newcommand{\mincir}{\raise
-2.truept\hbox{\rlap{\hbox{$\sim$}}\raise5.truept 
\hbox{$<$}\ }}
\newcommand{\magcir}{\raise
-2.truept\hbox{\rlap{\hbox{$\sim$}}\raise5.truept
\hbox{$>$}\ }}
\newcommand{\minmag}{\raise-2.truept\hbox{\rlap{\hbox{$<$}}\raise
6.truept\hbox
{$>$}\ }}

\newcommand{\be}{\begin{equation}}
\newcommand{\ee}{\end{equation}}
\newcommand{\ba}{\begin{eqnarray}}
\newcommand{\ea}{\end{eqnarray}}
\newcommand{\brr}{\begin{array}}
 
\newcommand{\err}{\end{array}}
\newcommand{\bc}{\begin{center}}
\newcommand{\ec}{\end{center}}

\DeclareMathAlphabet{\mathsc}{OT1}{cmr}{m}{sc}
\def\testbx{bx}%
\DeclareRobustCommand{\ion}[2]{%
\relax\ifmmode
\ifx\testbx\f@series
{\mathbf{#1\,\mathsc{#2}}}\else
{\mathrm{#1\,\mathsc{#2}}}\fi
\else\textup{#1\,{\mdseries\textsc{#2}}}
\fi}

\title[Galactic outflows at low redshift: SAMI vs EAGLE]{The SAMI
  Galaxy Survey: understanding observations of large-scale outflows at
  low redshift with EAGLE simulations}

\author[E. Tescari et al.]{E. Tescari$^{1,2,}$\footnote{},
  L. Cortese$^{3}$, C. Power$^{2,3}$, J. S. B. Wyithe$^{1,2}$,
  I.-T. Ho$^{4,5,6}$, R. A. Crain$^7$,
  \newauthor{J. Bland-Hawthorn$^{2,8}$, S. M. Croom$^{2,8}$,
    L. J. Kewley$^5$, J. Schaye$^9$, R. G. Bower$^{10}$,}
  \newauthor{T. Theuns$^{10}$, M. Schaller$^{10}$, L. Barnes$^8$,
    S. Brough$^{2,11}$, J. J. Bryant$^{2,8,11}$, M. Goodwin$^{11}$,}
  \newauthor{M. L. P. Gunawardhana$^{10}$, J. S. Lawrence$^{11}$,
    S. K. Leslie$^{2,5,6}$, \'A. R. L\'opez-S\'anchez$^{11,12}$,}
  \newauthor{N. P. F. Lorente$^{11}$,
    A. M. Medling$^{5,13,\triangledown}$,
    S. N. Richards$^{2,8,11}$, S. M. Sweet$^{5}$ and C. Tonini$^1$} \\
  \\ $^1$ School of Physics, The University of Melbourne, Parkville,
  VIC 3010, Australia \\ $^2$ ARC Centre of Excellence for All-Sky
  Astrophysics (CAASTRO) \\ $^3$ International Centre for Radio
  Astronomy Research (ICRAR), The University of Western Australia, 35
  Stirling Highway, Crawley, WA 6009, Australia \\ $^4$ Institute for
  Astronomy, University of Hawaii, 2680 Woodlawn Drive, Honolulu, HI
  96822, USA \\ $^5$ Research School of Astronomy and Astrophysics,
  Australian National University, Cotter Road, Weston Creek, ACT 2611,
  Australia \\ $^6$ Max Planck Institute for Astronomy, K\"onigstuhl
  17, 69117 Heidelberg, Germany \\ $^7$ Astrophysics Research
  Institute, Liverpool John Moores University, 146 Brownlow Hill,
  Liverpool L3 5RF, UK \\ $^8$ Sydney Institute for Astronomy, School
  of Physics, University of Sydney, NSW 2006, Australia \\ $^9$ Leiden
  Observatory, Leiden University, PO Box 9513, NL-2300 RA Leiden, the
  Netherlands \\ $^{10}$ Institute for Computational Cosmology,
  Department of Physics, Durham University, South Road, Durham DH1
  3LE, UK \\ $^{11}$ Australian Astronomical
  Observatory, PO Box 915, North Ryde, NSW 1670, Australia \\
  $^{12}$ Department of Physics and Astronomy, Macquarie University,
  NSW 2109, Australia \\ $^{13}$ Cahill Center for Astronomy and
  Astrophysics, California Institute
  of Technology, MS 249-17 Pasadena, CA 91125, USA \\ $^\triangledown$ Hubble Fellow \\
  \\ \color{blue} $^\star$ \color{black} E-mail:
  \color{blue}edoardo.tescari@unimelb.edu.au \\}

\begin{document}

\maketitle

\begin{abstract}

  This work presents a study of galactic outflows driven by stellar
  feedback. We extract main sequence disc galaxies with stellar mass
  $10^9\le$ M$_{\star}/$M$_{\odot} \le 5.7\times10^{10}$ at redshift
  $z=0$ from the highest resolution cosmological simulation of the
  Evolution and Assembly of GaLaxies and their Environments (EAGLE)
  set. Synthetic gas rotation velocity and velocity dispersion
  ($\sigma$) maps are created and compared to observations of disc
  galaxies obtained with the Sydney-AAO Multi-object Integral field
  spectrograph (SAMI), where $\sigma$-values greater than $150$ km
  s$^{-1}$ are most naturally explained by bipolar outflows powered by
  starburst activity. We find that the extension of the simulated
  edge-on (pixelated) velocity dispersion probability distribution
  depends on stellar mass and star formation rate surface density
  ($\Sigma_{\rm SFR}$), with low-M$_{\star}/$low-$\Sigma_{\rm SFR}$
  galaxies showing a narrow peak at low $\sigma$ ($\sim30$ km
  s$^{-1}$) and more active, high-M$_{\star}/$high-$\Sigma_{\rm SFR}$
  galaxies reaching $\sigma>150$ km s$^{-1}$. Although
  supernova-driven galactic winds in the EAGLE simulations may not
  entrain enough gas with T $<10^5$ K compared to observed galaxies,
  we find that gas temperature is a good proxy for the presence of
  outflows. There is a direct correlation between the thermal state of
  the gas and its state of motion as described by the
  $\sigma$-distribution. The following equivalence relations hold in
  EAGLE: $i)$ low-$\sigma$ peak $\,\Leftrightarrow\,$ disc of the
  galaxy $\,\Leftrightarrow\,$ gas with T $<10^5$ K; $ii)$
  high-$\sigma$ tail $\,\Leftrightarrow\,$ galactic winds
  $\,\Leftrightarrow\,$ gas with \mbox{T $\ge 10^5$ K}.

\end{abstract}


\section{Introduction}

In the last few years, low redshift galaxy surveys have been
transformed by the advent of integral field spectroscopy (IFS). IFS
enables astronomers to obtain spatially resolved spectroscopic
information across different regions of the same object. This
technological progress has considerably improved our understanding of
various galactic properties (previously estimated using only a single
aperture measurement per target) and changed the way we study and
classify galaxies \citep[see the review by][]{cappellari2016}.

A number of IFS surveys have already been carried out and others are
currently ongoing, e.g. DiskMass \citep{bershady2010},
ATLAS$^{\rm 3D}$ \citep{cappellari2011}, CALIFA \citep{sanchez2012}
and MaNGA \citep{bundy2015}. In this work, we utilize the results of
the SAMI Galaxy Survey \citep{bryant2015}. SAMI is the Sydney-AAO
(Australian Astronomical Observatory) Multi-object Integral field
spectrograph, mounted at the prime focus of the Anglo-Australian
Telescope (AAT) and attached to the AAOmega spectrograph
\citep{sharp2006}. It allows for the simultaneous observations of $12$
objects and one calibration star by means of fibre {\it hexabundles}
\citep{joss2011, bryant2014}, each composed of $61$ optical fibres
fused together for a field of view of $15$ arcsec
\citep{croom2012}. The aim of the survey is to observe $\sim3400$
galaxies in a wide range of stellar masses and environments within the
redshift interval $0.004<z<0.095$. The interested reader can find a
discussion of the SAMI data reduction in \citet{sharp2015}, the early
data release in \citet{allen2015} and the data release one details in
\citet{green2017}. One of the key scientific drivers is the study of
feedback processes related to star formation activity.

Galactic winds affect galaxies and their surrounding environments, by
regulating the star formation rate (SFR) and therefore shaping the
luminosity function, and by enriching the intergalactic medium
\citep[see the reviews by][]{veilleux2005,joss2007}. Bipolar
stellar$/$supernova (SN) driven outflows appear to be ubiquitous at
high redshift \citep{shapley2011}, while at low $z$ they are readily
detected only in starburst galaxies \citep{heckman2015}. Considerable
observational effort has been spent to constrain the thermodynamic and
kinematic properties of galactic winds \citep[e.g.][]{steidel010,
  nestor2011, martin2012, leitherer2013, rubin2014, glenn2015,
  zhu2015, cicone2016, chisholm2016, chisholm2016b, pereira2016,
  leslie2017}. Despite the plethora of data available, a completely
consistent picture of stellar feedback remains elusive\footnote{The
  same is true for feedback associated with active galactic nuclei
  (AGN), which is not considered in this work.}. For example, the
interplay between different energy and momentum injection mechanisms
(e.g. radiation pressure $-$ \citealt{murray2011}, thermal runaway $-$
\citealt{li2015}, cosmic rays $-$ \citealt{salem2014, wiener2016}) is
still poorly understood. Moreover, observations of cold and neutral
gas clouds entrained in hot, ionised outflows
\citep[e.g.][]{sarzi2016} are difficult to address theoretically
\citep{scannapieco2015, thompson2016, bruggen2016, zhang2015}.

Analytical and numerical models are essential to explore the physics
of galactic winds and to interpret observed data \citep[see the recent
works of][]{barai13,barai2015, rosdahl2015, keller2015, muratov2015,
  ceverino2015, busta2016, christensen2016, girichidis2016,
  martizzi2016, meiksin2016, tanner2016a, tanner2016b, ruszko2016,
  hayward2015, schneider2016, kim2016, alcazar2016, li2016,
  zhang2016}. Properly describing the variety of scales (from star
forming molecular clouds to the intergalactic medium) and complexity
of processes involved is a challenging task, especially in
cosmological simulations aimed at reproducing representative volumes
of the Universe. For this reason, phenomenological sub-resolution
prescriptions are usually adopted in simulations of galaxy formation
and evolution. Major progress has been made in this field during the
last five years \citep[cf. the reviews by][and references
therein]{dale2015, somervilledave2015}. In the future, increasingly
detailed observations will call for more sophisticated codes that
include additional physics implemented using advanced numerical
techniques. In particular, IFS data promise to play a critical role in
the investigation of galactic winds \citep{sharp2010}.

The potential of SAMI for this fundamental scientific research was
confirmed in the first commissioning run, when \citet{fogarty2012}
serendipitously discovered a spiral galaxy (ESO $185$-G$031$ at
$z=0.016$) showing diffuse emission along the minor axis consistent
with starburst-driven galactic winds. Later, \citet{ho2014}
investigated an isolated disc galaxy (SDSS J$090005.05$+$000446.7$ at
$z = 0.05386$) that exhibits emission line profiles differently skewed
in different regions. Accurate modeling revealed the presence of major
outflows affecting the velocity dispersion ($\sigma$) distribution of
gas by introducing shock excitation on top of stellar
photoionisation\footnote{When we refer to observations, {\it velocity
    dispersion} is equivalent to {\it line broadening}.}. Based on
this pilot study, \citet{ho2016} developed an empirical method to
identify wind-dominated galaxies and applied this to a sample of $40$
edge-on main sequence SAMI galaxies. The method quantifies the
asymmetry of the extraplanar gas and the relative importance of its
velocity dispersion over the maximum rotation velocity of the disc.

In this work, we repeat the analyses of \citet{ho2014} and
\citet{ho2016} on synthetic disc galaxies extracted from
state-of-the-art cosmological smoothed particle hydrodynamics (SPH)
simulations. Note that we use the terms {\it galactic winds} and {\it
  outflows} as synonyms to identify gas that is moving away from the
plane of our galaxies. This includes both material that is in the
process of leaving the galaxy and gas that will eventually stop and
fall back (i.e. galactic fountains). We adopt a different approach
with respect to other theoretical investigations designed to mimic the
observations of current IFS surveys and based on hydrodynamic
simulations. In particular, \cite{naab2014} for ATLAS$^{\rm 3D}$ and
\citet{guidi2016CALIFA} for CALIFA use ``zoom-in'' simulations of
individual galaxies, while our synthetic sample is extracted from a
large cosmological box. \citet{guidi2016CALIFA} also include radiative
transfer processes, while we estimate the kinematic state of the gas
directly from the SPH scheme. Differently from these two works, the
main goal of our analysis is reproducing the kinematic features seen
in the outflows of SAMI disc galaxies, rather than the exact setup of
the observations.

The paper is organized as follows. In Section \ref{thesims} we present
the simulations used for this work, our sample of synthetic galaxies
and the methodology adopted. In Sections \ref{gkin} and
\ref{veldistr1} we study the impact of various galactic properties on
the velocity dispersion distribution, and compare with SAMI
observations of galaxies with outflows. In Section \ref{asym} we apply
the empirical identification of wind-dominated SAMI galaxies of
\cite{ho2016} to our simulated sample. We discuss our results and
conclude in Section \ref{concl}. Finally, in Appendices
\ref{appendix_b} and \ref{constrsim} we present resolution tests and
verify our analysis using idealized simulations of disc galaxies.


\section{EAGLE simulations}
\label{thesims}

The simulations used in this work are part of the Virgo Consortium's
Evolution and Assembly of GaLaxies and their Environments (EAGLE)
project \citep{schaye2015, crain2015}. The EAGLE set includes
hydrodynamic high resolution$/$large box size cosmological
simulations, run with a modified version of the SPH code
\mbox{\small{GADGET-3}} \citep[][]{springel2005}. The adopted
cosmology is a standard $\Lambda$CDM model calibrated according to
\citet{Planck2014} data ($\Omega_{\rm \Lambda}=0.693$,
$\Omega_{\rm m}=0.307$, $\Omega_{\rm b}=0.04825$, $h=0.6777$,
$\sigma_8=0.8288$ and $n_{\rm s}=0.9611$). Parallel friends-of-friends
and {\small{SUBFIND}} algorithms \citep{subfind,klaus_subf09} identify
collapsed dark matter haloes and populate them with
galaxies$/$substructures.

\subsection{Star formation and feedback}
\label{sform_feedb}

Among the technical improvements implemented in EAGLE \citep[e.g. the
new {\small{ANARCHY}} formulation of SPH described
in][]{schaller2015}, the subgrid model for feedback associated with
star formation is particularly important for this study. Each star
particle represents a simple stellar population of stars with mass in
the range $0.1-100$ M$_\odot$, distributed according to a
\citet{chabrier03} initial mass function (IMF). Metallicity dependent
lifetimes are used to identify which stars reach the end of the main
sequence phase as the simulation evolves. Then, the fraction of mass
lost through core collapse \& type Ia supernovae and winds from
asymptotic giant branch \& massive stars is calculated for each
element that is important for radiative cooling
\citep{wiersma09,wiersma09b}.

SNe and stellar winds deposit energy, momentum and radiation into the
interstellar medium (ISM). When the star formation rate is high
enough, the associated feedback can expel considerable quantities of
gas from the ISM generating large-scale galactic winds. In EAGLE,
stellar feedback is implemented thermally without shutting off
radiative cooling or decoupling particles from the hydrodynamic
scheme\footnote{AGN feedback is also implemented thermally, but its
  role is negligible for the analysis presented in this paper.}. As a
result, galactic outflows develop via pressure gradients established
by the heating, without the need to specify wind velocities,
directions or mass loading factors. To avoid a rapid dissipation of
the injected SN energy due to efficient cooling, the temperature
increment of heated resolution elements is imposed to be
$\Delta$T$_{\rm SF}=10^{7.5}$ K \citep{dallavecchia2012}. The
efficiency of feedback scales negatively with metallicity and
positively with gas density\footnote{Physical thermal losses increase
  with metallicity, while the density scaling accounts for spurious,
  resolution dependent radiative losses \citep{schaye2015,
    crain2015}.}  and was calibrated to reproduce the observed galaxy
stellar mass function (GSMF) and mass-size relation of disc galaxies
at $z=0.1$ \citep{trayford2015,furlong2015}. The interested reader can
find an extensive discussion of the subgrid physics in
\citet{schaye2015} and of the feedback calibration in
\citet{crain2015}.

In this work, we use a single simulation snapshot at $z=0$. When
considering a static distribution of particles and their properties,
EAGLE's feedback scheme makes it harder to identify and track wind
particles outflowing from galaxies\footnote{In principle, using a
  series of high time-resolution snapshots would allow one to catch
  thermodynamic changes as they happen (Crain et al. in prep).}, with
respect to simulations based on {\small{GADGET-3}} where stellar
feedback is implemented kinetically (rather than thermally) and wind
particles are $a$) allowed to leave the galaxy by temporarily
disabling the hydrodynamic interactions \citep{springel2003} and $b$)
tagged during the time they spend decoupled from the hydrodynamics
\citep[see e.g. the {\textsc{Angus}} project,][]{tescari2014}. As
mentioned above, \mbox{EAGLE} simulations develop mass loading by
heating relatively few ISM particles and allowing winds to form via
pressure gradients, rather than directly ejecting a number of
particles specified by the subgrid scheme. This leads to entrainment:
most outflowing gas was never heated directly by the subgrid scheme,
but was heated by shocks$/$compression resulting from the initial
energy injection \citep{bahe2016}. At the relatively low masses
considered in this work (M$_{\star} \le 5.7\times10^{10}$ M$_{\odot}$,
see below), young stars and supernovae in EAGLE drive high entropy
winds that are more buoyant than any tenuous galaxy's corona: the
majority of gas leaves in a hot (T $>10^5$ K), diffuse form rather
than through ballistic winds \citep{bower2016}. There are pros and
cons. Advantages: gas temperature is a good first order proxy for the
presence or absence of galactic winds (Section
\ref{veldistr_temp}). Disadvantages: as noted by \citet{turner2016},
outflows driven by stellar feedback may not entrain enough gas with T
$<10^5$ K (Sections \ref{total_vs_ha} and \ref{veldistr3}).

Note that, following \citet{schayedv2008}, in EAGLE the star formation
rate depends on pressure (rather than density) through an equation of
state P $=$ P$_{\rm eos}(\rho_{\rm gas})$. Since the simulations do
not have enough resolution to model the interstellar, cold gas where
star formation occurs, a temperature floor (normalized to
T$_{\rm eos}=8000$ K at $n_{\rm H}=0.1$ cm$^{-3}$, which is typical of
the warm ISM) is imposed. We stress that the temperature of star
forming gas in EAGLE cannot be interpreted as a measure of the gas
kinetic energy, but simply reflects the effective pressure imposed on
the unresolved, multiphase ISM \citep{schayedv2008, schaye2015}. SFR
in galaxies correlates with the emission of H$\alpha$ radiation at T
$\sim10^4$ K \citep{kennicutt1998}. In our post-process analysis, we
assume that any gas particle with SFR $>0$ has T $=10^4$ K and use
EAGLE star forming gas as a proxy for gas that would be detected in
H$\alpha$ (see also Section \ref{binn}).

\subsection{Recal-L025N0752}
\label{conf}

High spatial$/$mass resolution is crucial to robustly sample both the
gas distribution in galaxies and, especially, the relatively small
fraction of gas outflowing from them. Hence, for this work we used the
highest resolution configuration available in the EAGLE set:
L025N0752. As the name suggests, a cubic volume of linear size L $=25$
comoving Mpc (cMpc) is sampled with N $=2\times752^3$ dark matter (DM)
+ gas particles with initial mass $1.21\times10^6$ M$_{\odot}$ and
$2.26\times10^5$ M$_{\odot}$, respectively. The comoving
Plummer-equivalent gravitational softening length is $1.33$ ckpc and
the maximum proper softening length is $0.35$ kpc.

\begin{figure*}
\centering 
\includegraphics[width=5.8cm, height=5.1cm]{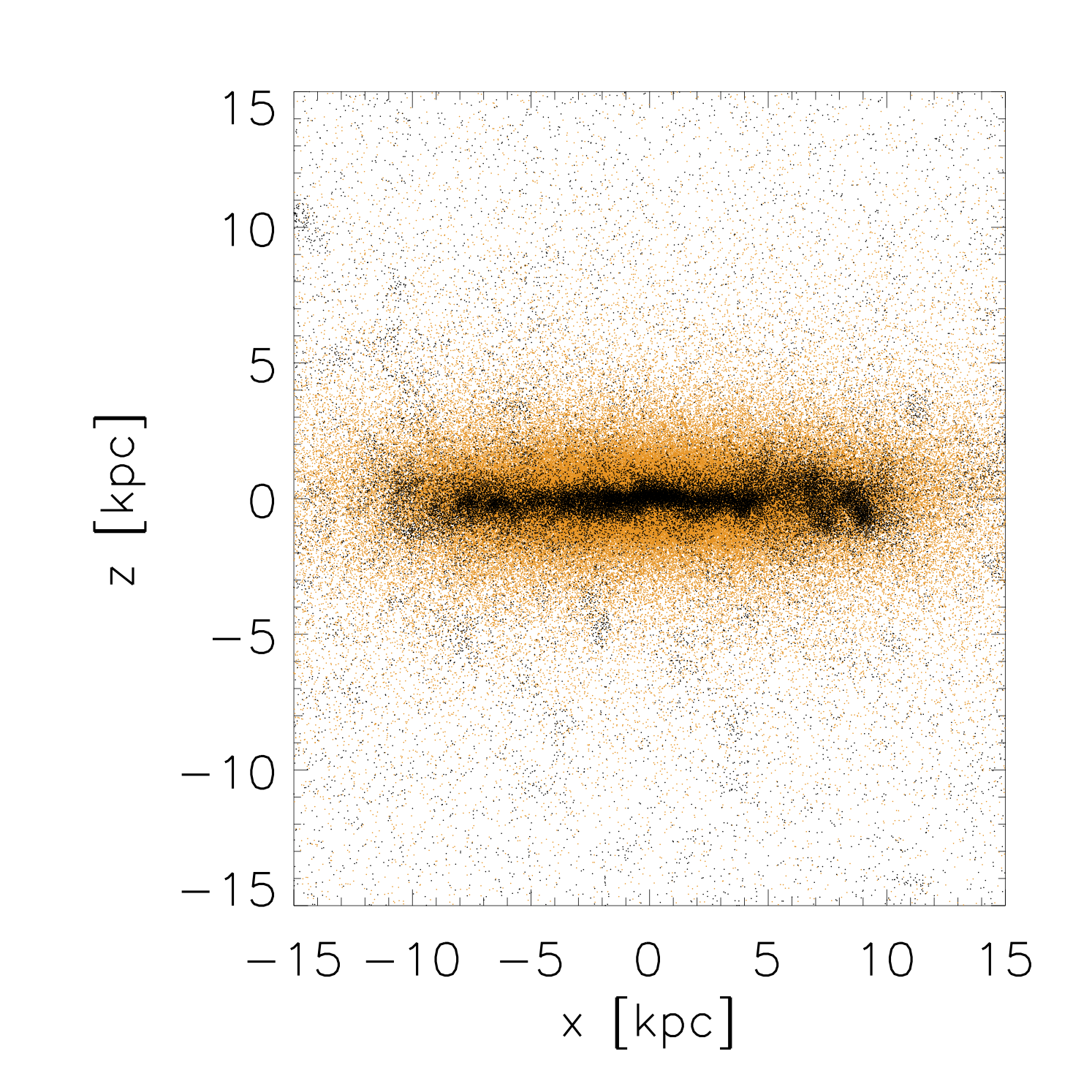}
\includegraphics[width=5.8cm, height=5.1cm]{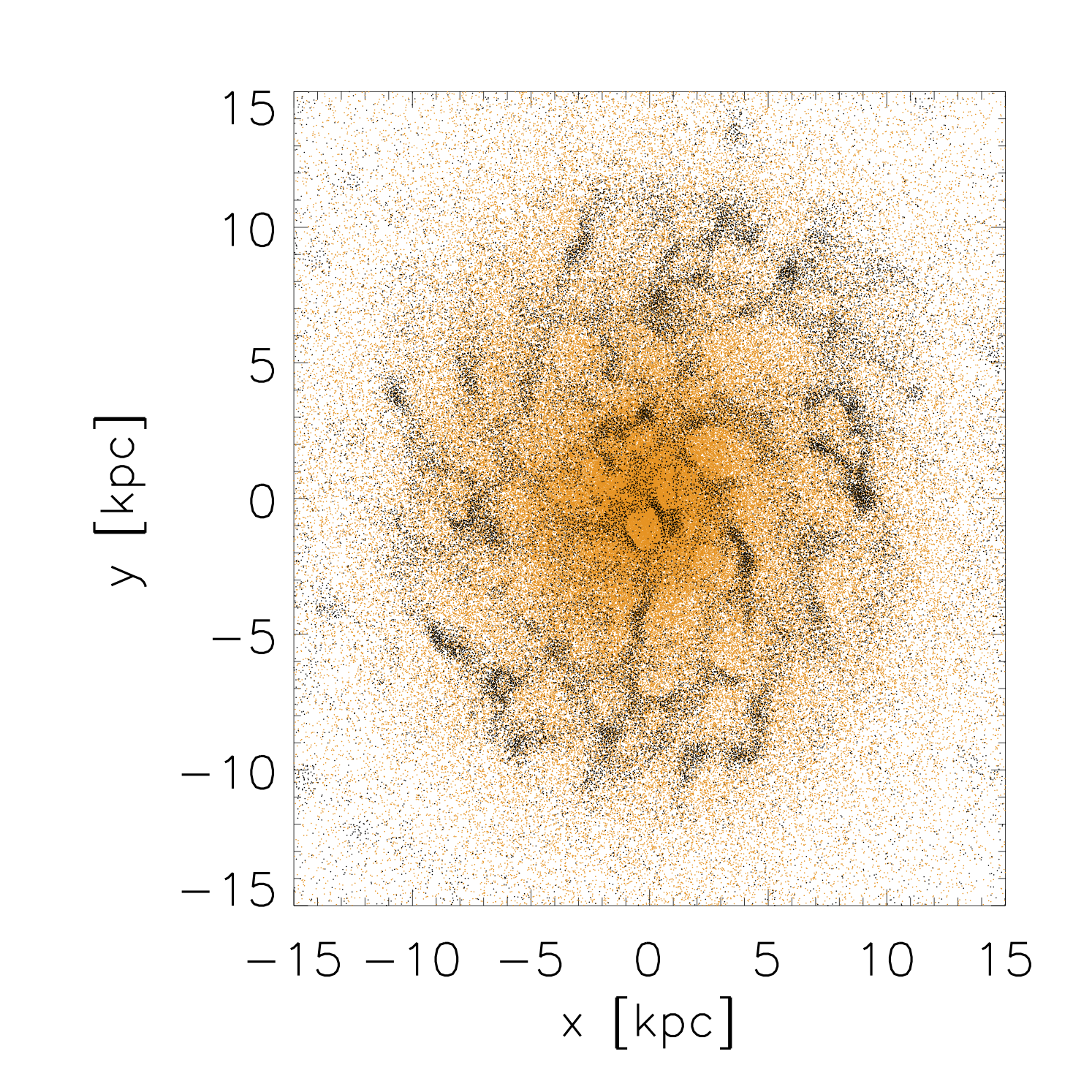}
\includegraphics[width=5.8cm, height=5.1cm]{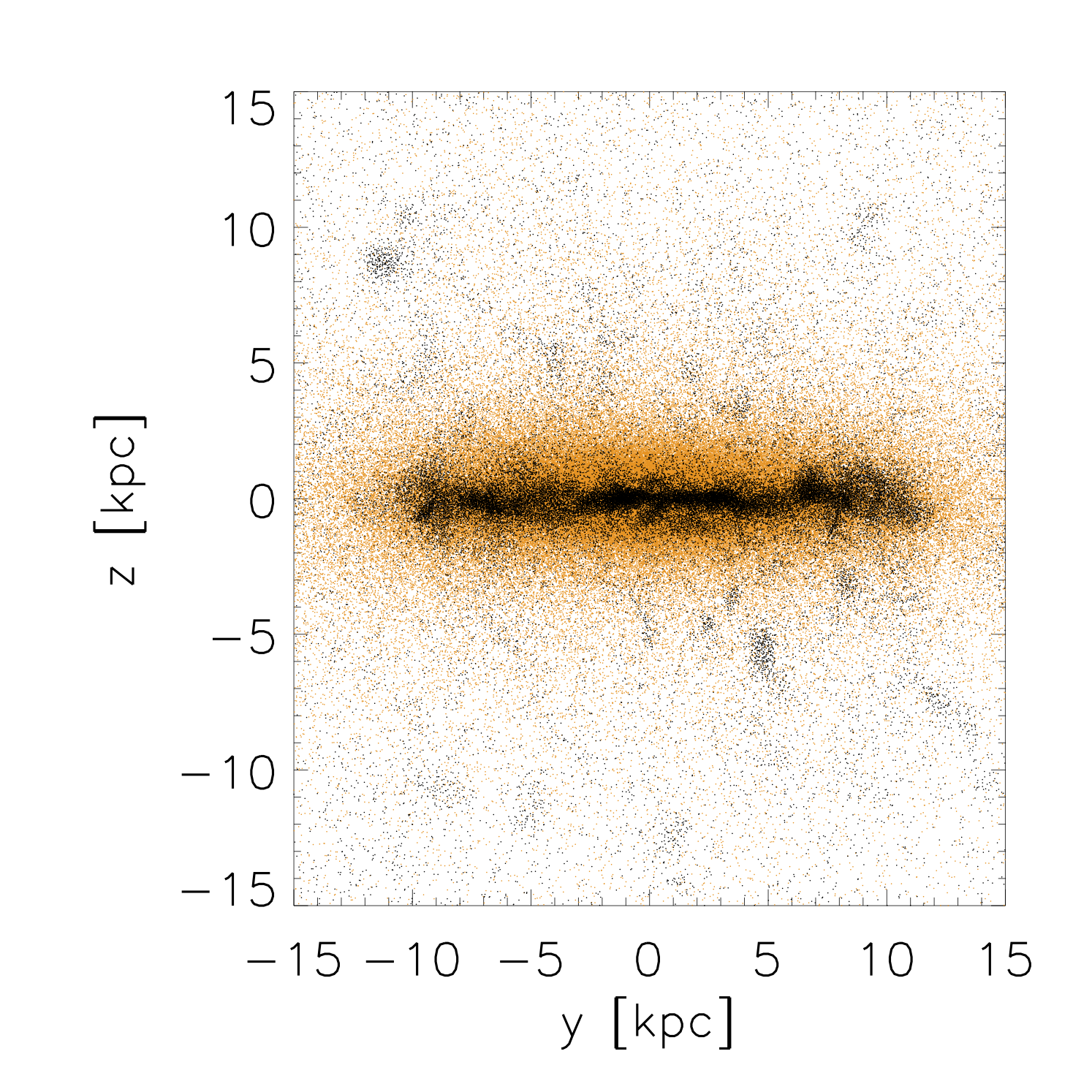}

\vspace{0.25cm}

\includegraphics[width=6.5cm]{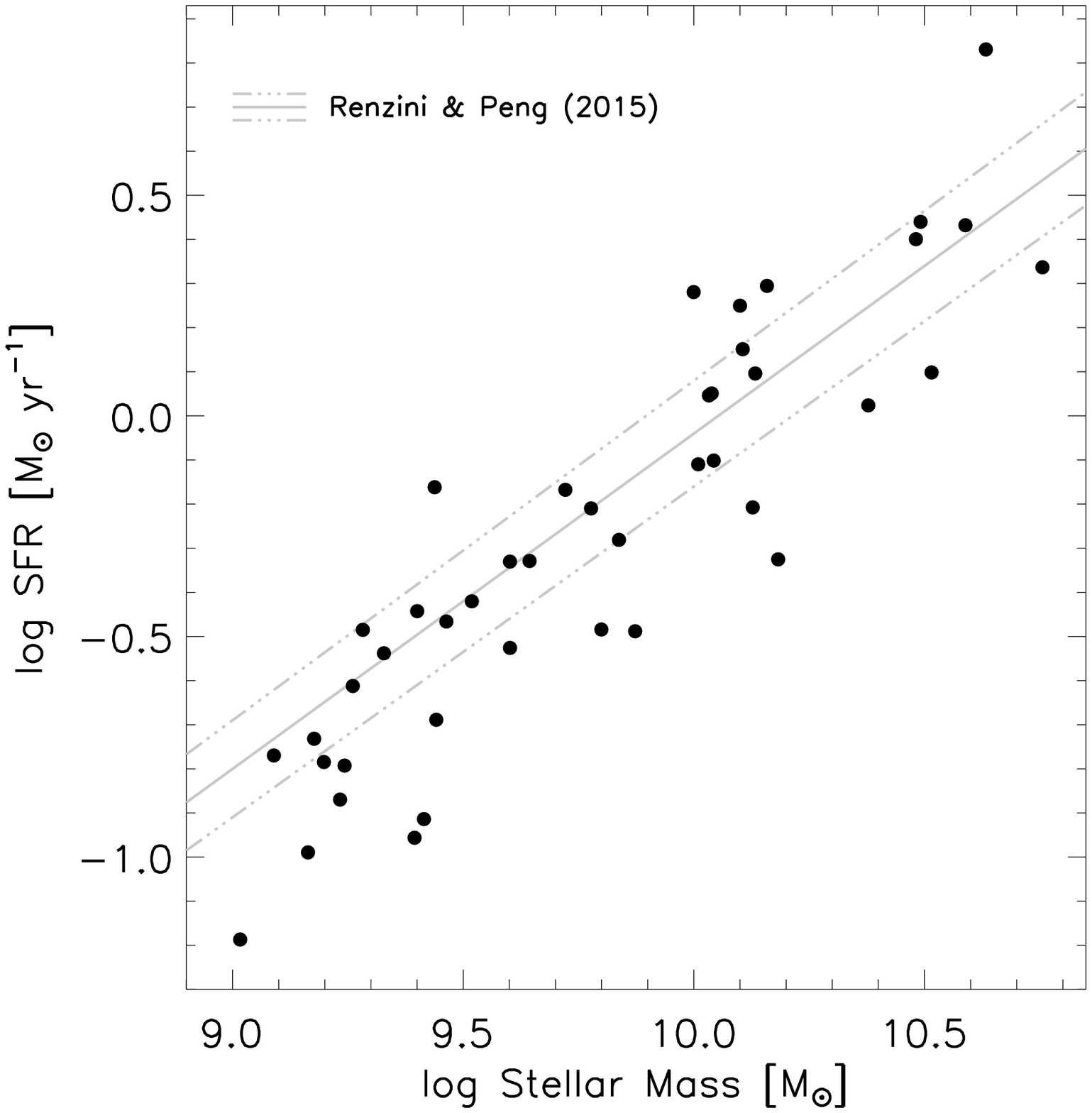}
\includegraphics[width=6.5cm]{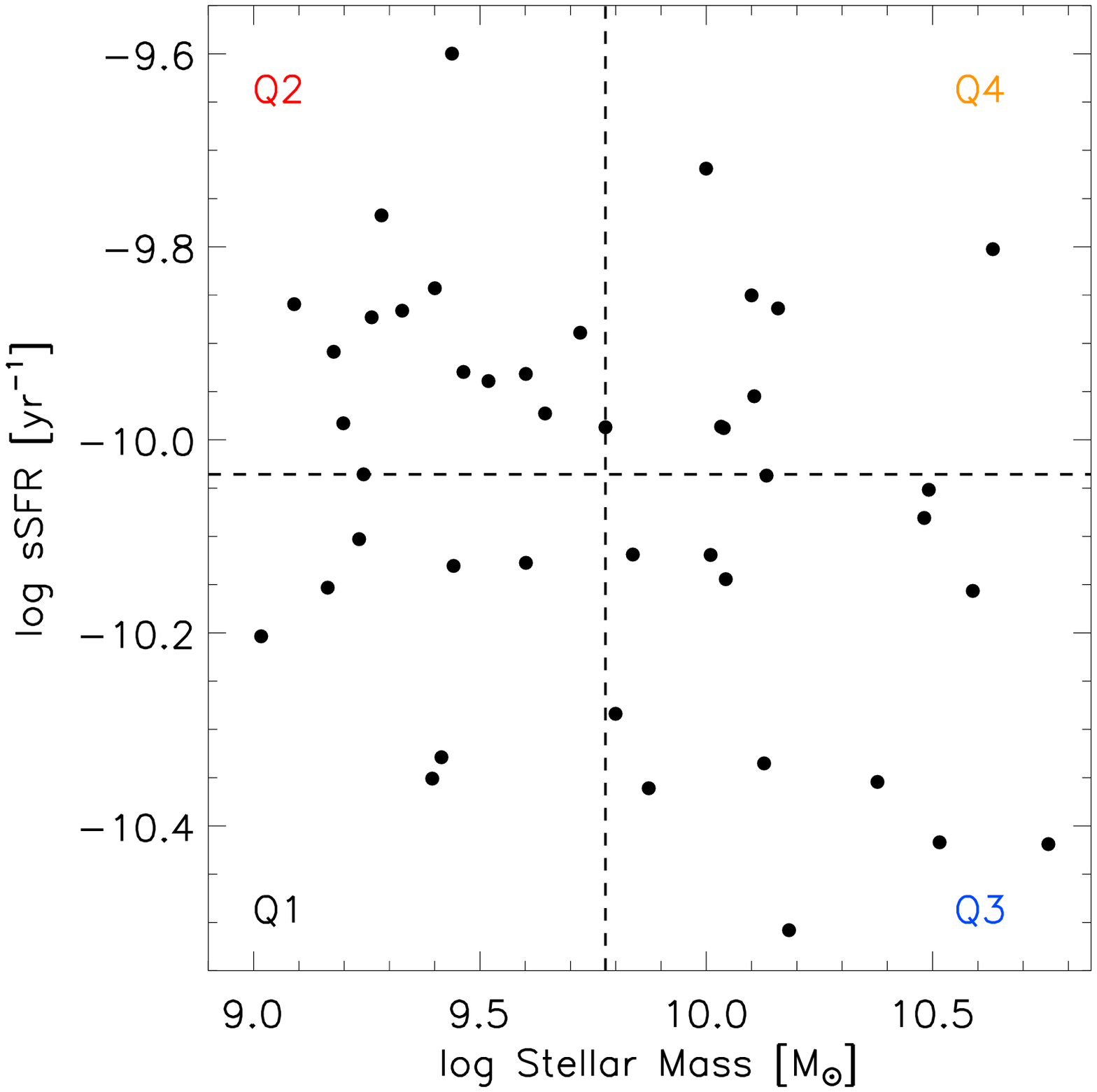}

\caption{{\bf Top row}: Positions in physical kpc of gas (black dots)
  and star (orange dots) particles for the galaxy of the simulated
  sample with the highest SFR. This object has total (i.e. as
  determined by {\small{SUBFIND}}) M$_{\star}=4.30\times10^{10}$
  M$_\odot$, M$_{\rm gas}=1.94\times10^{11}$ M$_\odot$,
  M$_{\rm tot}=1.89\times10^{12}$ M$_\odot$, SFR $=6.77$ M$_\odot$
  yr$^{-1}$ and $\log\big($sSFR$/$yr$^{-1}\big)=-9.80$. The stellar
  and gas mass inside the ($30$ kpc)$^3$ cube are, respectively,
  M$_{\rm\star,c}=3.51\times10^{10}$ M$_\odot$ and
  M$_{\rm gas,c}=1.19\times10^{10}$ M$_\odot$. Left panel: xz edge-on
  projection. Middle panel: xy face-on projection. Right panel: yz
  edge-on projection. {\bf Bottom left panel}: SFR$-$M$_{\star}$
  relation for our simulated galaxies. The grey solid (+ triple
  dot-dashed) line is the best fit SFR$-$M$_{\star}$ relation for star
  forming local galaxies of \citet{renzinipeng2015}. The synthetic
  sample represents disc galaxies on the main sequence. {\bf Bottom
    right panel}: simulated specific SFR$-$M$_{\star}$ relation. The
  vertical and horizontal dashed lines mark, respectively, the median
  M$_{\star}$ and sSFR of the final sample:
  $\log(\tilde{{\rm M}}_{\star}/$M$_\odot)=9.78$,
  $\log(\tilde{\rm sSFR}/$yr$^{-1})=-10.04$. To facilitate the
  subsequent analysis, we divided the plot in four quadrants: Q$1$ =
  low M$_{\star}-$ low sSFR, \color{red} Q$2$ \color{black}= low
  M$_{\star}-$ high sSFR, \color{blue} Q$3$ \color{black}= high
  M$_{\star}-$ low sSFR and \color{orange} Q$4$ \color{black}= high
  M$_{\star}-$ high sSFR (see text).}
\label{fig_gal11_proj}
\end{figure*}

This configuration comes in two different setups: {\textit
  {Ref-L025N0752}} and {\textit {Recal-L025N0752}}. The first one is
the initial reference setup, while in the second the subgrid stellar
and AGN feedback parameter values were re-calibrated to better match
the observed low redshift GSMF \citep{schaye2015}. In practice, the
main difference is that feedback is slightly more effective in
{\textit {Recal}}. This prevents overcooling problems and leads to
more realistic gas and stellar distributions in galaxies. For this
reason, we used {\textit {Recal-L025N0752}} for the analysis presented
in this paper.

\subsection{Galaxy Sample}
\label{gsample}

For this project, we have developed a pipeline to optimise the
analysis of EAGLE simulations. The pipeline first (and only once)
loads the original EAGLE snapshot and reads the output of {\small
  SUBFIND} (i.e. the catalogue of substructures within DM
haloes). Then it extracts and saves only the information needed by the
user (e.g. only galaxies with stellar mass, M$_{\star}$, and$/$or
specific star formation rate, sSFR $\equiv$ SFR/M$_{\star}$, in a
given range, et cetera). Since the original snapshot can be several
gigabytes in size, this procedure drastically reduces both memory and
time access requirements, speeding up considerably the subsequent
analysis.

We used the pipeline to extract cubes of size $30$ physical kpc around
all the galaxies with M$_{\star} \ge 10^{9}$ M$_\odot$ from {\textit
  {Recal-L025N0752}} at $z=0$ (taking into account periodic boundary
conditions and a buffer of extra $70$ kpc per side for SPH
interpolation, see Section \ref{binn}). Each cube is a {\small GADGET}
format file containing a new header and the following information for
gas ($g$) and star ($s$) particles: position ($g$+$s$), velocity
($g$+$s$), mass ($g$+$s$), temperature ($g$), density ($g$), smoothing
length ($g$), SFR ($g$), metallicity ($g$+$s$) and age ($s$). The
initial sample included $266$ galaxies.

\begin{figure*}
\centering 
\vspace{0.3cm}
\includegraphics[width=7.85cm, height=7.1cm]{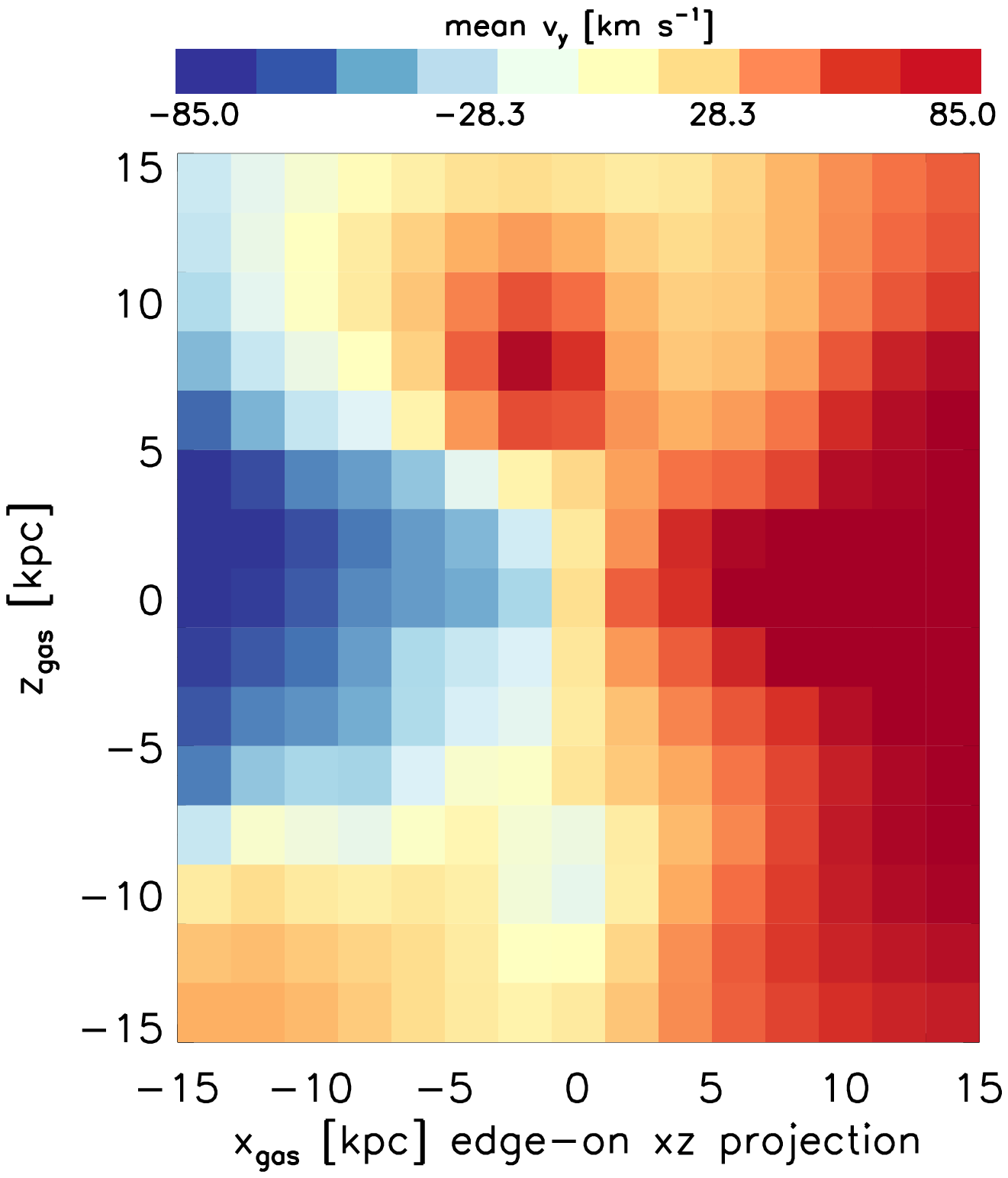}
\includegraphics[width=7.85cm, height=7.1cm]{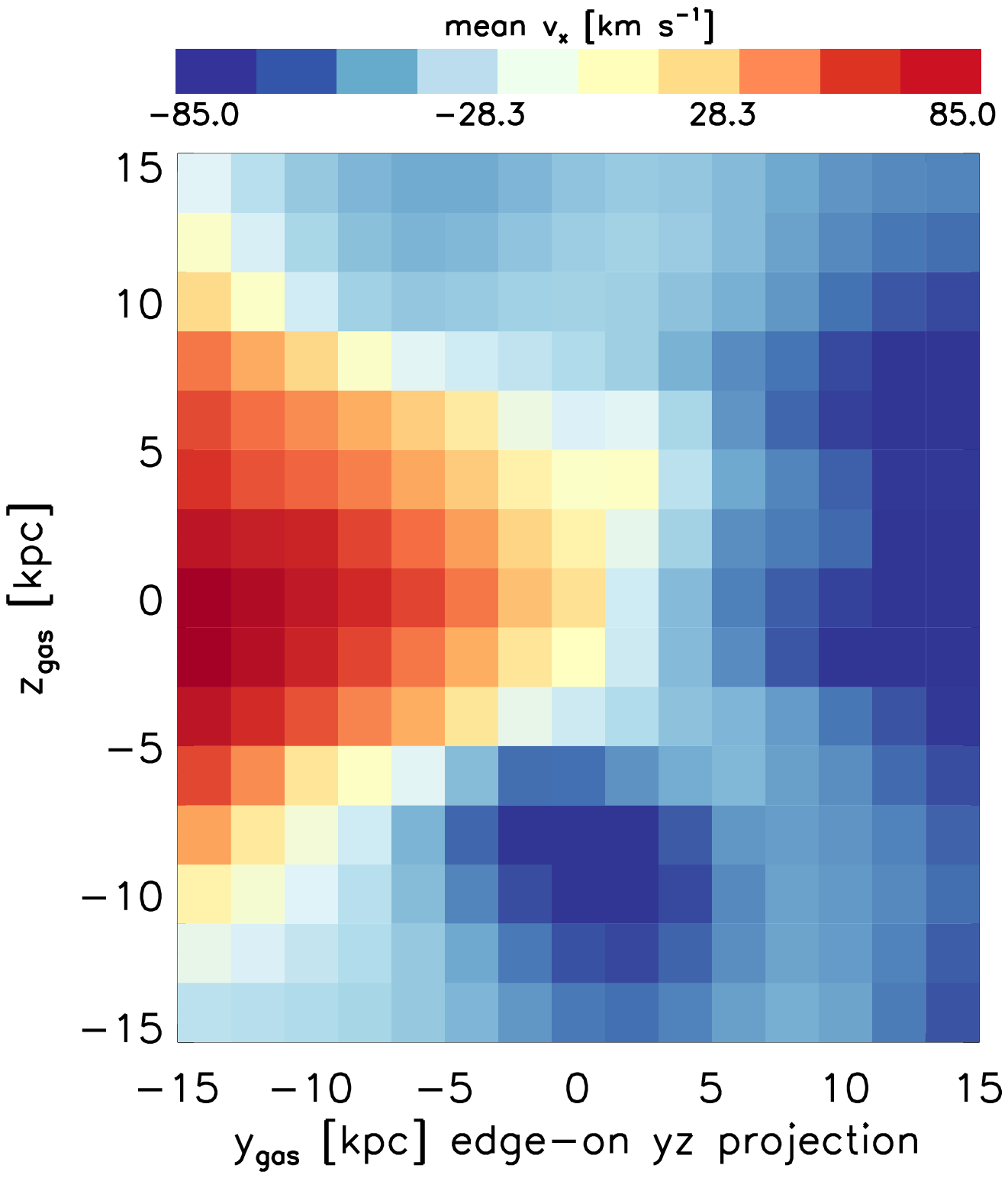}

\vspace{0.25cm}

\includegraphics[width=7.85cm, height=7.1cm]{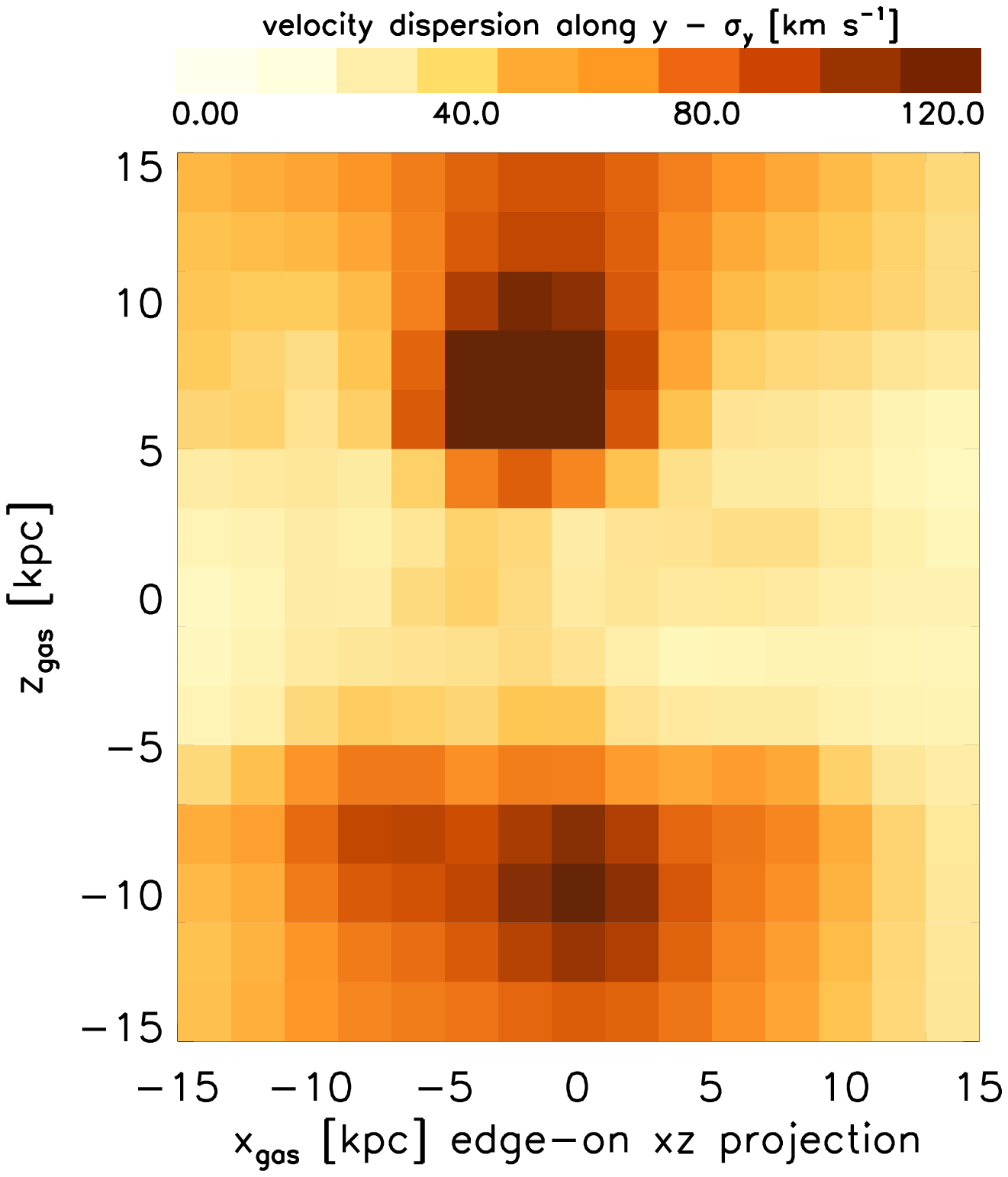}
\includegraphics[width=7.85cm, height=7.1cm]{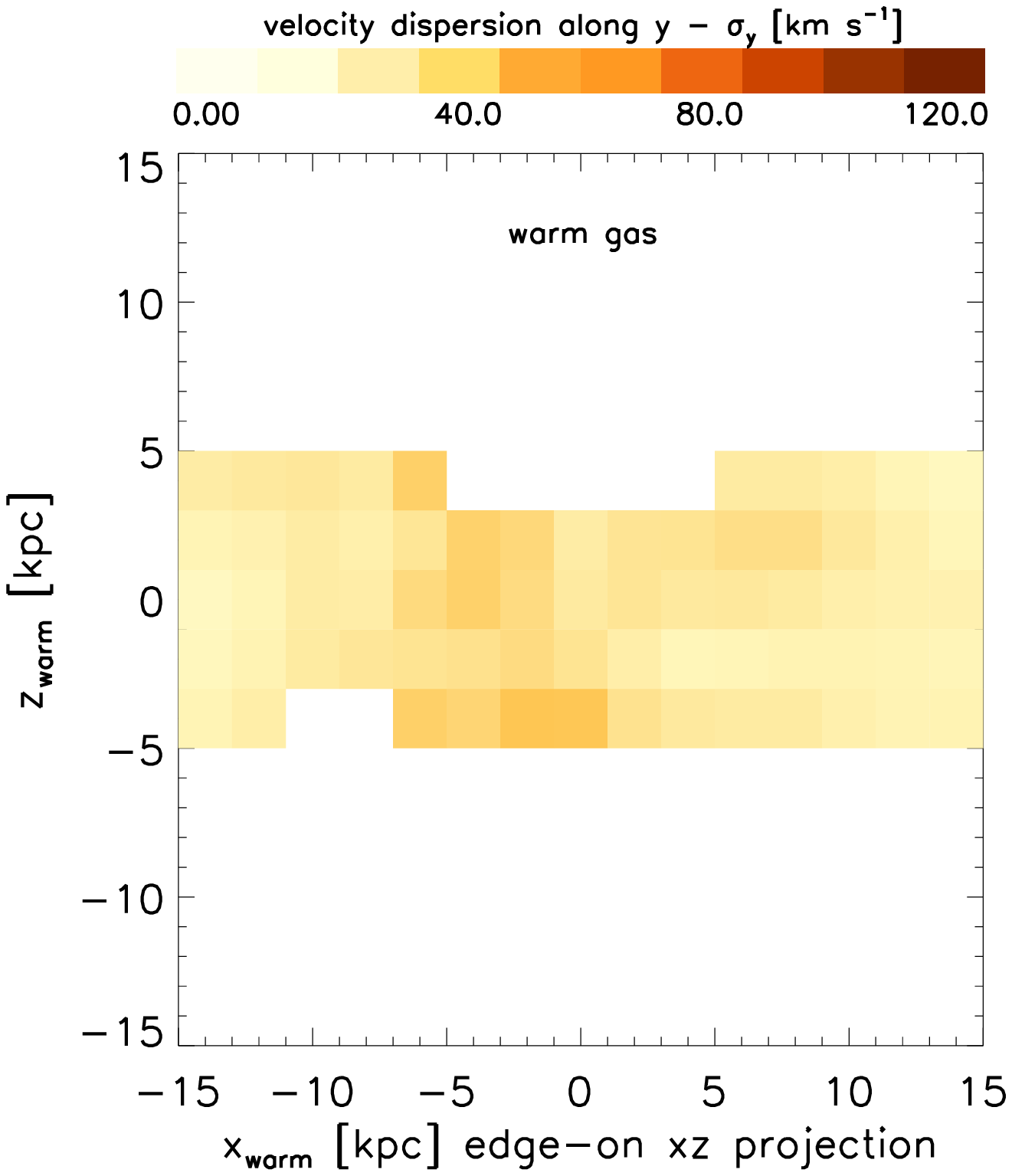}

\vspace{0.25cm}

\includegraphics[width=7.85cm, height=7.1cm]{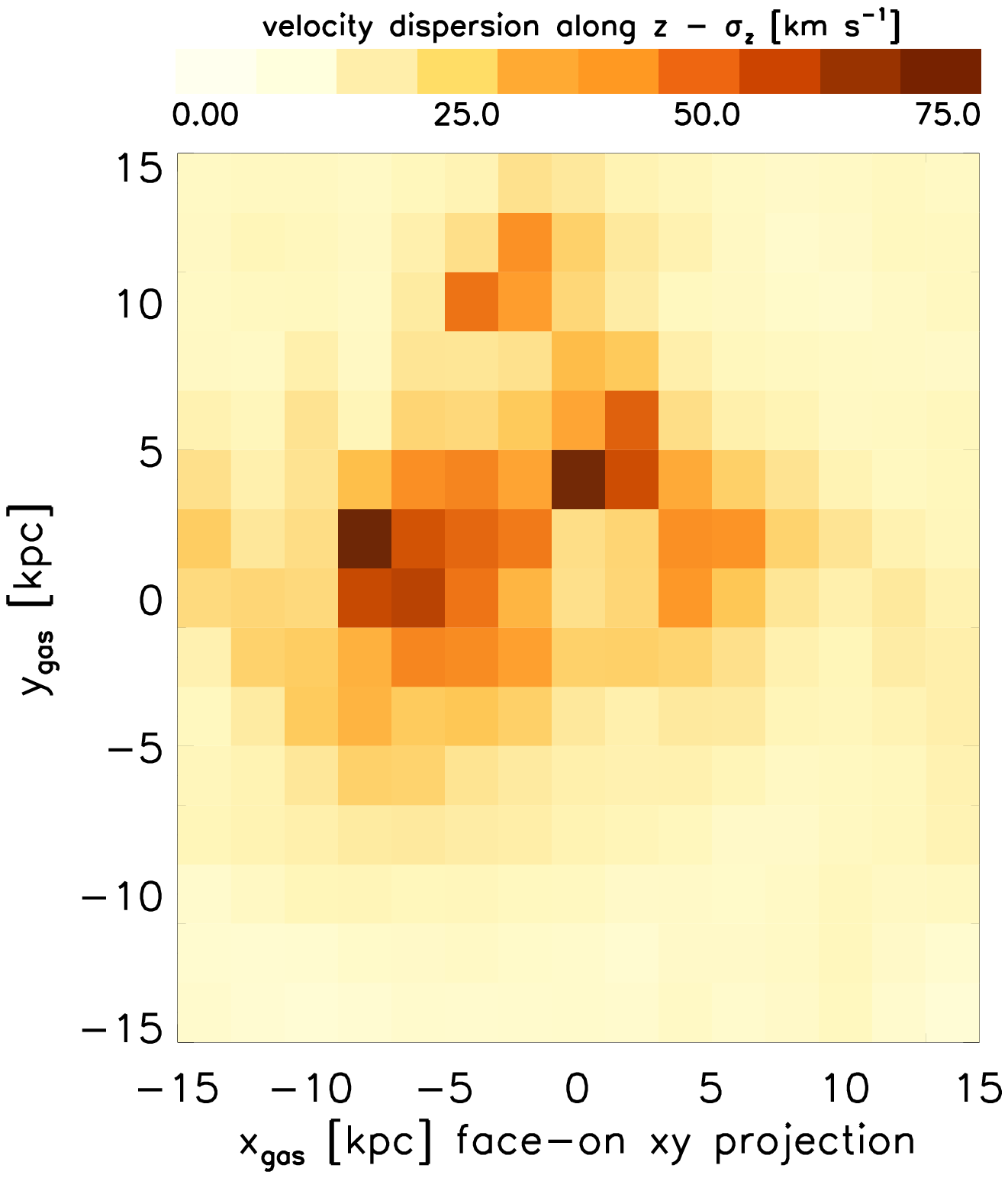}
\includegraphics[width=7.85cm, height=7.1cm]{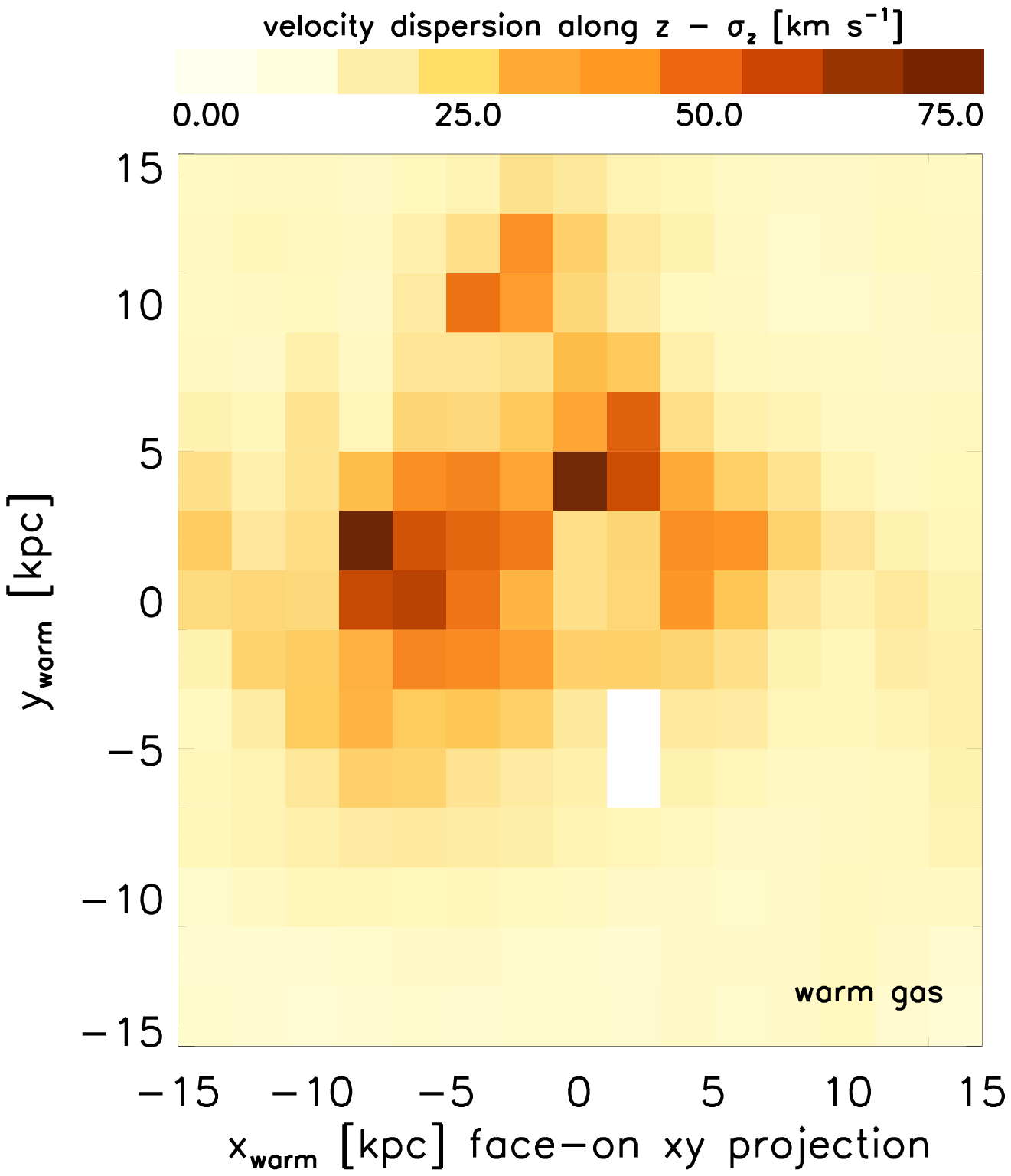}
\caption{{\bf Top row}: examples of mean (rotation) velocity maps
  created using all the gas in the cube. Left panel: edge-on xz
  projection $-$ mean $v_{\rm y}$. Right panel: edge-on yz projection
  $-$ mean $v_{\rm x}$. {\bf Middle row}: gas velocity dispersion maps
  in the edge-on xz projection $-$ $\sigma_{\rm y}$. Left panel: {\it
    all gas}. Right panel: {\it warm gas}, that is only pixels where
  $3.8\le\log\big(\langle$T$_{\rm gas}\rangle/$K$\big)\le4.2$ (see
  Section \ref{binn}). Warm gas is clearly associated with the
  galactic disc and virtually absent when moving away from the galaxy
  plane, where the all gas $\sigma_{\rm y}$-map peaks. {\bf Bottom
    row}: gas velocity dispersion maps in the face-on xy projection
  $-$ $\sigma_{\rm z}$. Left panel: {\it all gas}. Right panel: {\it
    warm gas}. Now, all the lines-of-sight pierce the disc (where warm
  gas dominates) and the two maps are almost identical.}
\label{fig_maps}
\end{figure*}

We then rotated the particle distribution ($g$+$s$) around the
gravitational potential minimum (using the angular momentum per unit
mass of star particles), in order to place each galaxy face-on in the
xy plane and edge-on in the xz and yz planes. Accordingly, we rotated
the velocity vector of each particle in the cube and used the velocity
of the (rotated) stellar centre of mass as the velocity reference
frame. At this point, we visually inspected the sample and only
selected unperturbed galaxies with a prominent disc structure in both
gas and stars, a number of gas particles (N$_{\rm gas}$) inside the
$30$+$70$ kpc cube sufficient to make our analysis robust, and fairly
regular$/$symmetric gas rotation velocity maps (constructed as
explained in Section \ref{binn}). The final sample includes $43$
galaxies with $9.02\le\log($M$_{\star}/$M$_\odot)\le10.76$ and
$7.8\times10^{3}\lesssim$ N$_{\rm gas}\lesssim9.4\times10^{4}$
($\langle$N$_{\rm gas}\rangle\approx3.8\times10^{4}$)\footnote{Note
  that a galaxy with M$_{\star}=10^{9}$ M$_\odot$ in {\textit
    {Recal-L025N0752}} contains more than $4.4\times10^{3}$ star
  particles.}. In the top row of Fig. \ref{fig_gal11_proj} we plot
positions of gas (black dots) and star (orange dots) particles for the
galaxy with the highest SFR. The left and right panels show the
edge-on xz and yz projections: the galactic disc is clearly visible in
both star and gas components. The middle panel shows the face-on xy
projection. The gas distribution is arranged in a complex clumpy
structure as a result of the interplay between star formation and
associated feedback processes.

The bottom left panel of Fig. \ref{fig_gal11_proj} shows the
distribution in the SFR$-$M$_{\star}$ plane of our final sample of
$43$ galaxies. The minimum and maximum star formation rates are
$0.065$ and $6.77$ M$_\odot$ yr$^{-1}$. The grey solid (+ triple
dot-dashed) line is the best fit SFR$-$M$_{\star}$ relation for star
forming local galaxies of \citet{renzinipeng2015}:
$\log\big($SFR$/[$M$_\odot$
yr$^{-1}]\big)=(0.76\pm0.01)\log($M$_{\star}/$M$_\odot)-
7.64\pm0.02$. The final sample represents disc galaxies on the main
sequence. We plot the simulated specific SFR$-$M$_{\star}$ relation in
the bottom right panel of Fig. \ref{fig_gal11_proj}. The vertical and
horizontal dashed lines mark, respectively, the median stellar mass,
$\log(\tilde{{\rm M}}_{\star}/$M$_\odot)=9.78$, and median specific
star formation rate, $\log(\tilde{\rm sSFR}/$yr$^{-1})=-10.04$. In the
following sections we will study how the gas velocity dispersion
distribution varies as a function of M$_{\star}$ and sSFR. For this
reason, we split the plot in four quadrants: Q$1$ = low M$_{\star}-$
low sSFR, \color{red} Q$2$ \color{black}= low M$_{\star}-$ high sSFR,
\color{blue} Q$3$ \color{black}= high M$_{\star}-$ low sSFR and
\color{orange} Q$4$ \color{black}= high M$_{\star}-$ high sSFR. Q$1$
and Q$4$ contain, respectively, $7$ and $8$ galaxies, while both Q$2$
and Q$3$ contain $14$ objects.

\subsection{Binning and warm gas}
\label{binn}

We binned the gas particles in each galactic cube on a $2$D spatial (+
$1$D depth) grid of pixels with linear size $2$ kpc ($15$ pixels per
cubic side), which is comparable to the effective resolution of SAMI
after accounting for a typical AAT seeing of $2.1$ arcsec\footnote{In
  Appendix \ref{appendix_b} we will explore the impact of cube size
  ($30$ and $60$ kpc) and grid resolution ($2$ and $3$ kpc) on our
  results.}. To obtain SPH quantities on the grid, we followed the
procedure described in Section \color{blue} 4 \color{black} of
\citet{altay2013}. We started by extracting a buffer of additional
$70$ kpc per side around the central cube of volume $(30$ kpc$)^3$, to
ensure that all the gas particles in the simulation whose $3$D SPH
kernels intercept one or more pixels of the grid in the $2$ spatial
directions and the cube margins in the line-of-sight direction were
taken into account. Then, we assigned a truncated Gaussian kernel to
each gas particle (Eqs. \color{blue} 8 \color{black} and \color{blue}
9 \color{black} of \citealt{altay2013}) and integrated it over the
square pixels. Using this procedure, we calculated the density
weighted (mean) velocity, velocity dispersion and temperature along
the line-of-sight in all the pixels.

We considered different projections. In the edge-on xz (yz)
projection, the line-of-sight direction is the direction y (x)
perpendicular to the xz (yz) plane. The corresponding mean velocity
and velocity dispersion are, respectively, $v_{\rm y}$ and
$\sigma_{\rm y}$ for the xz projection and $v_{\rm x}$ and
$\sigma_{\rm x}$ for the yz projection. The velocity dispersion for
the face-on projection xy is $\sigma_{\rm z}$. The two top panels of
Fig. \ref{fig_maps} show examples of mean (rotation) velocity maps
in the two edge-on projections created using all the gas in the cube.

SAMI observations of low-redshift galaxies with outflows are largely
based on the detection of H$\alpha$ emitting gas at T $\sim 10^4$
K. From now on, to better compare with these observations we will
distinguish between {\it all gas} and {\it warm gas}. In an SPH
simulation, the fluid conditions at any point are defined by
integrating over {\it all} particles, weighted by their
kernel. Selecting only a subset of them (e.g. only star forming or
cold$/$hot gas) would break mass$/$momentum$/$energy conservation
laws. Therefore, we define \\

\noindent{\it Warm gas}: pixels in a velocity$/$velocity
dispersion$/$temperature map where the density weighted gas
temperature (calculated using all the particles) is in the range
$3.8\le\log\big(\langle$T$_{\rm gas}\rangle/$K$\big)\le4.2$. \\

\noindent Gas with temperature around $10^4$ K is usually referred to
as warm to distinguish it from cold gas in molecular clouds (T $<100$
K) and hot gas in the halo or in supernova bubbles (T $>10^5$ K),
based on the model of a three-phase ISM medium
\citep{mckeeost1977}. H$\alpha$ emission in real galaxies is mostly
from \text{H\,\textsc{\lowercase{II}}} regions, and hence correlates
strongly with star formation rate \citep{kennicutt1998}. To a good
approximation, in our analysis warm pixels trace pixels with density
weighted SFR greater than zero. We introduced this temperature cut to
qualitatively compare with the kinematic signatures seen in SAMI
observations, without having to model complicated and uncertain
radiative transfer effects.

In the middle row of Fig. \ref{fig_maps} we plot edge-on velocity
dispersion maps (xz $-$ $\sigma_{\rm y}$) for $a$) all gas (left
panel) and $b$) warm gas (right panel). In $a$) the velocity
dispersion increases when moving away from the disc of the galaxy (in
both vertical directions) and peaks at abs(z$_{\rm gas}$) $\sim9$
kpc. On the other hand, in $b$) the high-$\sigma$ part is completely
suppressed and warm gas is mainly associated with the galactic
disc\footnote{We stress again that the map in $b$) is the map in $a$)
  with only pixels fulfilling the condition
  $3.8\le\log\big(\langle$T$_{\rm gas}\rangle/$K$\big)\le4.2$
  included.}. This has important consequences for our analysis. We
will explore them in Sections \ref{total_vs_ha} and \ref{veldistr3}.

The situation is different in the bottom two panels of
Fig. \ref{fig_maps}, which show face-on velocity dispersion maps (xy
$-$ $\sigma_{\rm z}$) for all gas (left panel) and warm gas (right
panel). This time the two maps are almost identical (except for two
pixels). This is due to the fact that now all the lines-of-sight
pierce the disc, where warm gas dominates. The $\sigma$-maps in
Fig. \ref{fig_maps} are the base of all our analyses and we will
discuss them more in the next sections.

\section{Signatures of outflows: the velocity dispersion distribution}
\label{gkin}

\begin{figure}
\centering 
\includegraphics[width=8.7cm]{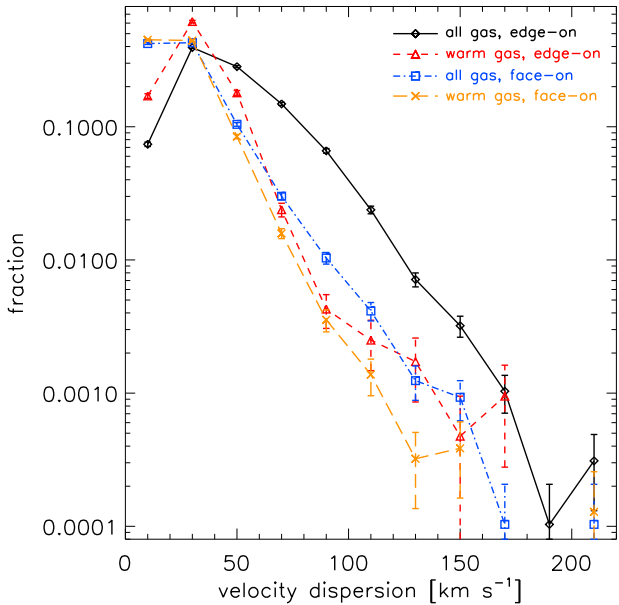}
\caption{Pixelated velocity dispersion probability distributions
  (i.e. the fraction of pixels $-$ with pixel size $= 2$ kpc $-$ per
  velocity dispersion bin) calculated using all gas and warm gas in
  different projections. The bin size is $20$ km s$^{-1}$. Black diamonds +
  solid line and red triangles + short dashed line: all gas and warm
  gas, respectively, in the edge-on xz projection $-$
  $\sigma_{\rm y}$. Blue squares + dot-dashed line and orange crosses
  + long dashed line: all gas and warm gas, respectively, in the
  face-on xy projection $-$ $\sigma_{\rm z}$. Each line was created by
  stacking the histograms of all the $43$ galaxies in the simulated
  sample. Errors are Poissonian. Since in the face-on projection the
  lines-of-sight in different pixels always pierce the galactic disc,
  where warm gas dominate, the all gas and warm gas
  $\sigma_{\rm z}$-distributions are very similar (see the bottom
  panels of Fig. \ref{fig_maps}). In the edge-on projection, the
  $\sigma_{\rm y}$-distribution for all gas is shifted to higher
  velocity dispersions than the warm gas distribution, which is
  associated with the galactic disc and therefore has a more prominent
  peak at low $\sigma_{\rm y}=30$ km s$^{-1}$ (cf. the two middle panels of
  Fig. \ref{fig_maps}).}
\label{fig_total_vs_ha}
\end{figure}

The aim of this work is to determine whether or not current and
upcoming IFS surveys can succesfully identify large-scale outflows and
provide meaningful constraints on the physical processes driving gas
out of galaxies. For this reason, we apply observationally-based
analysis techniques to the simulations. In particular, in the next
sections we compare with the observational investigations of SAMI
galaxies with outflows presented by \citet{ho2014} and
\citet{ho2016}. We stress that observational and numerical analyses
estimate the kinematic state of the gas in two different ways. We
track the motion of particles as sampled by the SPH scheme in the
simulation, without including any radiative transfer effects. Instead,
\citet{ho2014,ho2016} extract kinematic information from emission line
spectra of galaxies. Thus, while observations only probe ionised gas,
simulations take into account all the gas in the galactic halo. To
facilitate the comparison, we therefore introduced in the previous
section the definition of warm gas that we will use throughout the
paper.

\citet{ho2014} studied the nature of a prototypical low redshift
isolated disc galaxy with outflows: SDSS J$090005.05$+$000446.7$ (SDSS
J$0900$, $z = 0.05386$). Its emission line spectrum was decomposed
using the spectral fitting pipeline {\small{LZIFU}} \citep{LZIFU2016},
a likelihood ratio test and visual inspection. Emission lines were
modelled as Gaussians composed of up to three kinematic components
with small, intermediate and high velocity dispersion relative to each
other (i.e from narrow to broad features). Fig. \color{blue} 6
\color{black} of \citet{ho2014} shows the velocity dispersion
distribution of SDSS J$0900$. The statistically prominent narrow
component peaks at $\sim40$ km s$^{-1}$ and is associated with the
(rotationally supported) disc of the galaxy. More interestingly, the
$\sigma$-distribution extends to very high values ($450$ km s$^{-1}$) with
the broad kinematic component peaking at $\sim300$ km s$^{-1}$. The authors
argue that this high-$\sigma$ component traces shock excited emission
in biconical outflows, likely driven by starburst activity.

\subsection{All gas vs warm gas}
\label{total_vs_ha}

Following \citet{ho2014}, we begin our analysis by investigating the
velocity dispersion distribution of the simulated galaxies. Throughout
the paper, we adopt the following procedure. Using $\sigma$-maps like
those shown in the bottom four panels of Fig. \ref{fig_maps}, we first
calculate the histogram of the pixelated velocity dispersion for each
galaxy (i.e. the fraction of pixels per velocity dispersion bin),
where the pixel size is $2$ kpc, the upper limit is the maximum
$\sigma$ of the entire sample (that changes depending on the
projection and if all gas or warm gas is considered) and the bin size
is always $20$ km s$^{-1}$. Then, we stack these histograms and normalize by
the number of galaxies to obtain the final {\it pixelated velocity
  dispersion (probability) distribution} and its associated Poissonian
errors.

In Fig. \ref{fig_total_vs_ha} we examine the differences between the
pixelated velocity dispersions calculated using all gas and warm gas
in different projections for all the $43$ simulated galaxies. In the
face-on (xy view) projection, the $\sigma_{\rm z}$-distributions of
all gas (blue squares and dot-dashed line) and warm gas (orange
crosses and long dashed line) are very similar, with the highest
fractions at low $\sigma_{\rm z}$ and a declining trend at larger
velocity dispersions, and share the same $\max(\sigma_{\rm z})=201.31$
km s$^{-1}$. The only difference is that the all gas distribution shows
slightly larger statistics at high $\sigma_{\rm z}$. According to the
bottom panels of Fig. \ref{fig_maps}, this is not surpising. When the
lines-of-sight in different pixels pass through the galactic disc,
particles in the warm gas regime are the majority and dominate the
$\sigma_{\rm z}$-maps.

On the other hand, the edge-on (xz view) $\sigma_{\rm y}$-distribution
for all gas (black diamonds and solid line,
$\sigma_{\rm y,\max}=211.08$ km s$^{-1}$) is shifted to higher velocity
dispersions than the warm gas distribution (red triangles and short
dashed line, $\sigma_{\rm y,\max}=169.29$ km s$^{-1}$), which has a more
prominent peak at $30$ km s$^{-1}$ and then rapidly drops to lower fractions
at high $\sigma_{\rm y}$. The reasons for this are mentioned in
Section \ref{sform_feedb} and highlighted by the two middle panels of
Fig. \ref{fig_maps}. In our simulated disc galaxies, warm gas traces
the disc and is virtually absent when moving away from the galaxy
plane. In Section \ref{gal_winds}, we will show how the extraplanar
velocity dispersion is dominated by outflowing gas. Therefore, the
lack of warm gas outside the disc is a direct consequence of the
thermal stellar feedback implemented in EAGLE that heats outflowing
particles up to a temperature higher than the warm range (T
$\sim10^{4}$ K). Galactic winds in our simulations are mostly hot (T
$>10^5$ K) and, compared to observed galaxies, may entrain
insufficient gas with T $<10^5$ K. Note that such high-temperature gas
would not be visible in SAMI observations.

This issue with \mbox{EAGLE} was already noted by \citet{turner2016}
in a study of the $z\approx3.5$ intergalactic medium and has important
implications for our work too.  A direct comparison with
H$\alpha$-based SAMI observations should be done using warm gas (since
H$\alpha$ emitting gas has a temperature T $\sim10^4$ K). However, the
paucity of such gas in the outflows of our simulated galaxies could
lead to misleading results. Specifically, in edge-on projections:
underestimation of the outflowing gas mass$/$incidence of galactic
winds (that may be there but just too hot to be detected in H$\alpha$)
and poor sampling of the extraplanar gas. In the next section we will
show an example of this problem.


\subsection{Disc component and off-plane gas}
\label{veldistr3}

\begin{figure*}
\centering
\includegraphics[width=8.7cm]{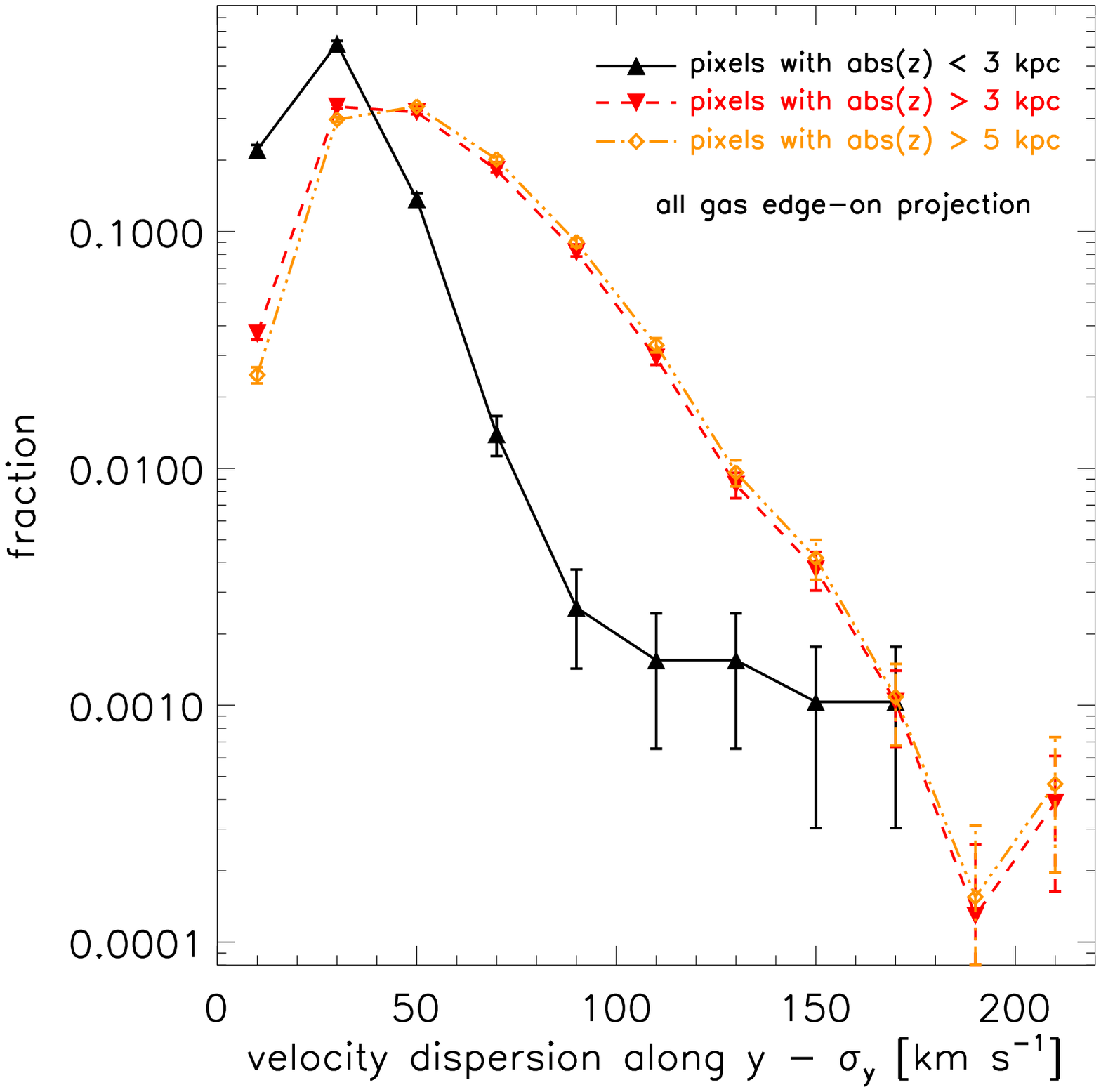}
\includegraphics[width=8.7cm]{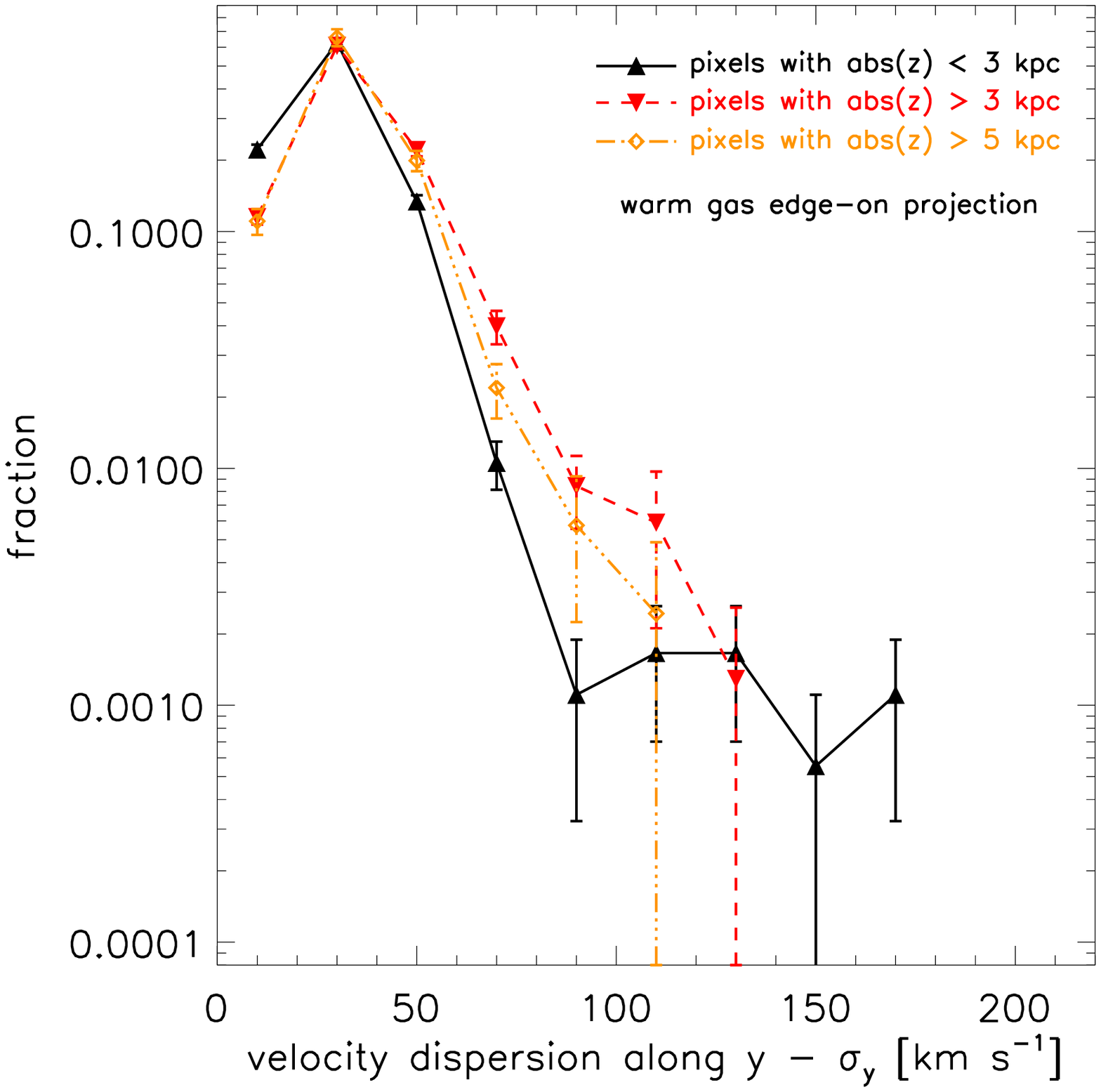}
\caption{Pixelated velocity dispersion probability distributions
  (cf. Fig. \ref{fig_total_vs_ha} and the beginning of Section
  \ref{total_vs_ha}) calculated by assuming different cuts in the z
  direction of the edge-on (xz view) $\sigma_{\rm y}$-maps (with pixel
  size $=2$ kpc) to separate the galactic disc component from the
  extraplanar one. {\bf Left panel}: all gas. {\bf Right panel}: warm
  gas. Black filled triangles and solid line, (internal) pixels with
  abs(z) $<$ $3$ kpc. Red filled inverted triangles and dashed line,
  (external) pixels with abs(z) $>$ $3$ kpc. Orange diamonds and
  triple dot-dashed line, (external) pixels with abs(z) $>$ $5$
  kpc. Errors are Poissonian. In the all gas case, excluding the
  galactic disc shifts the $\sigma_{\rm y}$-distribution to larger
  velocity dispersions. In the warm gas case, the off-plane
  distributions are poorly sampled and only marginally different from
  the disc one due to the scarcity of extraplanar gas at T $\sim10^4$
  K.}
\label{fig_hist_disc}
\end{figure*}

We focus on the difference between gas in the disc and extraplanar
gas. To do so, we take the edge-on $\sigma_{\rm y}$-maps (xz view) and
restrict the number of pixels in the direction perpendicular to the
disc plane by applying various vertical cuts. The resulting pixelated
velocity dispersion distributions (including Poissonian errors) are
shown in Fig. \ref{fig_hist_disc}. The two panels refer to all gas
(left) and warm gas (right).

In the left panel, when all gas and only pixels with abs(z) $<$ $3$
kpc (black filled triangles and solid line) are included\footnote{
  Here and throughout the paper, the vertical cuts are considered from
  the edge of the pixels, not the centre.}, the distribution shows a
sizeable peak at $\sigma_{\rm y}=30$ km s$^{-1}$ and then quickly
declines to low fractions. With this pixel selection, we are targeting
only a thin layer of gas in the edge-on galactic discs. Red filled
inverted triangles and the dashed line represent the complementary
distribution calculated using only pixels with abs(z) $>$ $3$ kpc. In
this case, the distribution is shifted to larger $\sigma$-values than
before, while the low-$\sigma$ part is greatly reduced. This
demonstrates that the low-$\sigma$ peak is indeed associated mainly
with the galactic disc, in qualitative agreement with
\citet{ho2014}. In the left panel of Fig. \ref{fig_hist_disc} we also
show the velocity dispersion distribution (for all gas) calculated
using only pixels with abs(z) $>$ $5$ kpc (orange diamonds and triple
dot-dashed lines). The low-$\sigma$ peak and high-$\sigma$ tail
become, respectively, slightly less and more important when moving
further away from the galactic disc, supporting our previous
conclusion.

The right panel of Fig. \ref{fig_hist_disc} illustrates how using only
warm gas can be misleading, in the framework of EAGLE simulations. All
three distributions are very similar (despite the fact that they probe
rather different environments) and more noisy at $\sigma_{\rm y}>50$
km s$^{-1}$ than the corresponding distributions for all gas. In the edge-on
warm $\sigma_{\rm y}$-maps there are only a few pixels with abs(z) $>$
$3$ and (especially) $5$ kpc, therefore poor sampling affects the
results in these two cases. Considering only warm gas in EAGLE would
lead to the wrong conclusion that planar and extraplanar gas
components are kinematically similar. This is a consequence of the
thermal implementation of stellar feedback that produces hot
outflows. When all gas (warm \& hot) is considered, the extraplanar
(mainly hot) gas is kinematically clearly distinct from the (mainly
warm) disc (left panel of Fig. \ref{fig_hist_disc}).

In the rest of the paper, whenever possible and appropriate (e.g. to
study face-on velocity dispersion distributions or the impact of
general galactic properties like M$_{\star}$ and sSFR) we will show
results obtained using warm gas. However, including all gas will be
necessary to ensure a robust description of galactic winds and outflow
signatures (see e.g. Section \ref{asym}).

\subsection{Outflows}
\label{gal_winds}

\begin{figure}
\centering
\includegraphics[width=8.7cm]{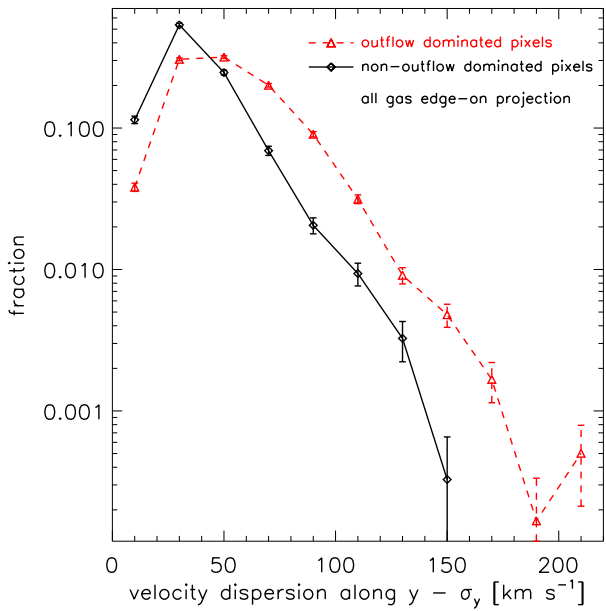}
\caption{Pixelated velocity dispersion probability distributions
  (cf. Fig. \ref{fig_total_vs_ha} and the beginning of Section
  \ref{total_vs_ha}) for outflow dominated (red triangles and dashed
  line) and non-outflow dominated (black diamonds and solid line)
  pixels. Errors are Poissonian and we consider all gas in the edge-on
  xz projection $-$ $\sigma_{\rm y}$. The velocity dispersion
  distribution based on outflow dominated pixels is more prominent at
  high-$\sigma$.}
\label{fig_v_esc}
\end{figure}

In the previous section we have demonstrated how the low-$\sigma$ part
of the edge-on velocity dispersion distribution is associated with the
galactic disc. Now, we study the origin of the high-$\sigma$ tail. In
the xz edge-on projection, we consider a pixel of the $\sigma$-map as
{\it outflow dominated} if the density weighted vertical velocity of
its gas particles, $\langle v_{\rm z}\rangle$, is positive in the
semi-plane with z$_{\rm gas}>0$ kpc or negative in the negative
z$_{\rm gas}$ semi-plane (i.e. if the gas particles contributing to
the pixel are predominantly moving away from the galactic
plane). Otherwise, a pixel is flagged as {\it non-outflow
  dominated}\footnote{We mask pixels in the galactic plane (i.e. those
  with $-1 < $ z$_{\rm gas}/$kpc $<1$), since their
  outflowing/non-outflowing status is undefined.}. The corresponding
$\sigma_{\rm y}$-distributions are shown in
Fig. \ref{fig_v_esc}. Errors are Poissonian.

The non-outflow dominated $\sigma$-distribution (black diamonds and
solid line) resembles the abs(z) $<$ $3$ kpc distribution
(i.e. associated with the galactic disc) visible in the left panel of
Fig. \ref{fig_hist_disc} (black filled triangles and solid line). With
respect to these two diagrams, the outflow dominated
$\sigma$-distribution of Fig. \ref{fig_v_esc} (red triangles and
dashed line) is shifted to higher velocity dispersions and has a more
statistically prominent high-$\sigma$ tail (in agreement with the
abs(z) $>$ $3$ and $5$ kpc distributions of
Fig. \ref{fig_hist_disc}). When only extraplanar gas is considered
(i.e. pixels with abs(z) $>$ $3$ kpc), we find that in $42$ out of
$43$ simulated galaxies, the number of outflow dominated pixels is
greater than the number of non-outflow dominated pixels. On average,
this excess is a factor of $\sim3.2$ times and can be up to $>15$
times.

\begin{figure*}
\centering 
\includegraphics[width=8.7cm]{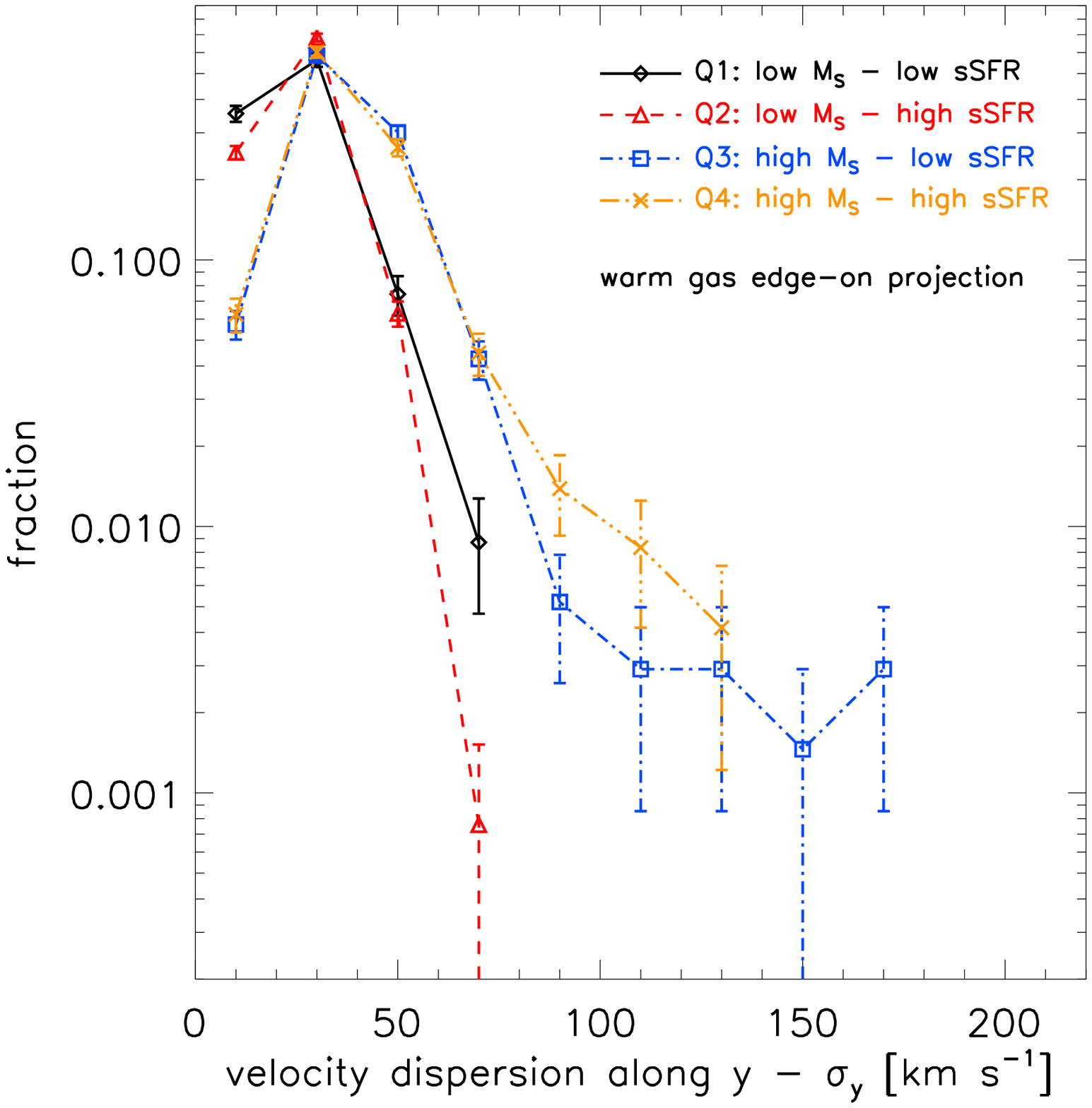}
\includegraphics[width=8.7cm]{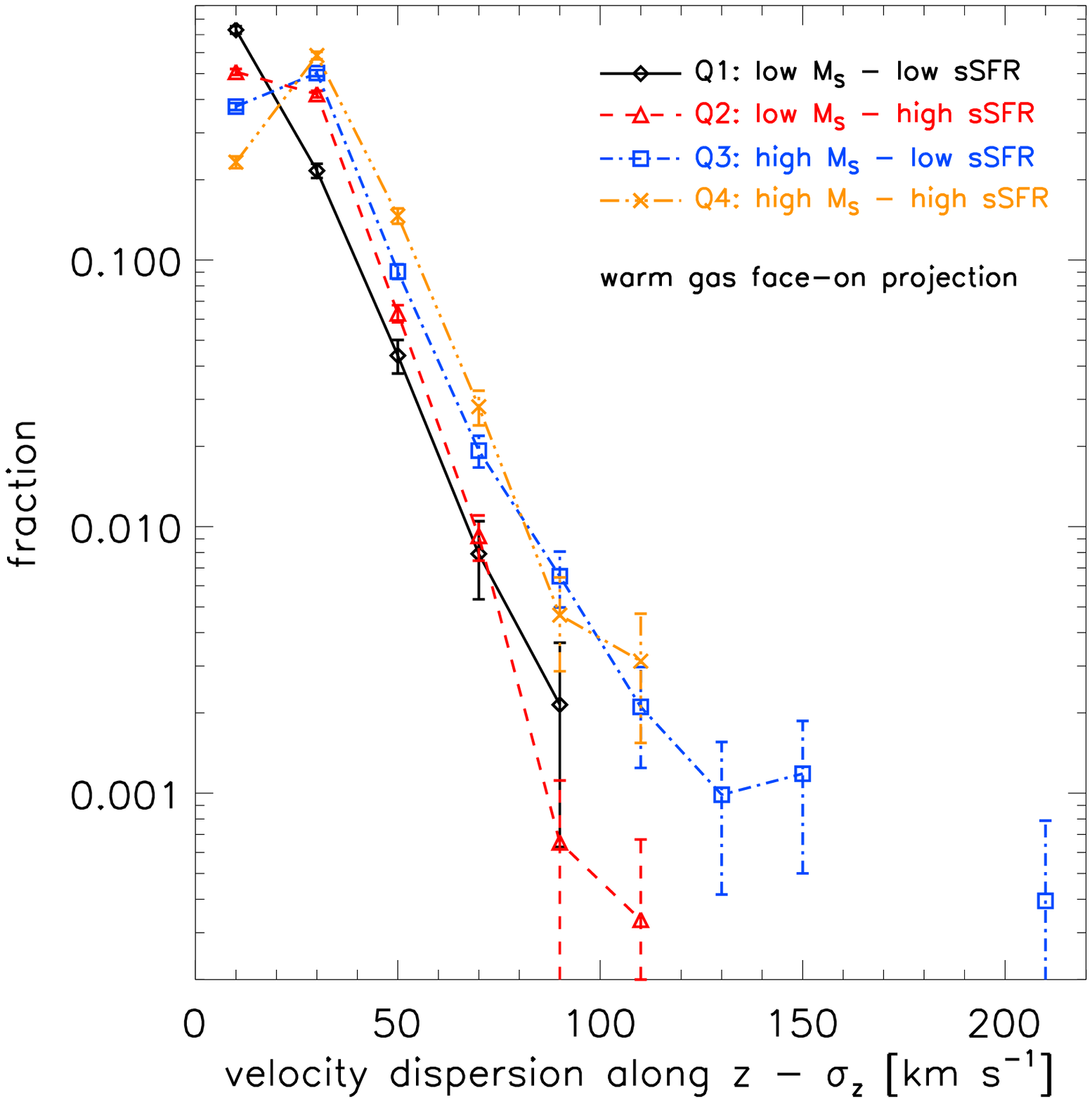}
\caption{Pixelated velocity dispersion probability distributions
  (cf. Fig. \ref{fig_total_vs_ha} and the beginning of Section
  \ref{total_vs_ha}) for galaxies in the four M$_{\star}-$sSFR
  quadrants defined in the bottom right panel of
  Fig. \ref{fig_gal11_proj}: Q$1$ = low M$_{\star}-$ low sSFR,
  \color{red} Q$2$ \color{black}= low M$_{\star}-$ high sSFR,
  \color{blue} Q$3$ \color{black}= high M$_{\star}-$ low sSFR and
  \color{orange} Q$4$ \color{black}= high M$_{\star}-$ high sSFR. {\bf
    Left panel}: edge-on xz projection $-$ $\sigma_{\rm y}$. {\bf
    Right panel}: face-on xy projection $-$ $\sigma_{\rm z}$. Errors
  are Poissonian and we consider only warm gas. In general, galaxies
  with higher M$_{\star}$ present a more extended
  $\sigma$-distribution, while the sSFR has a secondary effect with
  respect to stellar mass.}
\label{fig_hist_tot}
\end{figure*}

\begin{figure*}
  \centering
  \includegraphics[width=8.4cm]{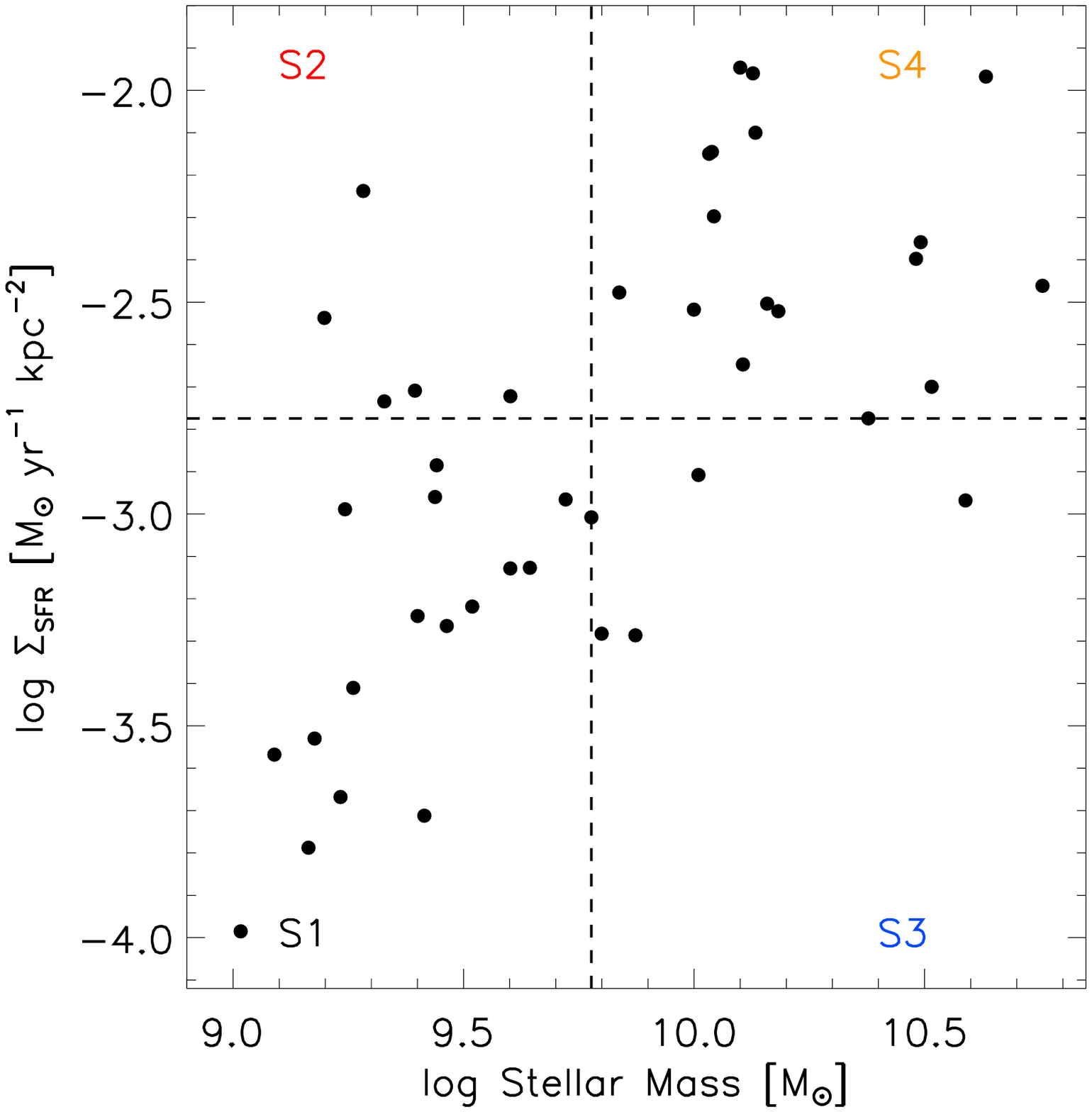}

  \includegraphics[width=8.4cm]{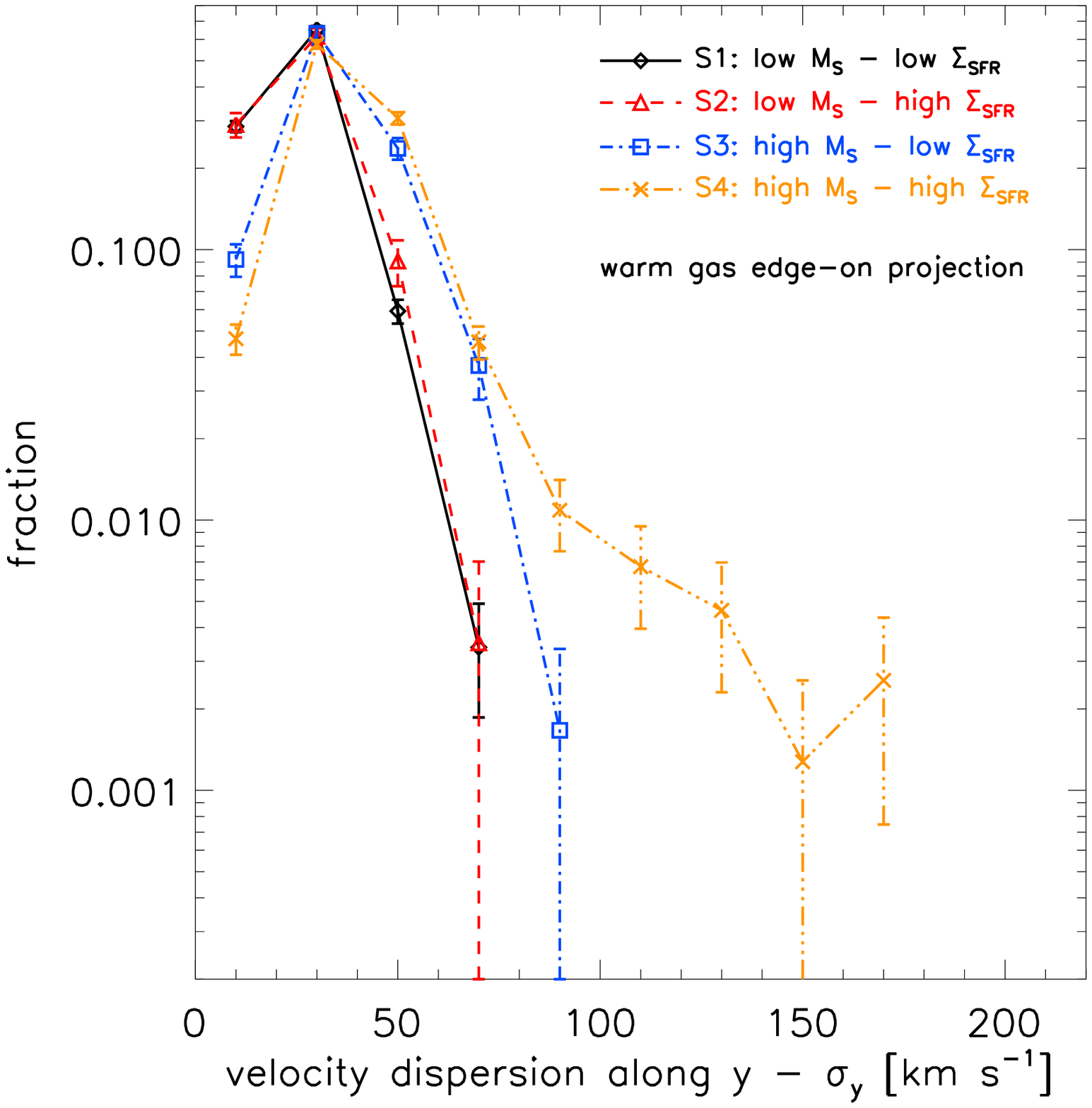}
  \includegraphics[width=8.4cm]{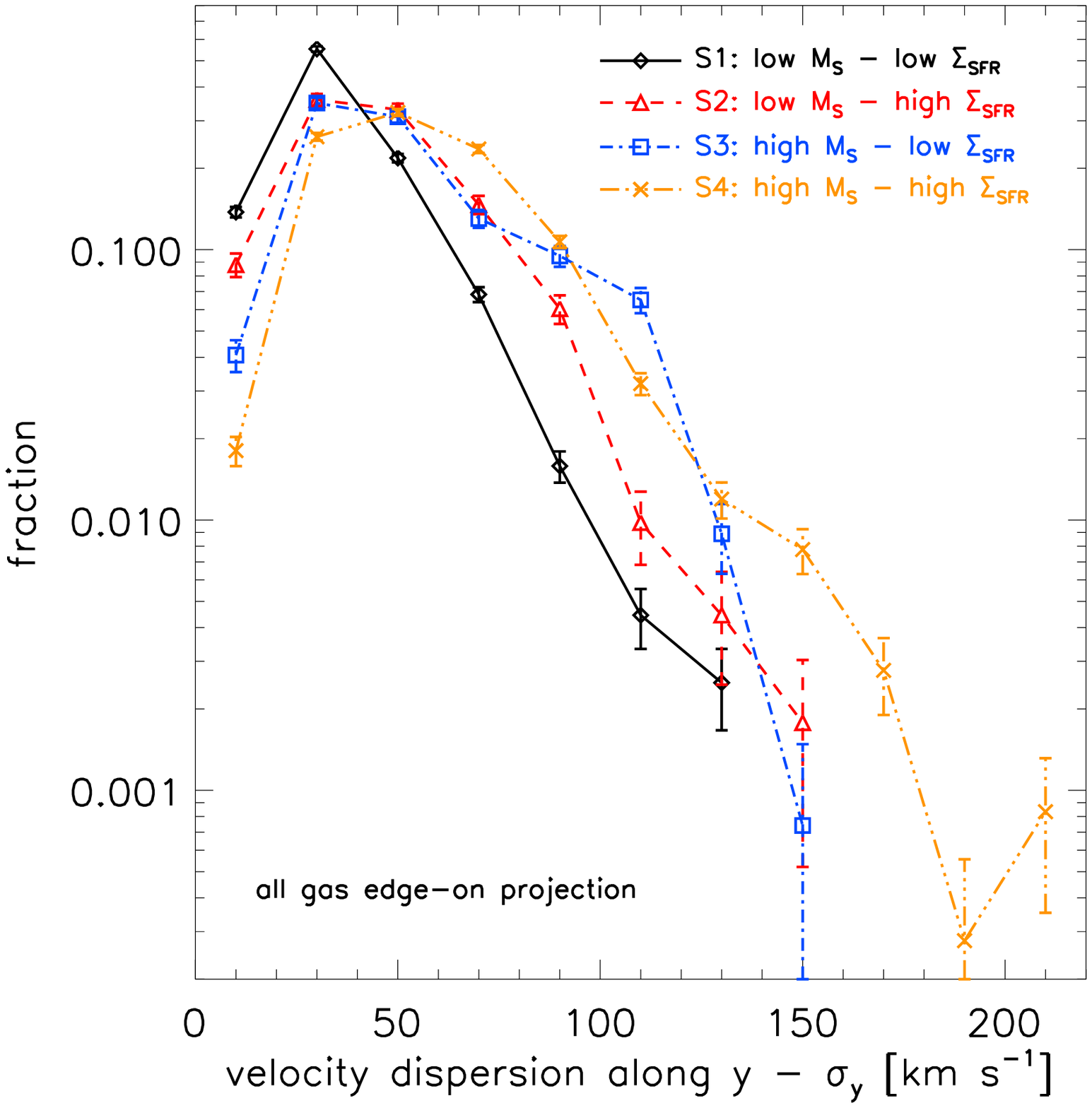}
  \caption{{\bf Top panel}: relation between star formation rate
    surface density, $\Sigma_{\rm SFR}$, and M$_{\star}$ for our
    simulated galaxies. $\Sigma_{\rm SFR}={\rm SFR}/(2\pi r_{50}^2)$,
    where SFR is the total star formation rate and $r_{50}$ is the
    radius within which half of the galaxy stellar mass is
    included. The plot is divided in four sectors: S$1$ = low
    M$_{\star}-$ low $\Sigma_{\rm SFR}$, \color{red} S$2$
    \color{black}= low M$_{\star}-$ high $\Sigma_{\rm SFR}$,
    \color{blue} S$3$ \color{black}= high M$_{\star}-$ low
    $\Sigma_{\rm SFR}$ and \color{orange} S$4$ \color{black}= high
    M$_{\star}-$ high $\Sigma_{\rm SFR}$ (see Section
    \ref{sigma_SFR}). {\bf Bottom panels}: pixelated velocity
    dispersion probability distributions
    (cf. Fig. \ref{fig_total_vs_ha} and the beginning of Section
    \ref{total_vs_ha}) for galaxies in the four sectors defined in the
    top panel. Errors are Poissonian. We only consider the edge-on xz
    projection $-$ $\sigma_{\rm y}$. Left panel: warm gas. As in the
    previous section, a trend with stellar mass is visible. Right
    panel: all gas. At fixed M$_{\star}$, galaxies with high
    $\Sigma_{\rm SFR}$ give rise to a $\sigma$-distribution shifted to
    larger velocity dispersions compared to galaxies with low
    $\Sigma_{\rm SFR}$.}
\label{fig_hist_sfrd}
\end{figure*}

Our results are qualitatively consistent with the observations of
\citet{ho2014}: in low redshift disc galaxies, the low-$\sigma$
component of the velocity dispersion distribution is associated with
the (rotationally supported) disc, while the high-$\sigma$ component
mainly traces extraplanar, outflowing gas. In the case of strong
disc-halo interactions through galactic winds, the velocity dispersion
of the extraplanar gas is broadened (up to $300$ km s$^{-1}$ in the extreme
case of M82) by both the turbulent motion of the outflowing gas and
line splitting caused by emissions from the approaching and receding
sides of the outflow cones \citep[][and references therein]{ho2016}.

There is an important caveat to consider here. In this work,
outflowing material includes both particles that are actually leaving
their host galaxy and particles that will eventually stop and fall
back to the disc. At this stage, our numerical analysis is not able to
differentiate between these two components (but this also applies to
observations). Theoretical predictions on how much gas actually
escapes from galaxies are crucial, since the escaping mass is very
hard to measure observationally \citep{joss2003, joss2007}. Recent
simulations run by different groups indicate that wind recycling
becomes particularly important at $z<1$ and galaxies of all masses
reaccrete more than $50$\% of the expelled gas
\citep[e.g.][]{oppe2010, nelson2015, christensen2016,
  alcazar2016}. This point will be addressed in an upcoming paper
(Crain et al. in prep). Note that the thermal$/$buoyant winds in our
EAGLE discs will allow particles without enough velocity$/$thermal
energy to escape to float up to the top of the galactic halo
\citep{bower2016}.

\section{Impact of stellar mass, specific SFR, SFR surface density and
  gas temperature}
\label{veldistr1}

In this section, we study the impact of different galactic properties
on the overall shape of the velocity dispersion distribution. Since we
do not focus primarily on the high-$\sigma$ tail associated with
outflows, results are presented for warm gas to better compare with
the observational analysis of \citet{ho2014, ho2016}. We begin with
stellar mass and the specific star formation
rate. Fig. \ref{fig_hist_tot} shows the result: the edge-on xz
projection $-$ $\sigma_{\rm y}$ in the left panel and the face-on xy
projection $-$ $\sigma_{\rm z}$ in the right panel (errors are
Poissonian). We divided our galaxies according to the four
M$_{\star}-$sSFR quadrants in the bottom right panel of
Fig. \ref{fig_gal11_proj} (the same colour code applies).

A clear trend with stellar mass is visible in both panels, while the
sSFR appears to have a secondary effect. It is interesting to note
how, especially in the edge-on projection (left panel), the velocity
dispersion distribution of low mass galaxies (black diamonds + solid
line and red triangles + dashed line, respectively associated with
Q$1$ and Q$2$) declines shortly after the peak at $30$ km s$^{-1}$
(associated with the disc) and drops to zero already at $70$ km s$^{-1}$. The
predominance of the disc component in Q$1$ and Q$2$ is present also
when all gas distributions (not shown here) are used. The
$\sigma$-distributions of galaxies with high-M$_{\star}$ (blue squares
+ dot-dashed line and orange crosses + triple dot-dashed line,
respectively associated with Q$3$ and Q$4$) are more extended, and
prominent at high-$\sigma$, than those of low mass galaxies
(regardless of the range in sSFR). At $\sigma<100$ km s$^{-1}$, this is due
to the fact that in the synthetic sample warm gas mainly traces the
galactic disc and objects in Q$3$ and Q$4$ are generally bigger. Due
to the SFR$-$M$_{\star}$ relation and the fact that in EAGLE there is
a direct connection between SFR and stellar feedback, these objects
also have higher outflowing activities than galaxies in Q$1$ and
Q$2$. Despite the lack of warm gas in EAGLE's galactic winds, this
causes the broadening of the $\sigma$-distributions to higher velocity
dispersion.


Our simulated face-on distributions, with
$\max(\sigma_{\rm z})=201.31$ km s$^{-1}$, do not extend as far as the
velocity dispersion diagram of the SDSS J$0900$ galaxy in
\citet{ho2014}, with $\max(\sigma_{\rm z})\sim 450$ km s$^{-1}$. This
might be in part due to the fact that SDSS J$0900$ is more massive,
$\log($M$_{\star}/$M$_\odot)=10.8$, and has a higher SFR ($\sim5-15$
M$_\odot$ yr$^{-1}$, depending on the adopted SFR indicator) than
objects in our synthetic sample, but could also indicate that the
effect of EAGLE stellar feedback on gas kinematics is too
weak. Simulated galaxies in Q$2$, Q$3$ and Q$4$ have distributions
that reach $\sigma_{\rm z}>100$ km s$^{-1}$\footnote{In the edge-on
  projection, only galaxies with high-M$_{\star}$ (Q$3$ and Q$4$) have
  distributions with $\max(\sigma_{\rm y})>100$ km s$^{-1}$.}, which
is the starting point of the broad kinematic component associated with
outflowing gas in SDSS J$0900$.

\subsection{SFR surface density}
\label{sigma_SFR}

\begin{figure*}
\centering
\includegraphics[width=8.7cm]{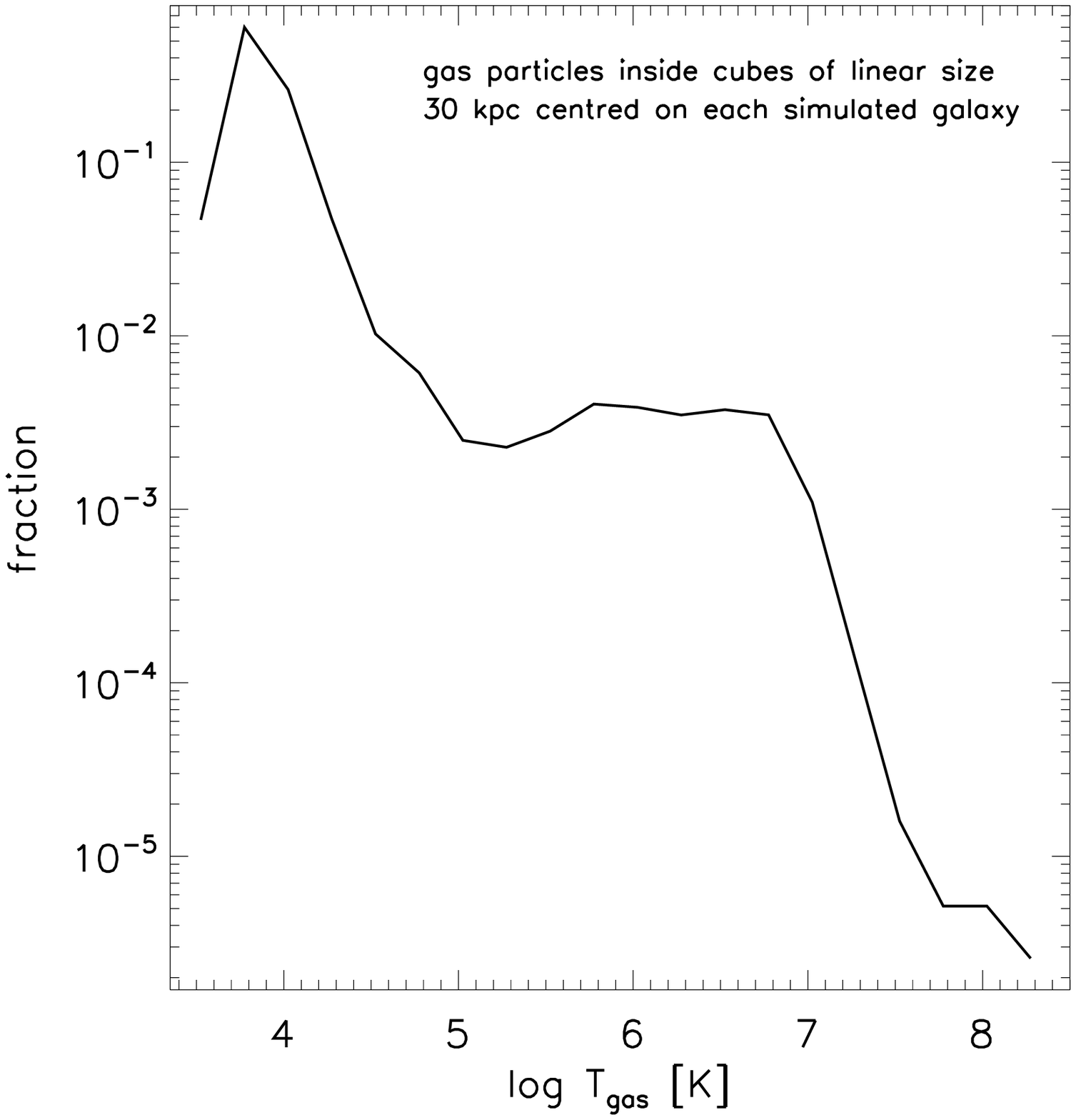}
\includegraphics[width=8.7cm]{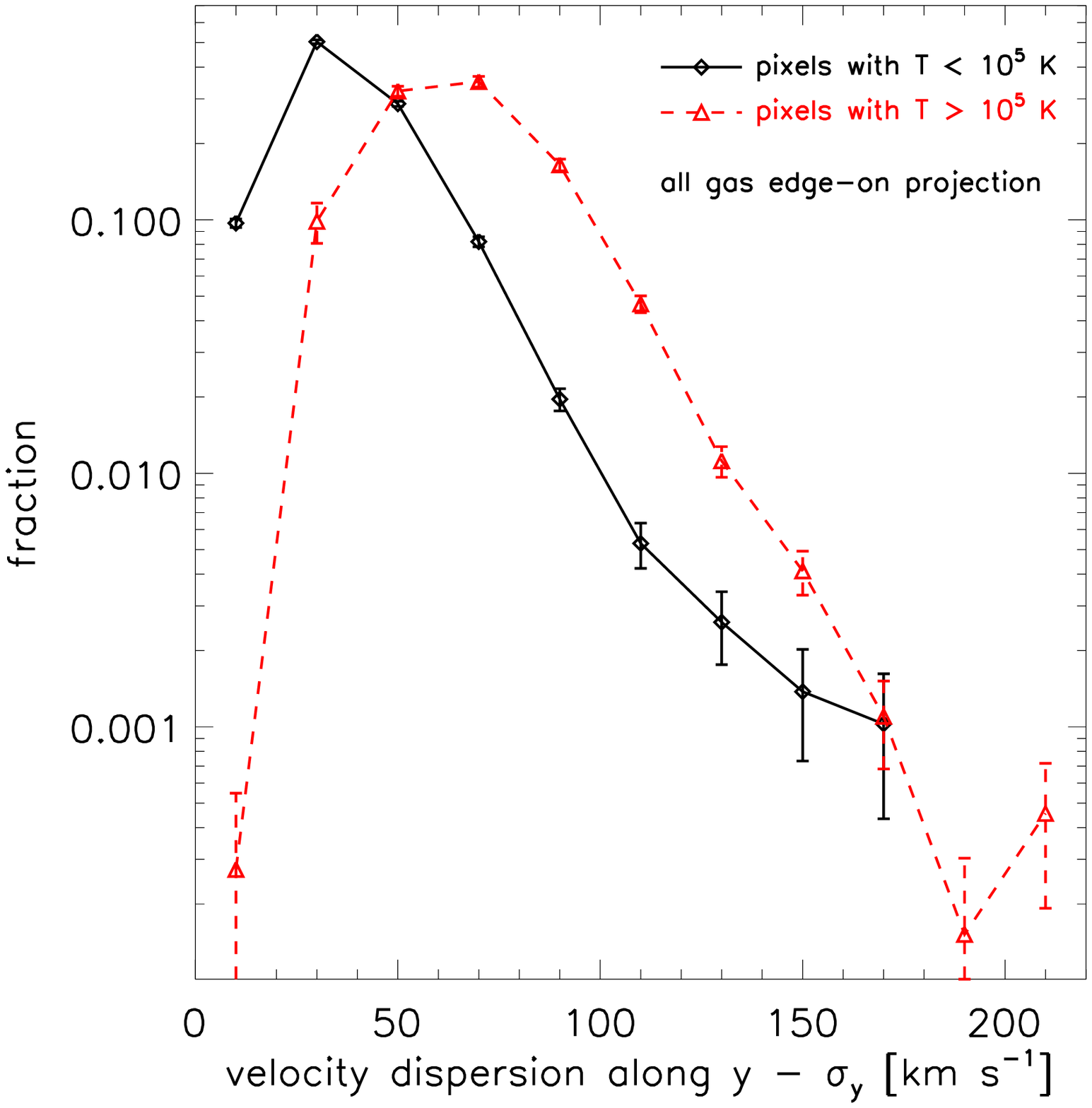}
\caption{{\bf Left panel}: average temperature distribution of gas
  particles inside cubes of linear size $30$ kpc centred on each
  simulated galaxy. Two distinct regions are visible in the gas
  temperature histogram: the bulk of particles with T $<10^5$ K, and a
  tail of particles with T $\ge 10^5$ K. {\bf Right panel}: effect of
  gas temperature on the pixelated velocity dispersion probability
  distribution (cf. Fig. \ref{fig_total_vs_ha} and the beginning of
  Section \ref{total_vs_ha}) of galaxies. Black diamonds and solid
  line: pixels where the density weighted gas temperature is T $<10^5$
  K. Red triangles and dashed line: pixels with T $\ge10^5$ K. Errors
  are Poissonian and we only consider the edge-on xz projection $-$
  $\sigma_{\rm y}$. Pixels with temperature T $<10^5$ K trace the
  galactic disc (low-$\sigma$), while those with T $\ge10^5$ K are
  associated with higher-$\sigma$ (i.e. extraplanar, outflowing gas).}
\label{fig_hist_temp_cut}
\end{figure*}

\citet{ho2016} found that, on average, wind galaxies have higher star
formation rate surface densities than those without strong wind
signatures. We checked this result with our simulated sample in
Fig. \ref{fig_hist_sfrd}. Following \citet{ho2016}, the star formation
rate surface density is defined as $\Sigma_{\rm SFR}=$
SFR$/(2\pi r_{50}^2)$, where SFR is the total star formation rate of
the object as determined by {\small SUBFIND} and $r_{50}$ is the
radius within which half of the galaxy stellar mass is included. The
top panel of Fig. \ref{fig_hist_sfrd} shows the
$\Sigma_{\rm SFR}-$M$_{\star}$ relation of EAGLE galaxies. As for SFR
and M$_{\star}$, the two quantities are positively correlated. The
vertical and horizontal dashed lines mark, respectively, the median
stellar mass, $\log(\tilde{{\rm M}}_{\star}/$M$_\odot)=9.78$, and
median star formation rate surface density,
$\log\big(\tilde{\rm \Sigma}_{\rm SFR}/[$M$_\odot$ yr$^{-1}$
kpc$^{-2}]\big)=-2.77$. We divided the plot in four sectors: S$1$ =
low M$_{\star}-$ low $\Sigma_{\rm SFR}$ (16 objects), \color{red} S$2$
\color{black}= low M$_{\star}-$ high $\Sigma_{\rm SFR}$ (5 objects),
\color{blue} S$3$ \color{black}= high M$_{\star}-$ low
$\Sigma_{\rm SFR}$ (6 objects) and \color{orange} S$4$ \color{black}=
high M$_{\star}-$ high $\Sigma_{\rm SFR}$ (16 objects). The
corresponding velocity dispersion distributions are shown in the
bottom panels of Fig.  \ref{fig_hist_sfrd}: warm gas on the left and
all gas on the right (we only consider the edge-on xz projection $-$
$\sigma_{\rm y}$).

We start by considering the warm gas case (left panel of Fig.
\ref{fig_hist_sfrd}). Trends are similar to those of the left panel of
Fig. \ref{fig_hist_tot}. At low masses (S$1$ and S$2$), the velocity
dispersion distributions of galaxies with low and high
$\Sigma_{\rm SFR}$ are almost identical (black diamonds + solid line
and red triangles + dashed line), with a narrow peak at $30$ km s$^{-1}$.
These galaxies have relatively low SFRs and weak outflowing
activities, therefore warm gas mainly traces their galactic discs (of
similar size). The probability distributions of high-mass galaxies
(blue squares + dot-dashed line and orange crosses + triple dot-dashed
line, respectively associated with S$3$ and S$4$) are shifted to
larger values than those of low-mass galaxies. As discussed in the
previous section, part of the shift is driven by the increase in
stellar mass, but a correlation with the SFR surface density is now
visible (a more extended high-$\sigma$ tail for objects in S$4$ with
high $\Sigma_{\rm SFR}$).

Patterns are different in the all gas case (right panel of Fig.
\ref{fig_hist_sfrd}). A trend with stellar mass is still present
(black \& red vs blue \& orange points and lines), but, at fixed
M$_{\star}$, galaxies with high $\Sigma_{\rm SFR}$ give rise to a
$\sigma$-distribution shifted to larger velocity dispersions compared
to galaxies with low $\Sigma_{\rm SFR}$ (red \& orange vs black \&
blue points and lines). As in the observations of \citet{ho2016}, this
result, only partially visible before due to the lack of warm gas in
the outflows of EAGLE galaxies, indicates that the star formation rate
surface density correlates with the outflowing activity even when
$\Sigma_{\rm SFR}$ is rather low, as it is the case of our simulated
galaxies (we will explore the correlation in more detail in Section
\ref{asym}). According to \citet{heckman2002}, starburst-driven winds
are observed to be ubiquitous in galaxies with
$\log\big(\Sigma_{\rm SFR}/[$M$_\odot$ yr$^{-1}$ kpc$^{-2}]\big) > -1$
\citep[see also the results of][]{sharma2017}. In our sample,
$\log\big(\Sigma_{\rm SFR,\max}/[$M$_\odot$ yr$^{-1}$
kpc$^{-2}]\big) = -1.95$, almost a dex lower\footnote{In
  \citet{ho2016}, winds are seen at
  $-3\lesssim\log\big(\Sigma_{\rm SFR}/[$M$_\odot$ yr$^{-1}$
  kpc$^{-2}]\big) \lesssim -1.5$.}.

\subsection{The role of gas temperature}
\label{veldistr_temp}

Since the predictive power of realistic numerical simulations allows
us to study the effect of additional galactic properties, which are
usually hard to measure observationally, we looked for a way to better
relate the shape of the velocity dispersion distribution to feedback
processes. In EAGLE, stellar feedback is implemented thermally, and we
found that temperature is a good proxy to distinguish low- and
high-$\sigma$ parts in our simulated galaxies. The average temperature
histogram of gas particles inside cubes of volume ($30$ kpc)$^3$
centred on each simulated galaxy splits into two regions: the bulk of
particles with T $<10^5$ K, and a tail of particles with T $\ge 10^5$
K (see the left panel of Fig. \ref{fig_hist_temp_cut}). For this
reason, we divided pixels in the velocity dispersion distribution
where the density weighted gas temperature is above and below $10^5$
K. The result is visible in the right panel of
Fig. \ref{fig_hist_temp_cut} (we consider all gas in the edge-on xz
projection $-$ $\sigma_{\rm y}$, errors are Poissonian). The condition
on temperature produces two very different probability
distributions. When pixels with T $<10^5$ K are selected (black
diamonds and solid line), the distribution peaks at
$\sigma_{\rm y}=30$ km s$^{-1}$, then quickly drops to low
fractions. On the other hand, pixels with T $\ge 10^5$ K give rise to
a distribution shifted to larger velocity dispersions and where the
low-$\sigma$ section is less prominent (red triangles and dashed
line). These trends, which we find are also visible in the face-on xy
projection $-$ $\sigma_{\rm z}$, support the results of the previous
sections (see in particular the left panel of
Fig. \ref{fig_hist_disc} and Fig. \ref{fig_v_esc}).

Thus, EAGLE simulations of low redshift disc galaxies indicate a
direct correlation between the thermal state of the gas and its state
of motion as described by the velocity dispersion distribution:
\begin{itemize}
  \renewcommand{\labelitemi}{$\bullet$}
\item Low-$\sigma$ peak $\,\,\Leftrightarrow\,$ galactic disc$/$gas
  with T $<10^5$ K;
\item High-$\sigma$ tail $\,\Leftrightarrow\,$ outflows$/$gas with T
  $\ge 10^5$ K.
\end{itemize}
The real picture is certainly more complicated than this. For example,
blobs of cold gas at relatively high density could be entrained in
hot, diffuse winds \citep{veilleux2005, cooper2008,
  cooper2009}. Despite the simplifications made in our analysis, the
predicted correlation between EAGLE's thermal$/$buoyant outflows and
high temperature gas can be very useful to guide and interpret real
observations.

\section{SIGNATURES OF WINDS: GAS KINEMATICS IN EAGLE AND SAMI
  GALAXIES}
\label{asym}

\citet{ho2016} proposed an empirical identification of wind-dominated
SAMI galaxies. In this section, we apply the same methodology to our
simulated sample. The authors defined two dimensionless quantities to
measure the (ionised gas) extraplanar velocity dispersion and
asymmetry of the velocity field. The first quantity is the velocity
dispersion to rotation ratio parameter:
\begin{equation}
  \eta_{50}=\sigma_{50}/v_{\rm rot},
\end{equation}
where $\sigma_{50}$ is the median velocity dispersion of all pixels
outside $\tilde{r}_{\rm e}$ (the $r$-band effective radius increased
by approximately 1 arcsec to reduce the effect of beam smearing) with
signal-to-noise in H$\alpha$ $-$ S$/$N(H$\alpha$) $-$ greater than
$5$.  $v_{\rm rot}$ is the maximum rotation velocity measured from the
pixels along the optical major axis (for galaxies without sufficient
spatial coverage, they used the stellar mass Tully-Fisher relation to
infer $v_{\rm rot}$). The second quantity is the asymmetry parameter:
\begin{equation}
\label{eq_xi}
\xi={\rm std}\,\bigg( \frac{v_{\rm gas}-v_{\rm
    gas,flipped}}{\sqrt{{\rm Err}(v_{\rm gas})^2+{\rm Err}(v_{\rm gas,flipped})^2}} \bigg),
\end{equation}
where ${\rm std}=$ standard deviation. To obtain $\xi$, the authors
first flipped the line-of-sight velocity map over the galaxy major
axis, $v_{\rm gas,flipped}$, and then subtracted the flipped map from
the original one, $v_{\rm gas}$.  ${\rm Err}(v_{\rm gas})$ and
${\rm Err}(v_{\rm gas,flipped})$ are the corresponding $1\sigma$ error
maps from {\small{LZIFU}}. The standard deviation is again calculated
taking into account only pixels outside $\tilde{r}_{\rm e}$ with
S$/$N(H$\alpha$) $>5$.

\citet{ho2016} used $\eta_{50}$ and $\xi$ to quantify the strength of
disc-halo interactions and to distinguish galactic winds from extended
diffuse ionised gas (eDIG) in a sample of $40$ low redshift disc
galaxies. Since galactic winds both perturb the symmetry of the
extraplanar gas velocity and increase the extraplanar emission line
widths, they should show high $\xi$ and high $\eta_{50}$. On the other
hand, eDIG is more closely tied to the velocity field of the galaxy
and therefore should result in low $\xi$ and low $\eta_{50}$.

\begin{figure*}
\centering
\vspace{0.6cm}
\includegraphics[width=8.7cm]{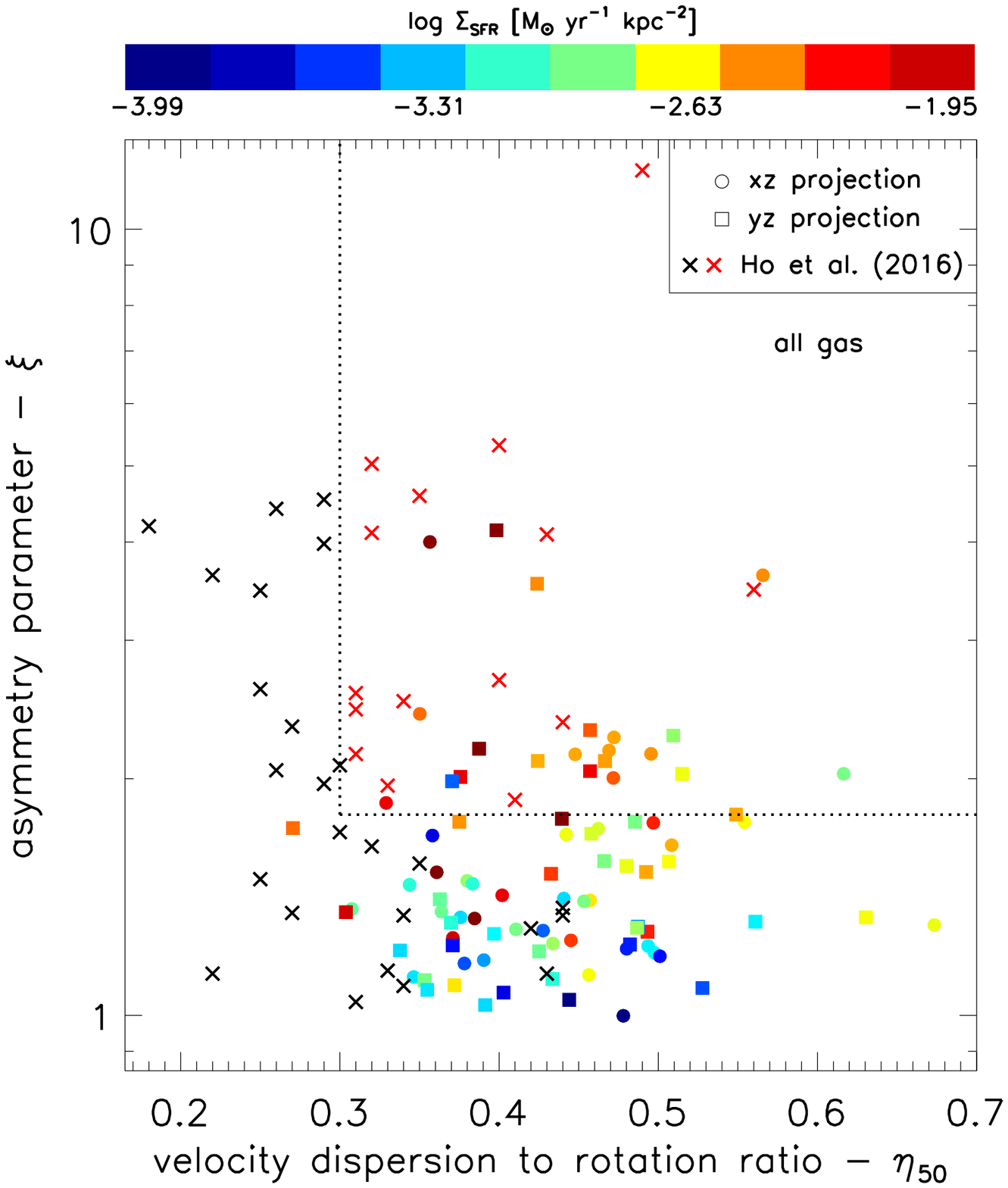}
\includegraphics[width=8.7cm]{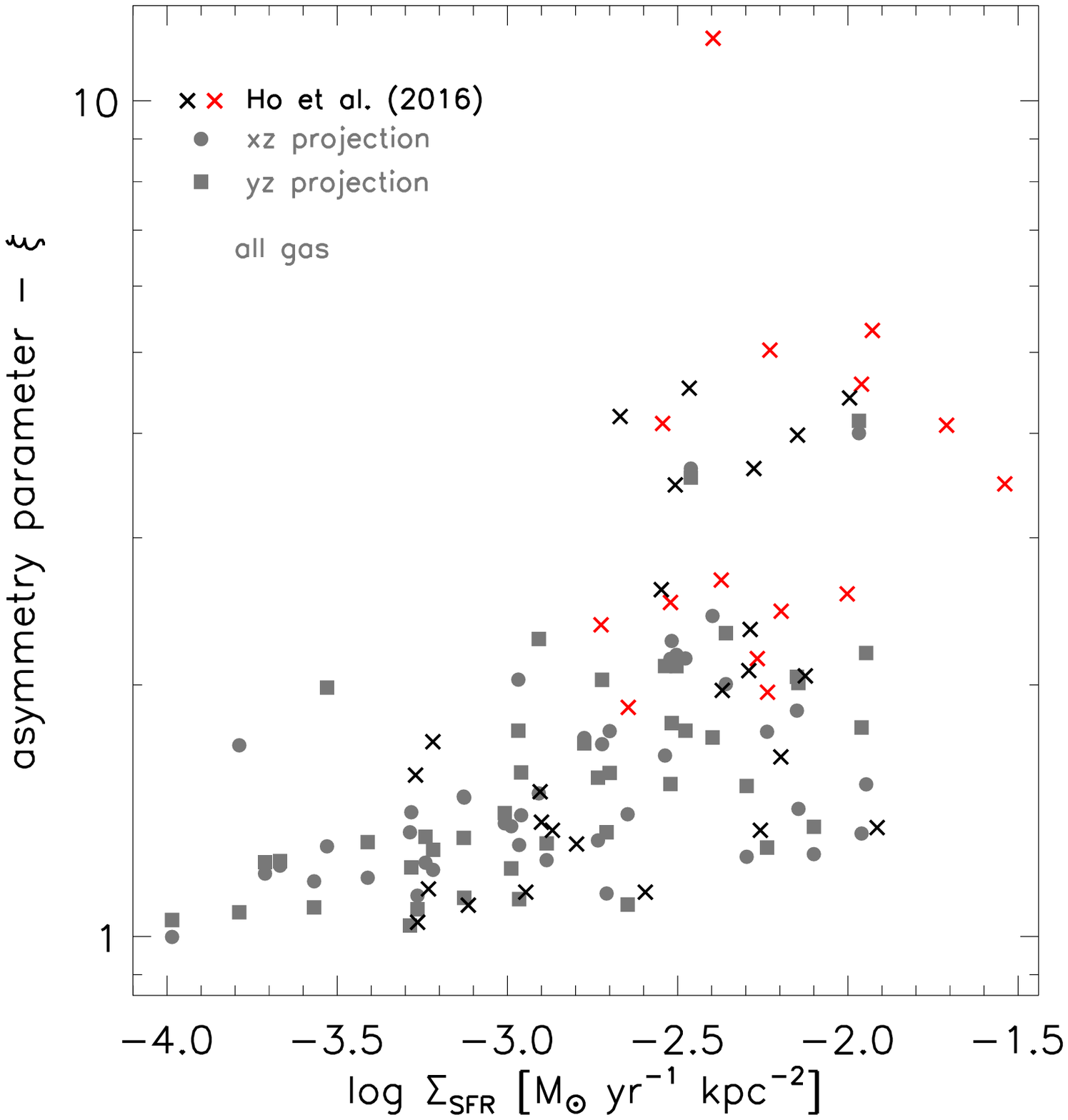}
\caption{{\bf Left panel}: asymmetry parameter $\xi$ vs velocity
  dispersion to rotation ratio parameter $\eta_{50}$. Dots: simulated
  edge-on xz projection. Squares: simulated edge-on yz
  projection. Synthetic data are colour coded according to the SFR
  surface density of the parent galaxy,
  $\Sigma_{\rm SFR}={\rm SFR}/(2\pi r_{50}^2)$. The vertical and
  horizontal dotted lines mark the observational limits above which
  galaxies show strong disc-halo interactions according to
  \citet{ho2016}, $\eta_{50}>0.3$ and $\xi>1.8$. Crosses:
  observational SAMI data, where red \color{red} \textsf{x}
  \color{black} symbols represent wind-dominated galaxies and black
  \textsf{x} symbols represent galaxies without strong wind
  signatures. {\bf Right panel}: correlation between the asymmetry
  parameter $\xi$ and $\Sigma_{\rm SFR}$. Grey dots and squares:
  simulated galaxies in the edge-on xz and yz projections,
  respectively. Crosses: observational data from \citet{ho2016}. In
  both panels, the range of the simulated parameters (determined using
  all gas) is broadly consistent with the observations. A clear
  positive $\xi-\Sigma_{\rm SFR}$ correlation is visible in both EAGLE
  and SAMI data.}
\label{fig_asym}
\end{figure*}

We calculated $\eta_{50}$ and $\xi$ for our simulated galaxies and
plot them (dots and squares) along with the observational data
(black$/$red crosses) in Fig. \ref{fig_asym}. We remind the reader
that there are some differences between the two analyses. The
underlying assumptions are the same (i.e. regular, not warped discs
and rotation maps, exclusion of mergers and systems undergoing major
interactions) but, for example, ${\rm Err}(v_{\rm gas})$ and
${\rm Err}(v_{\rm gas,flipped})$ in Eq. \ref{eq_xi} are undefined in
our analysis of EAGLE galaxies, since we rely on the SPH scheme to
directly determine the kinematic state of the gas, without performing
any emission line fitting. Instead, to obtain an estimate of the noise
consistent with SAMI data, we fit a polynomial to \citet{ho2016}'s
${\rm Err_{\rm obs}}(v_{\rm gas})-$z [kpc] scatter plot and then add
the observational noise to the simulated velocity maps. Note that the
error on $v_{\rm gas}$ increases with the distance from the galaxy
plane because the S$/$N of extraplanar H$\alpha$ emission is lower
than that of H$\alpha$ gas in the disc. Furthermore, observations only
consider ionised gas, while we take into account all (warm + hot) gas
in the galactic cubelet to better sample any outflowing activity (see
Sections \ref{total_vs_ha} and \ref{veldistr3}). Despite these
differences, the range in $\eta_{50}$ and $\xi$ is similar in
simulations and observations.

\citet{ho2016} empirically defined as {\it wind-dominated} those
galaxies with $\eta_{50}>0.3$ and $\xi>1.8$. These limits are visible
in the left panel of Fig. \ref{fig_asym} as the vertical and
horizontal black dotted lines. Accordingly, red \color{red} \textsf{x}
\color{black} symbols in the figure represent wind-dominated SAMI
galaxies ($15$ out of $40$), while black \textsf{x} symbols are
associated with ($25$) observed objects without strong wind
signatures. Among the EAGLE galaxies, just one object has
$\eta_{50}<0.3$ (and only in the yz projection value). Although this
would be consistent with a wind-dominated galaxy sample, the range in
$\xi$ does not support this conclusion. The mean asymmetry parameter
is $1.63$ (below the threshold of \citealt{ho2016}) and only five
discs have $\xi>1.8$ in both projections.

We are interested in studying the interdependency between $\eta_{50}$,
$\xi$ and relevant galactic properties. As for the observational
sample, the Spearman rank correlation test indicates no significant
correlation between the two parameters: $\rho=0.12$ with a
significance of $0.27$. \citet{ho2016} argue that if winds are the
only mechanism disturbing the extraplanar gas, then a trend between
$\eta_{50}$ and $\xi$ should be expected. The fact that both works
fail to find a significant correlation suggests that, when applied to
current data and simulations, the $\xi-\eta_{50}$ plot might not be
accurate enough for the identification of wind-dominated
galaxies. \citet{ho2016} speculate that gas accreted on to galaxies
through satellite accretion would cause a large velocity asymmetry of
the extraplanar gas without affecting much the off-plane velocity
dispersion (i.e. large $\xi$ and small $\eta_{50}$), and therefore
complicate the interpretation of the $\xi-\eta_{50}$ plot in terms of
galactic winds and eDIG. We saw in Section \ref{gal_winds} how the
extraplanar gas distribution of EAGLE objects is dominated by outflows
(the ratio of outflow to non-outflow dominated pixels is on average
$\sim3.2$). Unfortunately, this does not lead to a significant
$\xi-\eta_{50}$ correlation in the simulated sample.

However, Fig. \ref{fig_asym} highlights a different trend within EAGLE
and SAMI data. In the left panel, simulations are colour coded
according to the SFR surface density of the parent galaxy,
$\Sigma_{\rm SFR}={\rm SFR}/(2\pi r_{50}^2)$. In general, objects with
low$/$high $\xi$ have low$/$high $\Sigma_{\rm SFR}$ (i.e. bluish
points are below reddish points). This positive correlation is even
more visible in the right panel of Fig. \ref{fig_asym}, where we plot
$\xi$ as a function of $\Sigma_{\rm SFR}$. The Spearman rank
correlation test indicates a significant correlation between the two
parameters both in simulations\footnote{For each simulated galaxy,
  there are two values of $\xi$, corresponding to the edge-on
  projections xz and yz. We averaged the two values into a single one
  to calculate the asymmetry$-\Sigma_{\rm SFR}$ correlation.} (grey
dots and squares, $\rho=0.67$ with a significance of $\sim10^{-6}$)
and observations\footnote{\citet{ho2016} quoted both spectral energy
  distribution (SED) and H$\alpha$ based SFRs for their sample. Here
  we use SFR$_{\rm H\alpha}$ to calculate the observed
  $\Sigma_{\rm SFR}$. We checked that our conclusions do not change
  when using SFR$_{\rm SED}$.} (black$/$red crosses, $\rho=0.56$ with
a significance of $1.6\times10^{-4}$). Since the asymmetry parameter
marks the incidence of galactic winds in disc galaxies, this result is
in qualitative agreement with Section \ref{sigma_SFR} and the
conclusions of \citet{ho2016}: the star formation rate surface density
correlates with outflowing activity. \\

At this point, it is important to remark that numerical results could
be affected by statistical noise in various ways. For example, in low
mass, small galaxies high asymmetry could be due to poor sampling of
the extraplanar gas distribution, rather than galactic
winds. Moreover, since our discs are fairly regular, the two sets of
$\eta_{50}$ and $\xi$ parameters associated with the edge-on xz and yz
projections should contain the same information, and therefore be
statistically similar. However, the analysis is limited by the small
size of the synthetic sample ($43$ galaxies), which is due to the
small box size of the cosmological simulation used ($25$ cMpc). In
principle, increasing resolution and box size of the simulations will
alleviate the impact of statistical noise.

Note that the position of SAMI galaxies in the $\xi-\eta_{50}$ plot
can also be biased in different ways. For example, see the discussion
in \citet{ho2016} on how, for galaxies without a direct measurement of
$v_{\rm rot}$, the lack of correlation between $\xi$ and $\eta_{50}$
is partially due to the scatter in the Tully-Fisher
relation. Furthermore, the high asymmetry seen in some of the observed
galaxies might be accentuated by inclination effects (while all the
simulated galaxies are perfectly edge-on).

In summary, according to our analysis, the $\xi-\eta_{50}$ plot does
not seem to provide a clear, unambiguous tool to identify
wind-dominated galaxies. Although using simulations with higher
resolution (and better quality observational data) could improve the
results, different observables (e.g. the velocity dispersion
distribution) appear to estimate the outflowing activity of star
forming galaxies more accurately than $\xi$ and $\eta_{50}$.

\section{Conclusions}
\label{concl}

In this paper we have presented an analysis of stellar feedback-driven
galactic outflows based on the comparison between hydrodynamic
simulations and IFS observations from the SAMI survey
\citep{bryant2015}. We have extracted cubes of $30$ physical kpc
around unperturbed disc galaxies from the highest resolution
cosmological simulation of the EAGLE set \citep{schaye2015}. Our final
sample includes $43$ main sequence objects with
$9.02\le\log($M$_{\star}/$M$_\odot)\le10.76$ and
$\log(\tilde{\rm sSFR}/$yr$^{-1})=-10.04$
(Fig. \ref{fig_gal11_proj}). We have divided each cubelet into a grid
of pixel size $2$ kpc (which is comparable to the effective
resolution of SAMI) and created gas rotation velocity and velocity
dispersion maps \mbox{(Fig. \ref{fig_maps})}. In our study the terms
{\it galactic winds} and {\it outflows} are synonyms that we have used
to identify both gas in the process of leaving a galaxy and gas that
will eventually stop and fall back to the disc.

This work is the theoretical counterpart of the observational analyses
on SAMI galaxies with outflows presented in \citet{ho2014} and
\citet{ho2016}. In the first part of the paper, we have focussed on
the pixelated velocity dispersion (probability) distribution as a
tracer of galactic wind signatures. To better compare with SAMI
observations that mainly target H$\alpha$ emitting gas at T $\sim10^4$
K, we have distinguished between {\it all gas} and {\it warm gas}. The
latter identifies pixels in a velocity$/$velocity
dispersion$/$temperature map where the density weighted gas
temperature (calculated using all the particles) is in the range
$3.8\le\log\big(\langle$T$_{\rm gas}\rangle/$K$\big)\le4.2$. We have
found that, in EAGLE galaxies, warm gas traces the disc and is
virtually absent when moving away from the galaxy plane (middle two
panels of Fig. \ref{fig_maps}). For this reason, the edge-on velocity
dispersion for warm gas has a less prominent high-$\sigma$ tail (that
is mostly associated with outflows) and a lower $\sigma_{\max}$ than
the all gas distribution, while the two face-on distributions are very
similar (Fig. \ref{fig_total_vs_ha}). The lack of warm gas outside the
disc is a direct consequence of the thermal stellar feedback
implemented in EAGLE. Galactic winds in our simulations are mostly hot
(with a \mbox{T $>10^5$ K} they would not be seen in SAMI
observations), buoyant \citep[rather than ballistic, cf.][]{bower2016}
and, compared to observed galaxies, may entrain insufficient gas with
T $<10^5$ K \citep[as pointed out also by][]{turner2016}. Throughout
the paper, whenever possible and appropriate we have performed our
analysis using warm gas. However, when studying outflowing material
and galactic wind signatures we have included all gas to obtain more
reliable results.

We have targeted the galactic disc by taking into account only pixels
with abs(z) $<$ $3$ kpc in the edge-on xz projection (with the disc
lying in the xy plane). The $\sigma$-distribution peaks at $30$ km
s$^{-1}$ and then quickly declines to low fractions. The complementary
distributions (pixels with abs(z) $>$ $3$, $5$ kpc) extend to
progressively larger values and have a less prominent low-$\sigma$
peak (Fig. \ref{fig_hist_disc}). This demonstrates that the
low-$\sigma$ part of the velocity dispersion distribution is
associated mainly with the galactic disc. On the other hand,
(extraplanar) outflowing gas dominates the high-$\sigma$ tail
(Fig. \ref{fig_v_esc}). Both these results are in qualitative
agreement with the observations of \citet{ho2014}.

In general, galaxies with higher stellar mass present a more extended
$\sigma$-distribution (in both the edge-on and face-on projections),
while the specific star formation rate has a secondary effect with
respect to M$_{\star}$ (Fig. \ref{fig_hist_tot}). At fixed stellar
mass and when all gas is used, the edge-on probability distribution of
galaxies with high $\Sigma_{\rm SFR}=$ SFR$/(2\pi r_{50}^2)$ (where
$r_{50}$ is the radius within which half of the galaxy stellar mass is
included) is shifted towards larger velocity dispersions (and more
extended) than the low-$\Sigma_{\rm SFR}$ one
(Fig. \ref{fig_hist_sfrd}). As in the SAMI observations of
\citet{ho2016}, this indicates that the star formation rate surface
density correlates with the outflowing activity even at the low
$\Sigma_{\rm SFR}$ seen in our simulated galaxies:
$\log\big(\Sigma_{\rm SFR}/[$M$_\odot$ yr$^{-1}$
kpc$^{-2}]\big) \le -1.95$.

We have studied the impact of temperature on the velocity dispersion
distribution and found that there is a direct correlation between the
thermal state of the gas and its state of motion. Gas with temperature
T $<10^5$ K traces the low-$\sigma$ galactic disc, while gas with T
$\ge10^5$ K is mainly associated with higher-$\sigma$
(i.e. extraplanar, outflowing gas, Fig. \ref{fig_hist_temp_cut}). Our
results imply the following relations for low redshift disc galaxies
in EAGLE: \renewcommand{\labelitemi}{$\bullet$}
\begin{itemize}
\item Low-$\sigma$ peak $\,\,\Leftrightarrow\,\,$ galactic disc
  $\,\,\Leftrightarrow\,\,$ gas with \mbox{T $<10^5$ K};
\item High-$\sigma$ tail $\,\,\Leftrightarrow\,\,$ galactic winds
  $\,\,\Leftrightarrow\,\,$ gas with \mbox{T $\ge 10^5$ K}.
\end{itemize}

We have applied the empirical identification of wind-dominated SAMI
galaxies proposed by \citet{ho2016} to the simulated sample. The
ranges of the two parameters used to measure the strength of disc-halo
interactions, $\eta_{50}$ (velocity dispersion to rotation ratio) and
$\xi$ (asymmetry of the extraplanar gas), are similar in simulations
and observations, and most of EAGLE galaxies have $\eta_{50}>0.3$ (one
of the thresholds introduced by \citealt{ho2016} to define when
outflows become important, the other being $\xi>1.8$). Although this
would be consistent with a wind-dominated synthetic sample, the range
in $\xi$ does not support this conclusion: only five discs have
$\xi>1.8$. Moreover, the lack of a significant correlation between
$\eta_{50}$ and $\xi$ suggests that the $\xi-\eta_{50}$ plot is
currently not accurate enough to provide a clear, unambiguous way to
identify wind-dominated galaxies (left panel of
Fig. \ref{fig_asym}). However, we have found a significant correlation
between the asymmetry parameter $\xi$ and the SFR surface density of
the parent galaxy (right panel of Fig. \ref{fig_asym}) that
qualitatively confirms our result that $\Sigma_{\rm SFR}$ correlates
with the outflowing activity.


In Appendix \ref{appendix_b} we have tested the impact of different
cube sizes and grid resolutions. Increasing the pixel size from $2$ to
$3$ kpc has a minimum impact on the $\sigma$-distribution, while
changing the cube size from $30$ to $60$ kpc reduces the statistical
relevance of the high-$\sigma$ tail (Fig. \ref{fig_hist_res_test}).
Finally, in Appendix \ref{constrsim} we have tested our analysis on
idealized simulations of isolated disc galaxies
(Fig. \ref{fig_cdiscc_maps}). Results support the picture emerging
from the analysis of EAGLE galaxies, where outflows are responsible
for the high-$\sigma$ tail of the velocity dispersion distribution
(Fig. \ref{fig_hist_discs}). \\

In summary, the velocity dispersion distribution has the potential to
provide valuable information on the outflowing activity in
galaxies. Notably, shape and extension of the high-$\sigma$ tail
correlate with the strength of galactic winds. The comparison with
SAMI observations has highlighted a limitation of EAGLE's stellar
feedback: there is a dearth of cold and warm gas in the (mostly hot)
simulated galactic outflows. Our results emphasise the double benefit
of comparing simulations and observations: simulations are a valuable
tool to interpret IFS data, and detailed observations guide the
development of more realistic simulations. The next steps are to
calculate how much gas actually escapes from galaxies and probe star
formation activity and galactic winds in the time domain, following
thermodynamic and kinematic changes as they happen. This is where the
predictive power of simulations can play a pivotal role. As pointed
out by \citet{guidi2016SDSS}, it will be important moving towards an
unbiased, consistent comparison between simulated and real galaxies.

The work presented in this paper extends the investigation of
\citet{ho2014,ho2016} and contributes to placing SAMI observations of
wind-dominated galaxies in a physical context. Together, these studies
provide important constraints for ongoing IFS surveys and, in the
longer term, will help with planning for the next generation
multi-object integral field units including HECTOR, the successor of
SAMI \citep{lawrence2012,joss2015,bryant2016}.

\section*{Acknowledgements}

ET would like to thank Lemmy Kilmister for constant inspiration during
the writing of this paper and Jeremy Mould \& Paul Geil for many
insightful discussions. RAC is a Royal Society University Research
Fellow. SMC acknowledges the support of an Australian Research Council
Future Fellowship (FT100100457). Support for AMM is provided by NASA
through Hubble Fellowship grant \#HST-HF2-51377 awarded by the Space
Telescope Science Institute, which is operated by the Association of
Universities for Research in Astronomy, Inc., for NASA, under contract
NAS5-26555. The SAMI Galaxy Survey is based on observations made at
the Anglo-Australian Telescope. The Sydney-AAO Multi-object Integral
field spectrograph (SAMI) was developed jointly by the University of
Sydney and the Australian Astronomical Observatory. The SAMI input
catalogue is based on data taken from the Sloan Digital Sky Survey,
the GAMA Survey and the VST ATLAS Survey. The SAMI Galaxy Survey is
funded by the Australian Research Council Centre of Excellence for
All-sky Astrophysics (CAASTRO), through project number CE110001020,
and other participating institutions. The SAMI Galaxy Survey website
is \url{http://sami-survey.org/}. Part of this research was done using
the National Computational Infrastructure (NCI) {\it Raijin}
distributed-memory cluster and supported by the Flagship Allocation
Scheme of the NCI National Facility at the Australian National
University (ANU). For the post-processing we used the {\it Edward} and
{\it Spartan} High Performance Computing (HPC) clusters at the
University of Melbourne.  This work also used the DiRAC Data Centric
system at Durham University, operated by the Institute for
Computational Cosmology on behalf of the STFC DiRAC HPC Facility
(\url{http://dirac.ac.uk}). The DiRAC system was funded by BIS
National E-Infrastructure capital grant ST/K00042X/1, STFC capital
grants ST/H008519/1 and ST/K00087X/1, STFC DiRAC Operations grant
ST/K003267/1 and Durham University. DiRAC is part of the National
E-Infrastructure. This work was supported by STFC grant ST/L00075X/1.

\bibliographystyle{mnras}
\bibliography{tescari_sami_eagle_winds_resub2}

\appendix

\section{variations of cube size and grid resolution}
\label{appendix_b}

\begin{figure}
\centering
\includegraphics[width=8.7cm]{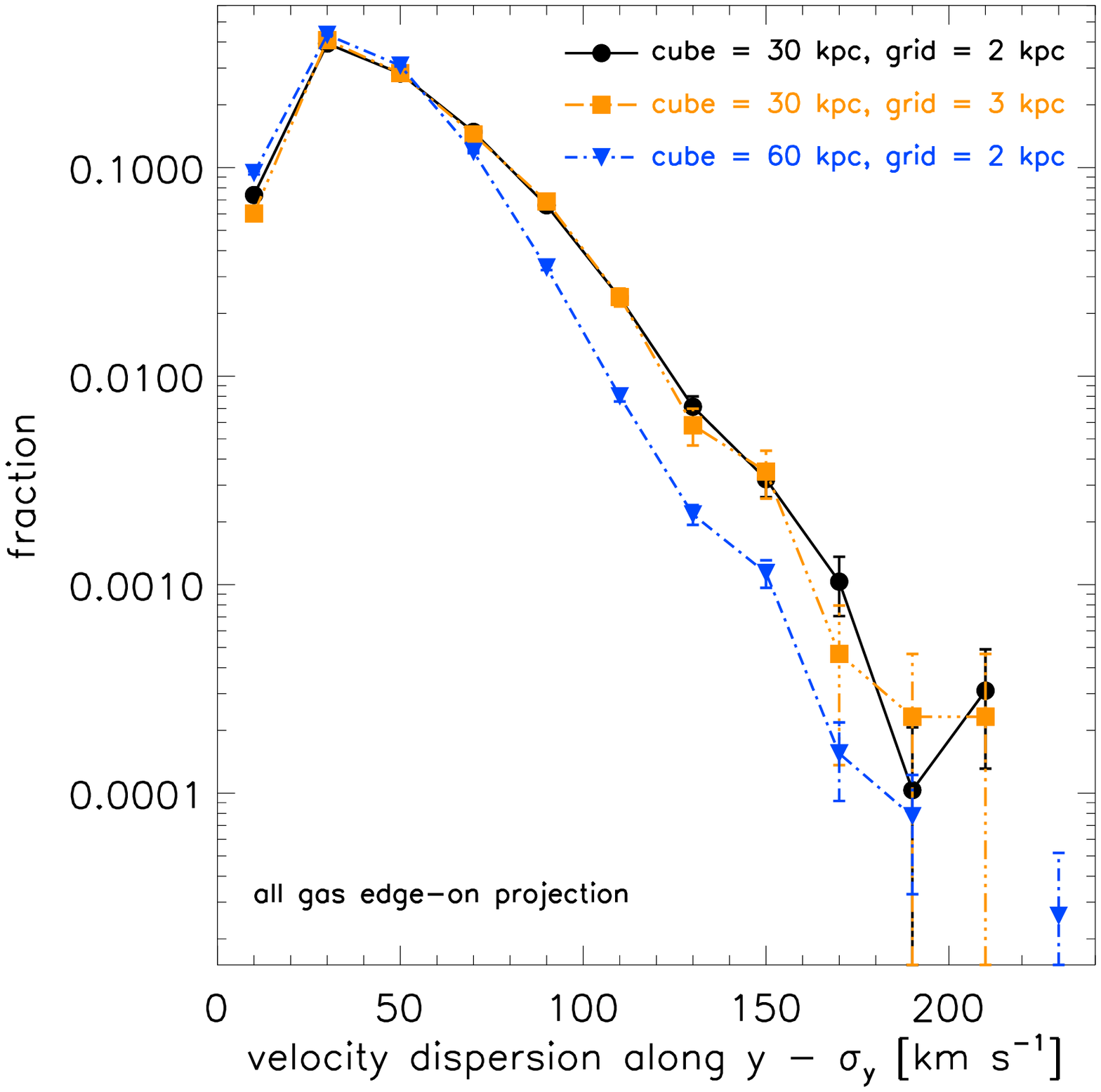}
\caption{Pixelated velocity dispersion probability distribution
  (cf. Fig. \ref{fig_total_vs_ha} and the beginning of Section
  \ref{total_vs_ha}) of galaxies: resolution tests. Black circles and
  solid line: fiducial model with cube size = $30$ kpc and pixel size =
  $2$ kpc. Orange squares and triple dot-dashed line: cube size = $30$
  kpc, \color{orange} pixel size = $3$ kpc\color{black}. Blue inverted
  triangles and dot-dashed line: \color{blue} cube size = $60$
  kpc\color{black}, pixel size = $2$ kpc. Errors are Poissonian. We
  consider all gas in the edge-on xz projection $-$ $\sigma_{\rm
    y}$. Decreasing the resolution of the grid does not affect the
  $\sigma$-distribution, while changing the cube size from $30$ to
  $60$ kpc reduces the statistical relevance of the high-$\sigma$
  tail.}
\label{fig_hist_res_test}
\end{figure}

Our galactic cubelets have a linear dimension of $30$ kpc and the
pixel size of the velocity dispersion maps is $2$ kpc. In this
appendix, we investigate the effect of different cube sizes and grid
resolutions on our analysis. The result is shown in
Fig. \ref{fig_hist_res_test}. Black circles and solid line represent
the pixelated velocity dispersion distribution for the fiducial choice
of parameters metioned above (we consider all gas in the edge-on xz
projection $-$ $\sigma_{\rm y}$).

To obtain the distribution described by the orange squares and triple
dot-dashed line, we kept fixed the cube size to $30$ kpc and increased
the pixel size to $3$ kpc. From Fig. \ref{fig_hist_res_test}, it is
clear that the decrement in grid resolution does not affect much the
new $\sigma$-distribution, which is essentially a smoothed version of
the fiducial model.

On the other hand, changing the cube size from $30$ to $60$ kpc (while
keeping the grid resolution fixed to $2$ kpc) has a large impact (blue
inverted triangles and dot-dashed line). The new velocity dispersion
distribution has a similar low-$\sigma$ part as the previous two, but
declines more rapidly at $\sigma_{\rm y}>70$ km s$^{-1}$. This is due
to the fact that the larger cubes include gas in the outskirts of
galaxies, where the velocity dispersion is generally small. The
additional pixels contribute mostly to the low-$\sigma$ part, while
the increased total number of bins reduces the statistical relevance
of the high-$\sigma$ tail.

\section{testing the methodology with idealized
  simulations of disc galaxies}
\label{constrsim}

\begin{figure*}
\centering 
\includegraphics[width=4.37cm, height=3.77cm]{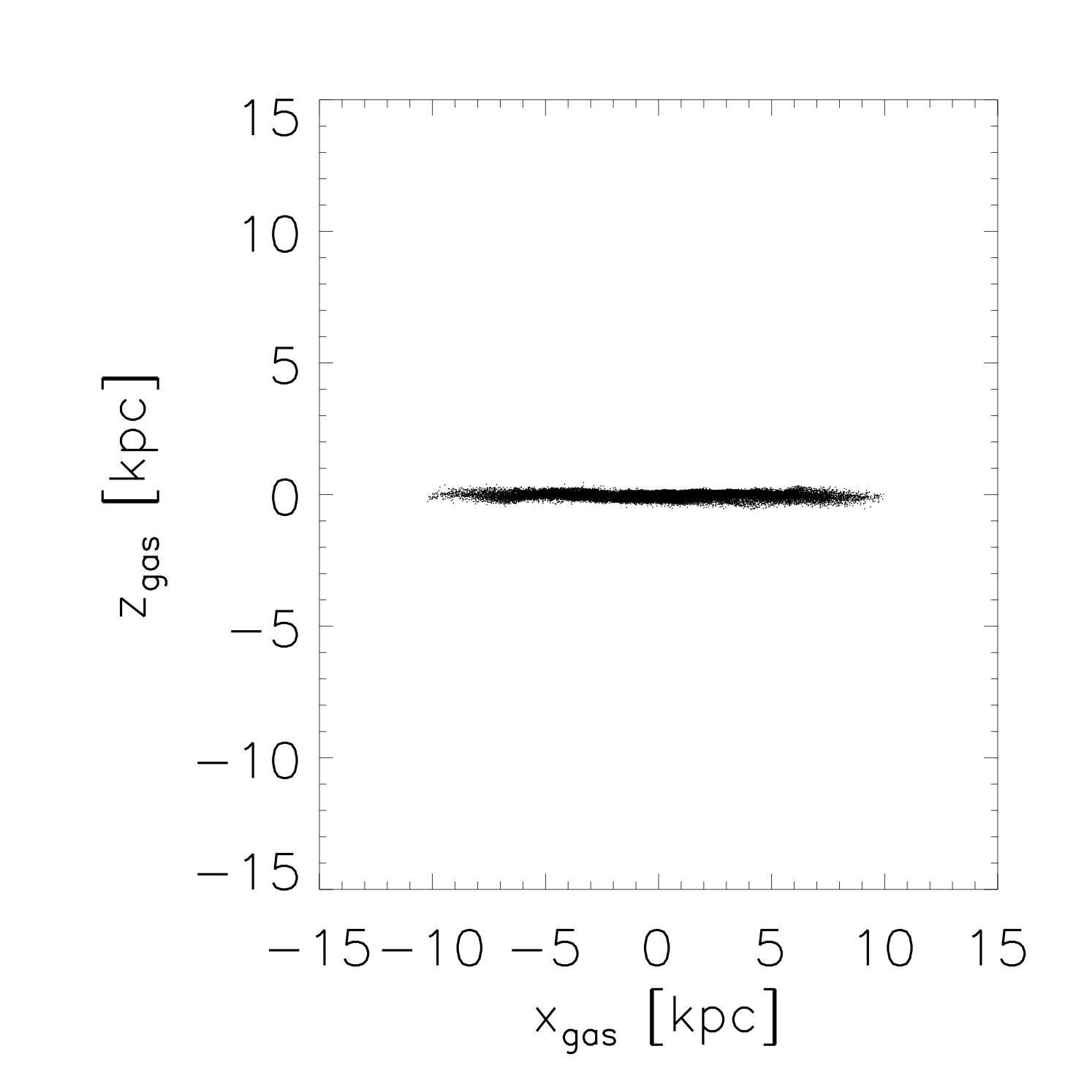}
\includegraphics[width=4.37cm, height=3.77cm]{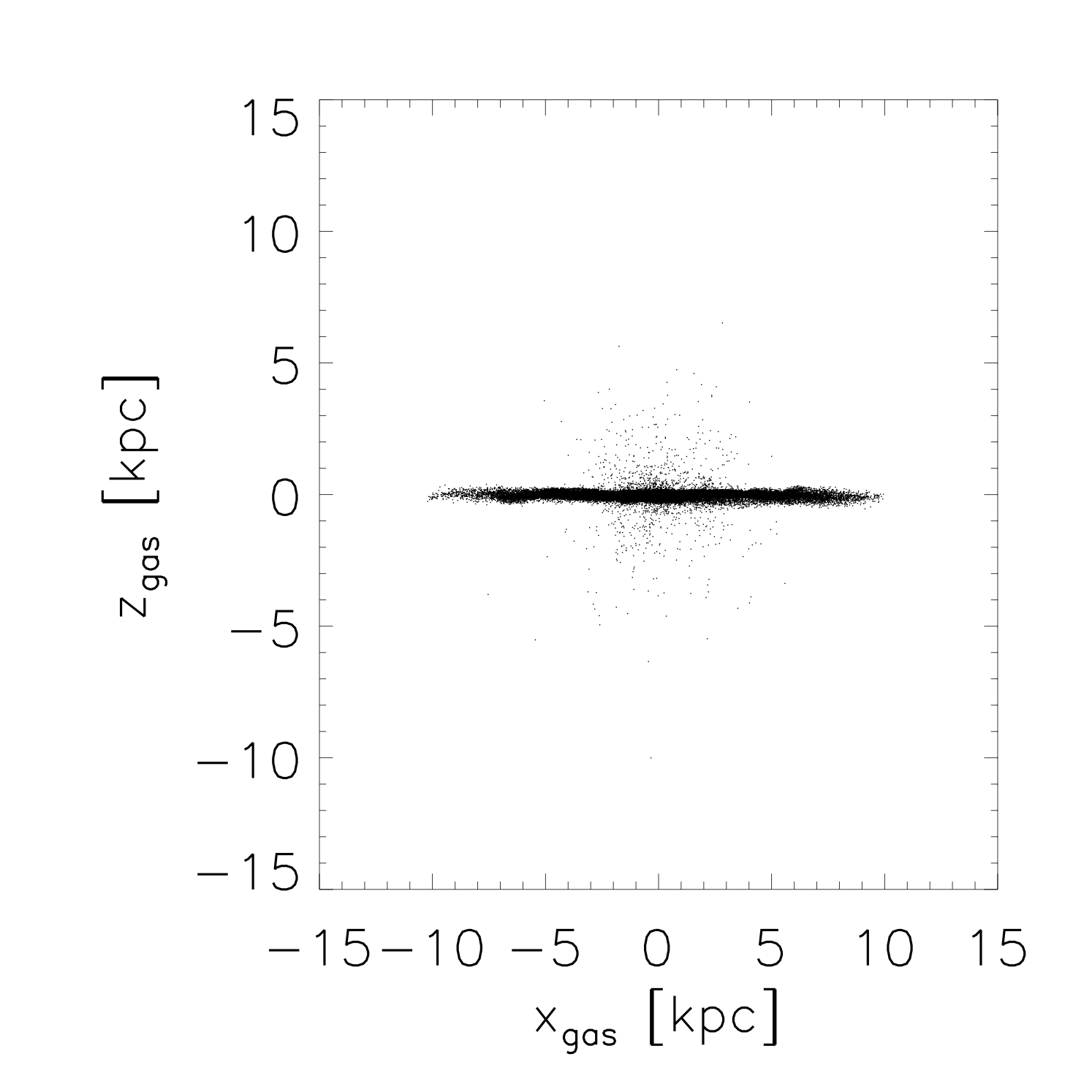}
\includegraphics[width=4.37cm, height=3.77cm]{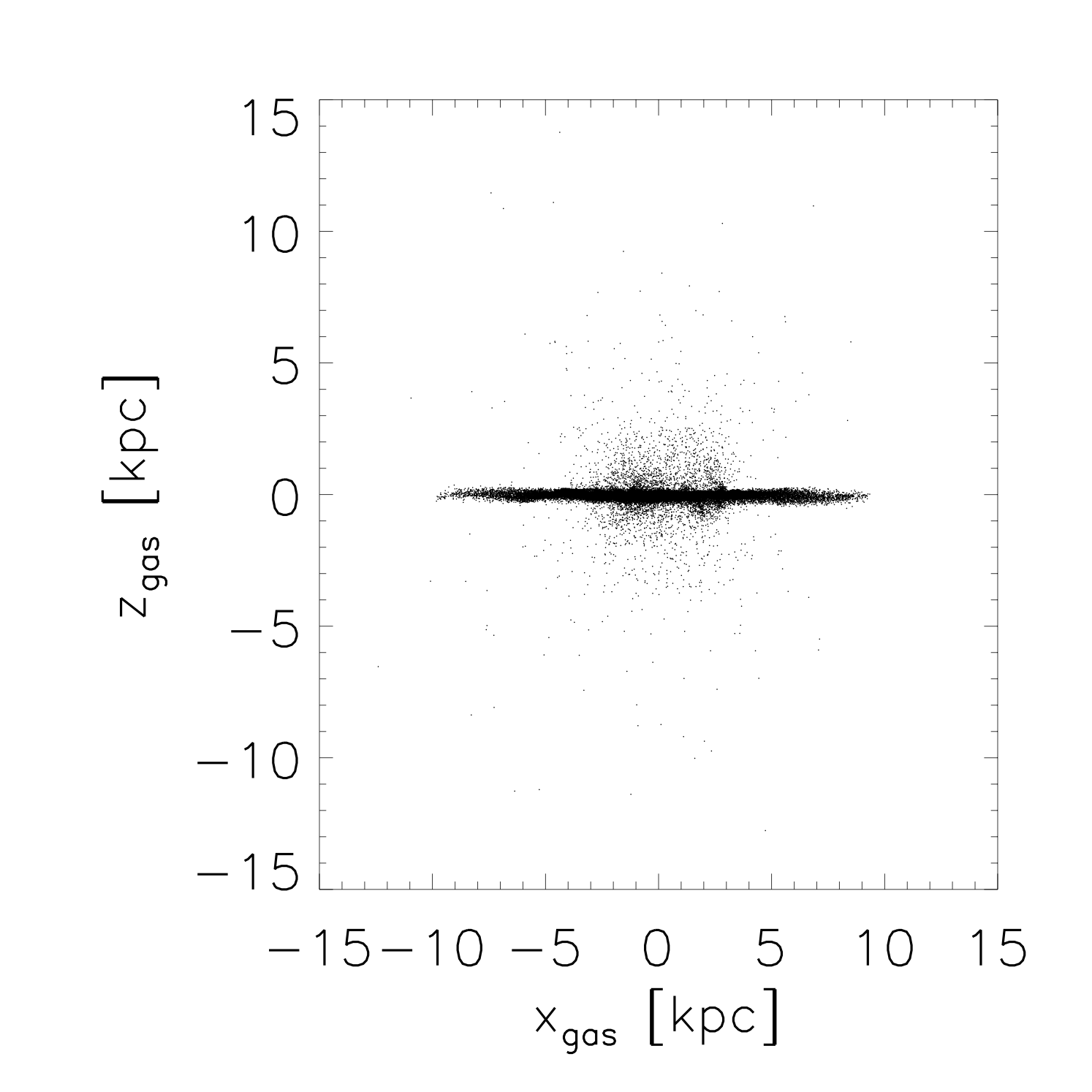}
\includegraphics[width=4.37cm, height=3.77cm]{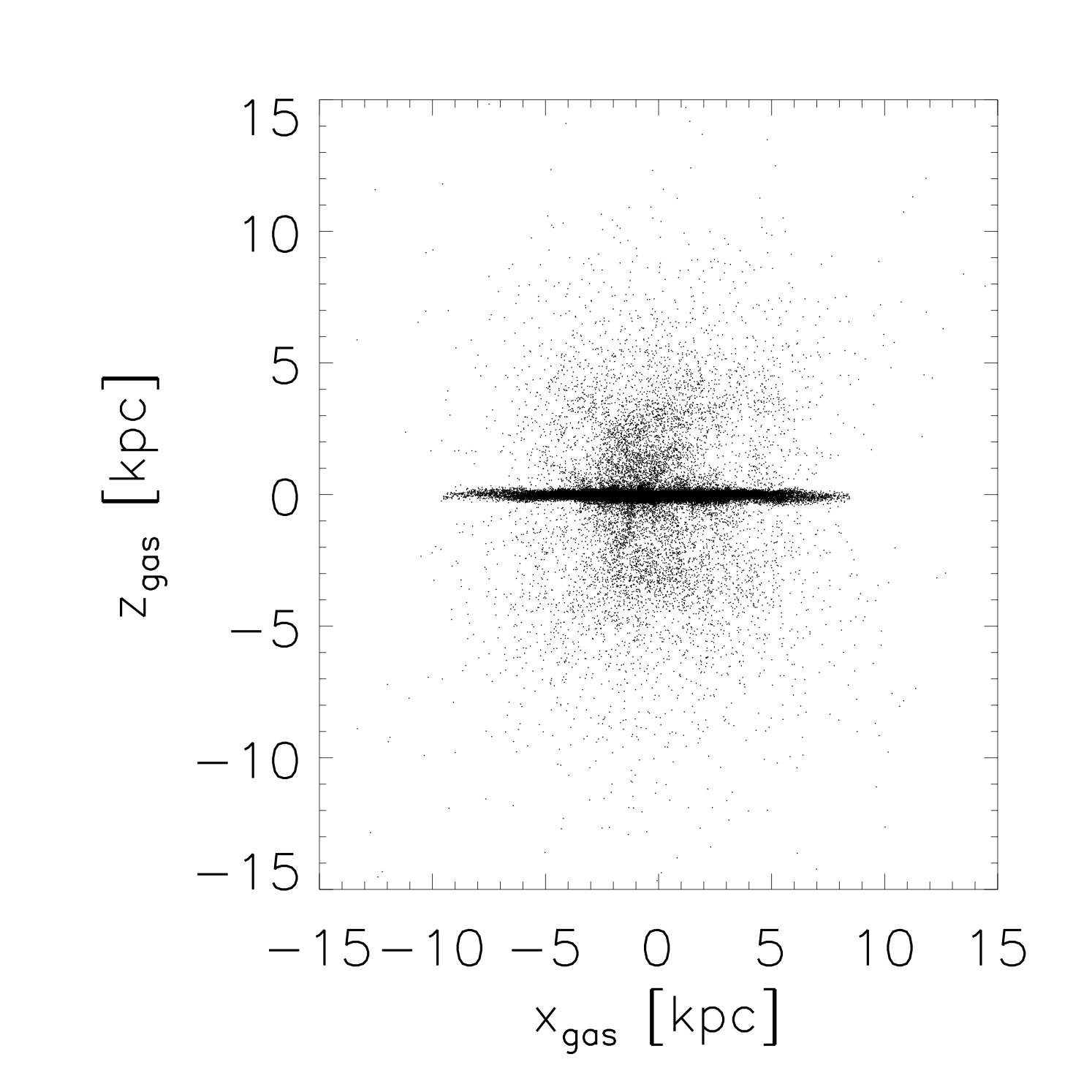}

\vspace{0.25cm}

\includegraphics[width=4.37cm, height=3.97cm]{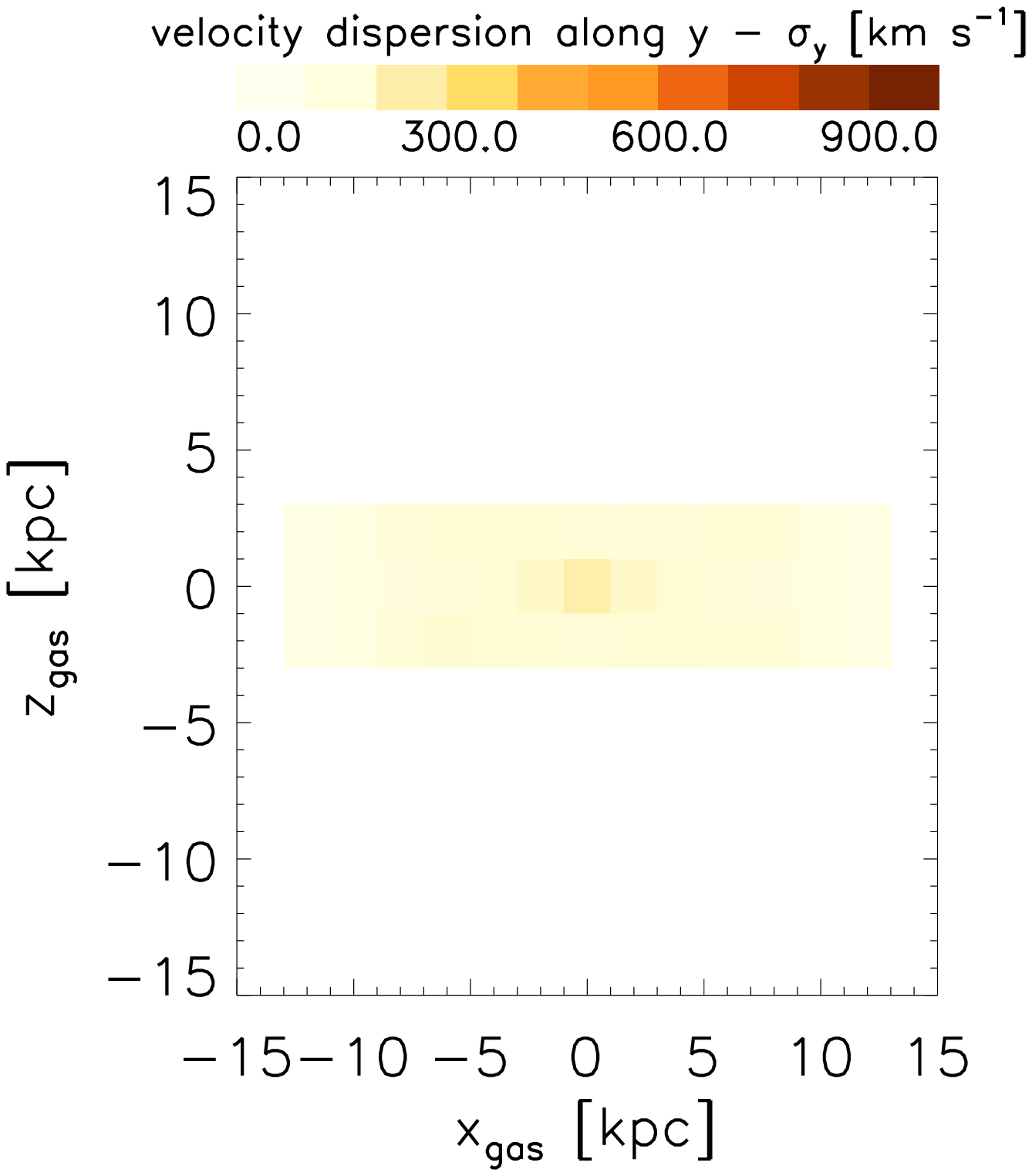}
\includegraphics[width=4.37cm, height=3.97cm]{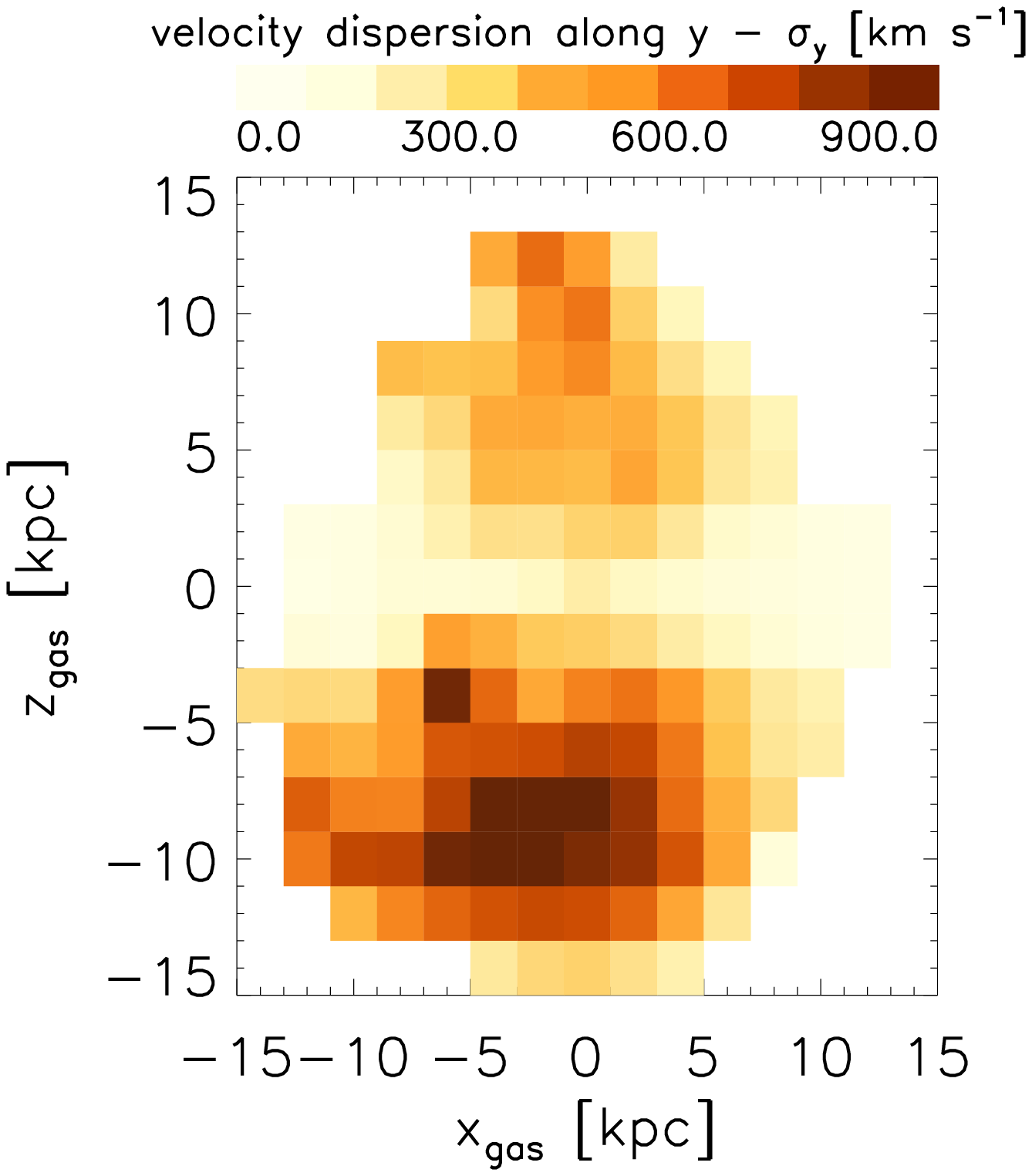}
\includegraphics[width=4.37cm, height=3.97cm]{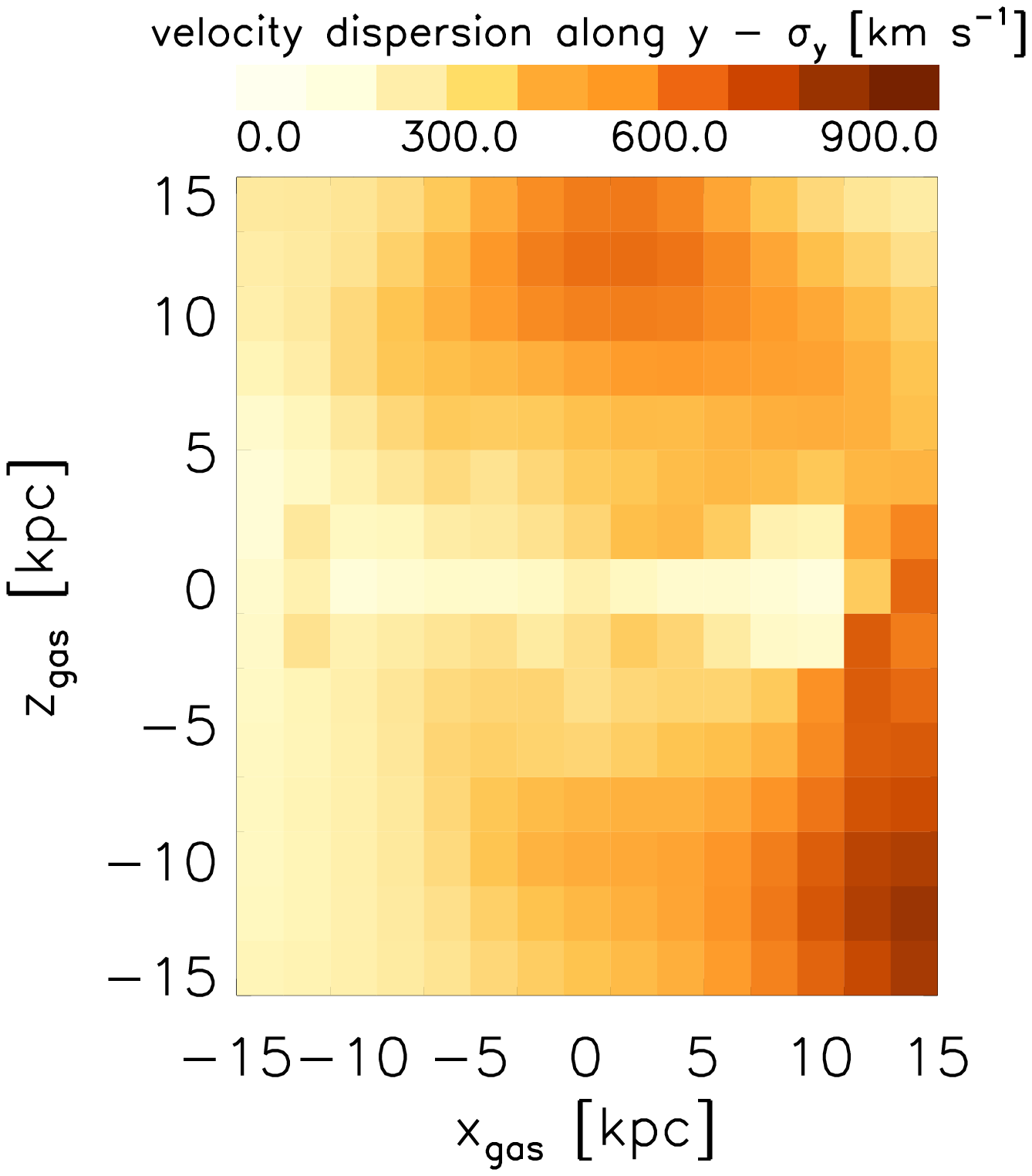}
\includegraphics[width=4.37cm, height=3.97cm]{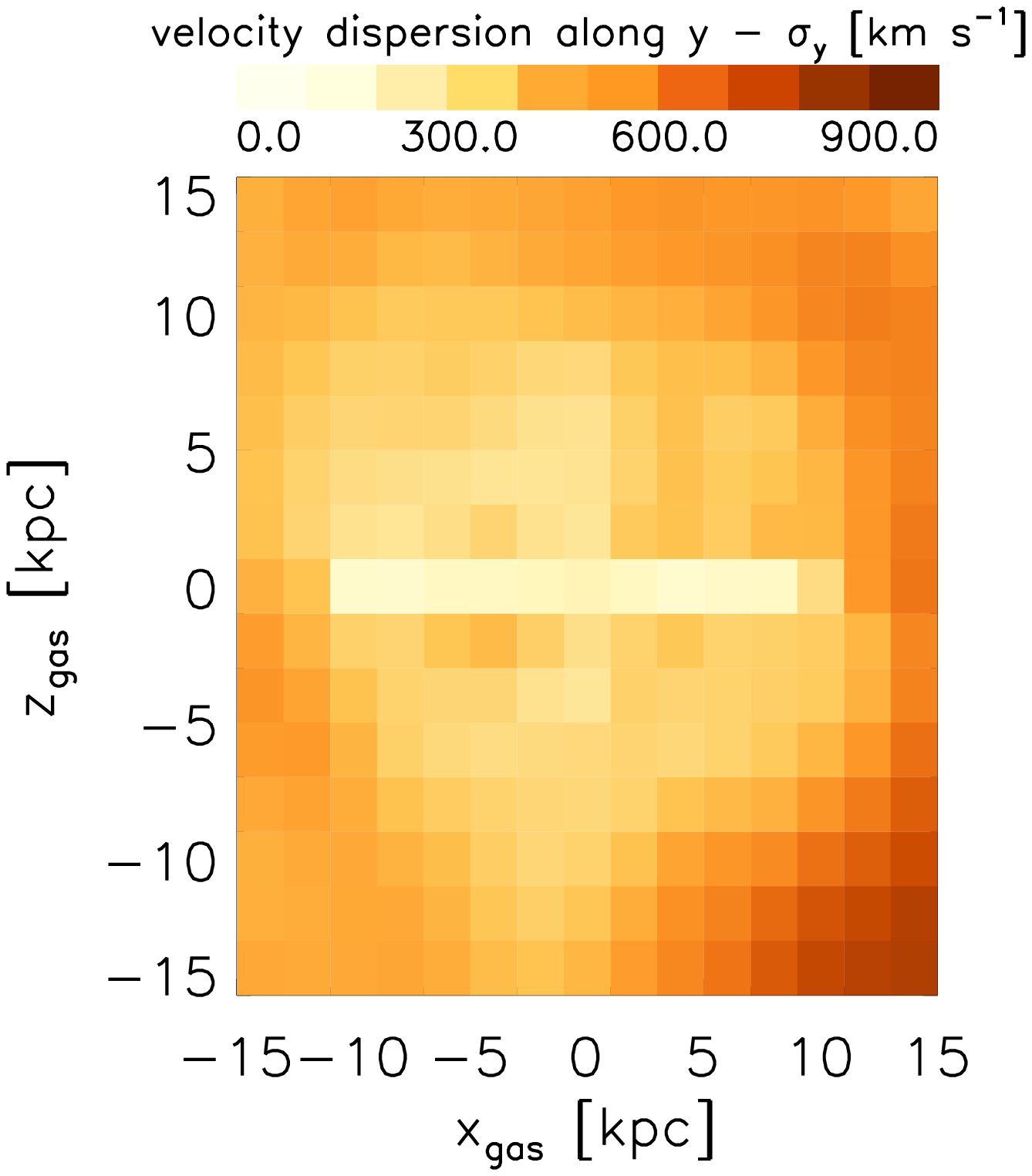}
\caption{{\bf First row}: positions in physical kpc of gas particles
  for our constrained disc galaxies. {\bf Second row}: corresponding
  gas velocity dispersion maps. From left to right: the first column
  shows the no feedback run; second column $-$ weak feedback run;
  third column $-$ fiducial model; fourth column $-$ strong feedback
  run. In each panel, we consider all gas in the edge-on xz projection
  $-$ $\sigma_{\rm y}$.}
\label{fig_cdiscc_maps}
\end{figure*}

Throughout this work, we have analysed galaxies extracted from one
particular run of the EAGLE simulations:
{\textit{Recal-L025N0752}}. As discussed in Section \ref{conf}, the
L025N0752 configuration has the highest resolution of the set, and
therefore comes in only two slightly different setups: \textit{Ref}
and \textit{Recal}. Fully exploring the parameter space around the
fiducial model would have been numerically too demanding, and was done
instead using configurations with lower resolution \citep[L050N0752
and L025N0376, see][]{crain2015}. Consequently, we have not been able
to study the impact of variations in the feedback strength, which are
important for interpreting real data and also validating our previous
results.

For this reason, we ran four simulations of an isolated disc
galaxy. We used a different version of the {\small{GADGET-3}} code
that includes smoothed particle hydrodynamics with a higher order
dissipation switch (SPHS) than classic SPH. The main advantage of SPHS
is that it suppresses spurious numerical errors in the calculation of
fluid quantities before they can propagate \citep{hobbs2013,
  power2014}. At temperatures T$_{\rm gas}\ge10^4$ K, the gas is
assumed to be of primordial composition and cools radiatively
following \citet{KWH}. At T$_{\rm floor}=100$ K $\le$
T$_{\rm gas}<10^4$ K, the gas cools following the prescription of
\citet{mash2008} for gas of solar abundance. Gas is also prevented
from cooling to the point at which the Jeans mass for gravitational
collapse becomes unresolved. Gas above a fixed density threshold forms
stars with a star formation efficiency of $10$\%
\citep{power2016}. The subgrid model for star formation is calibrated
to follow the Kennicutt-Schmidt law \citep{kennicutt1998}.

\begin{figure}
\centering
\includegraphics[width=8.37cm]{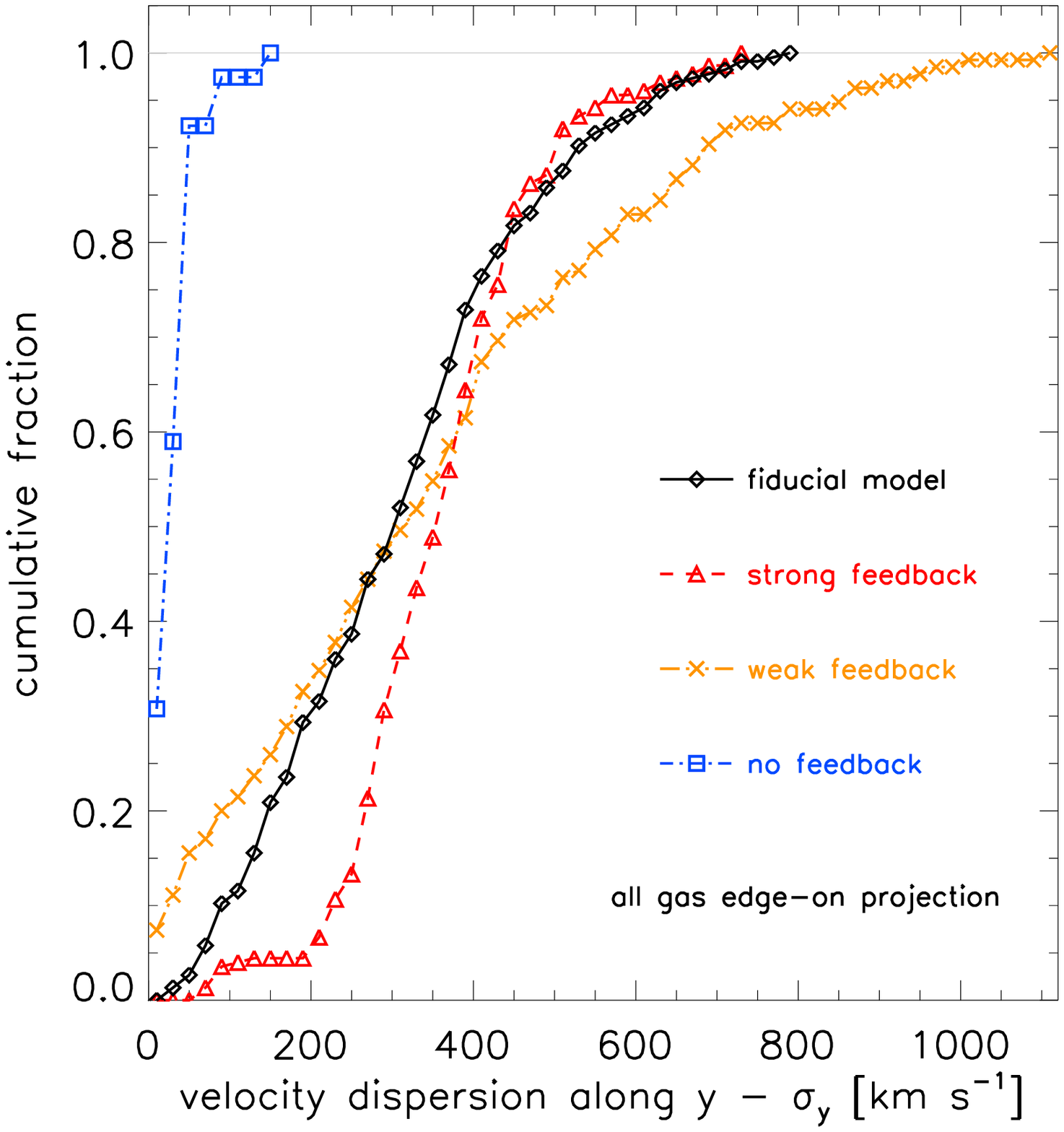}
\caption{Cumulative fraction of the pixelated velocity dispersion
  probability distribution (cf. Fig. \ref{fig_total_vs_ha} and the
  beginning of Section \ref{total_vs_ha}) for our constrained disc
  galaxies. Black diamonds and solid line: fiducial model. Red
  triangles and dashed line: strong feedback. Orange crosses and
  triple dot-dashed line: weak feedback. Blue squares and dot-dashed
  line: no feedback. The thin grey horizontal line marks the
  saturation point of each distribution (cumulative fraction $=1$). We
  consider all gas in the edge-on xz projection $-$ $\sigma_{\rm
    y}$. As soon as feedback is introduced, the distributions shift to
  larger velocity dispersions.}
\label{fig_hist_discs}
\end{figure}

The four {\it constrained} discs are identical in all but the strength
of stellar feedback. The latter is quantified by the feedback factor
$f_{\rm f}$, which simply (linearly) scales the amount of thermal
energy deposited by supernovae into neighbouring gas particles. To
calculate the energy released by SNe at any given time (where only SNe
II are considered and $E_{\rm SN} = 10^{51}$ erg), the code integrates
over a \citet{salpeter55} IMF in the range $8-100$ M$_{\odot}$ (to
determine the number of SNe II) and adopts an approximate
main-sequence time of $t_{\rm MS}\approx11.8$ Myr. At $t_{\rm MS}$
after the formation of the star particle, the energy injection is
implemented as a delta function in time \citep{hobbs2013}. We varied
the feedback factor from zero to strong feedback ($f_{\rm f}=1.50$),
the fiducial model being the one with $f_{\rm f}=1.00$ (see Table
\ref{tab_asym_discs}). Each galaxy is composed of a (gas + stellar)
disc of radius $10$ kpc and a stellar bulge embedded in a DM halo
(with a NFW $-$ \citealt{NFW1996} $-$ concentration parameter $c=10$),
sampled with $308,012$ total particles (of which
N$_{\rm DM} = 100,000$). The mass and spatial resolutions are
$1.6\times10^5$ M$_{\odot}$ and $0.01$ kpc, respectively. All systems
rotate (the DM halo spin parameter is $0.04$) and have the same total
mass M$_{\rm tot}=2.57\times10^{12}$ M$_{\odot}$. Gas and stellar
masses change according to $f_{\rm f}$ (as well as N$_{\rm gas}$ and
N$_{\star}$), with $\langle$M$_{\star}\rangle=7.42\times10^{10}$
M$_{\odot}$. We stress that these discs are not meant to faithfully
reproduce physical and morphological properties of the EAGLE galaxies
previously used. Our main goal here is testing the methodology.

\begin{table}
\centering
\begin{tabular}{llcc}
  \\ \hline & Run & Feedback & $\max(\sigma_{\rm y})$ \\
            & & factor ($f_{\rm f}$) & [km s$^{-1}$] \\ \hline
            & strong feedback & 1.50 & 738.74 \\
            & fiducial model & 1.00 & 786.67 \\
            & weak feedback & 0.75 & 1117.06 \\
            & no feedback & 0.00 & 157.30 \\
  \hline \\
\end{tabular}
\caption{Run name, feedback factor $f_{\rm f}$ (that regulates the amount of
  thermal energy injected by SNe into nearby gas particles) and
  gas maximum velocity dispersion $\sigma_{\rm y}$ for the edge-on
  projection of our idealized disc galaxies.}
\label{tab_asym_discs}
\end{table}

In the first row of Fig. \ref{fig_cdiscc_maps} we plot positions of
gas particles for each single disc. From left to right, the following
runs are shown: no feedback, weak feedback, fiducial model and strong
feedback. We consider all gas in the edge-on xz projection. All
systems are plotted at the same evolutionary stage of $t\sim50$
Myr. As expected, the gas disc becomes more and more perturbed and the
amount of outflowing material increases when moving from the no
feedback to the strong feedback case. This trend is partially visible
also in the second row of Fig. \ref{fig_cdiscc_maps}, where we show
the corresponding gas velocity dispersion maps. It is interesting to
note how the weak feedback run produces the largest velocity
dispersions, $\max(\sigma_{\rm y})\approx1120$ km s$^{-1}$ (see Table
\ref{tab_asym_discs}). This somewhat counterintuitive result is due to
the fact that, compared to the fiducial and strong feedback
simulations, only a few gas particles left the disc at the
evolutionary stage considered and therefore the weak feedback
$\sigma$-map is subjected to large statistical fluctuations.

In Fig. \ref{fig_hist_discs} the cumulative pixelated velocity
dispersion probability distributions of the constrained discs are
visible. We consider all gas in the edge-on xz projection $-$
$\sigma_{\rm y}$. The no feedback run quickly saturates to one at
$\sigma_{\rm y}=150$ km s$^{-1}$. In this case, only the disc
component is present (according to the leftmost panel in the second
row of Fig. \ref{fig_cdiscc_maps}). This $\sigma$-distribution is
statistically different from the other three. As soon as feedback is
introduced, the distributions shift to progressively larger velocity
dispersions following the increase in feedback strength. This is
particularly visible up to $\sigma_{\rm y}\sim 400$ km s$^{-1}$, where
there is a crossover between the weak feedback (orange crosses and
triple dot-dashed line) and the strong feedback (red triangles and
dashed line) runs for the reason explained in the previous paragraph.

These trends fully support the low-$\sigma$ $\Leftrightarrow$ disc +
high-$\sigma$ $\Leftrightarrow$ outflows correlations emerging from
the analysis of EAGLE galaxies presented in the paper. We refrained
from calculating the $\xi$ and $\eta_{50}$ parameters for the
constrained discs because the results would be completely dominated by
statistical noise$/$fluctuations.

\label{lastpage}
\end{document}